%% file: main_DF.tex
\documentclass[onecolumn,preprint,preprintnumbers,amsfonts,amsmath,amssymb,nofootinbib,superscriptaddress]{revtex4}
\preprint{MIT-CTP/4616, MITP/14-100}

\pdfoutput=1
\usepackage{amsmath}
\usepackage{amsfonts}
\usepackage{amssymb}
\usepackage{array}
\usepackage{latexsym}
\usepackage{graphicx}
\usepackage{epstopdf}
\usepackage[english]{babel}
\usepackage{listings}
\usepackage{natbib}
\usepackage{url}
\usepackage{bbold}
\usepackage{bm}
\usepackage{hyperref}
\usepackage{color}
\usepackage{comment}
\usepackage{longtable}

\excludecomment{toexclude}
\includecomment{toinclude}

\newcommand{\be}{\begin{equation}}
\newcommand{\ee}{\end{equation}}
\newcommand{\ba}{\begin{array}}
\newcommand{\ea}{\end{array}}
\newcommand{\bqa}{\begin{eqnarray}}
\newcommand{\eqa}{\end{eqnarray}}
\renewcommand{\d}{\mathrm{d}}

\newcommand{\GeV}{\text{GeV}}
\newcommand{\gev}{\text{GeV}}
\newcommand{\mev}{\text{MeV}}

\usepackage{comment}

\def\bea{\begin{eqnarray}}
\def\eea{\end{eqnarray}}
\def\ltap{\ \raise.3ex\hbox{$<$\kern-.75em\lower1ex\hbox{$\sim$}}\ }
\def\gtap{\ \raise.3ex\hbox{$>$\kern-.75em\lower1ex\hbox{$\sim$}}\ }
\def\lsim{\ \raise.3ex\hbox{$<$\kern-.75em\lower1ex\hbox{$\sim$}}\ }
\def\gsim{\ \raise.3ex\hbox{$>$\kern-.75em\lower1ex\hbox{$\sim$}}\ }

\usepackage{slashed}

\usepackage{color}

\begin{document}

\title{Signals of a Light Dark Force in the Galactic Center}
\date{\today}

\author{Jia Liu}
\email{liuj@uni-mainz.de}
\affiliation{PRISMA Cluster of Excellence and Mainz Institute for Theoretical Physics, Johannes Gutenberg University, 55099 Mainz, Germany}
\affiliation{Center for Cosmology and Particle Physics, Department of Physics, New York University, New York, NY 10003, USA}

\author{Neal Weiner}
\email{nw32@nyu.edu}
\affiliation{Center for Cosmology and Particle Physics, Department of Physics, New York University, New York, NY 10003, USA}

\author{Wei Xue}
\email{weixue@mit.edu}
\affiliation{Center for Theoretical Physics, Massachusetts Institute of Technology, Cambridge, MA 02139, USA}

\begin{abstract}
Recent evidence for an excess of gamma rays in the GeV energy range about the Galactic Center have refocused attention on models of dark matter in the low mass regime ($m_\chi \lesssim m_Z/2$). Because this is an experimentally well-trod energy range, it can be a challenge to develop simple models that explain this excess, consistent with other experimental constraints.
We reconsider models where the dark matter couples to dark photon, which has a weak kinetic mixing to the Standard Model photon, or scalars with a weak mixing with the Higgs boson. We focus on the light  ($\lesssim 1.5 \gev$) dark mediator mass regime. Annihilations into the dark mediators can produce observable gamma rays through decays to $\pi^0$, through radiative processes when decaying to charged particles ($e^+e^-, \mu^+\mu^-,...$), and subsequent interactions of high energy $e^+e^-$ with gas and light. However, these models have no signals of $\bar p$ production, which is kinematically forbidden. We find that in these models, the shape of resulting gamma-ray spectrum can provide a good fit to the excess at Galactic Center. We discuss further constraints from AMS-02 and the CMB, and find regions of compatibility.

\end{abstract}
 \maketitle


\input{Introduction}
\input{Model}

\input{fitting}

\input{Constraints}

\input{AMS02}

\input{CMB}

\input{conclusion}

\begin{acknowledgments}
We thanks Tracy Slatyer, Jesse Thaler, Alfredo Urbano, Daniele Gaggero,
Satyanarayan Mukhopadhyay for useful discussion. NW is supported by the NSF under grants PHY-0947827 and PHY-1316753. JL is supported by the PRISMA Cluster of Excellence and the DFG Grant KO4820/1-1.
\end{acknowledgments}

\appendix

\input{branchingratios}
\input{photon-spectrum}

\input{spectrum2}

\begin{toexclude}

\end{toexclude}

\newpage

\bibliography{ref}
\bibliographystyle{h-physrev}

\end{document}

%% file: Introduction.tex
\section{Introduction}
The search for dark matter~(DM) remains one of the cornerstone components in the
search for physics Beyond the Standard Model~(BSM). While arguments of naturalness,
both of the weak scale and the QCD $\theta$-parameter point us to new physics,
DM remains unique in being an {\em experimental} indication of new physics, and
likely of a particle type.\footnote{Neutrino physics also provides an
experimental motivation for new physics, but with the most natural scale for
the new physics near the GUT scale, at least with our current understanding.}
DM appears within many BSM scenarios, with candidates such as the axion and
the WIMP well explored in their potential signals. If DM is one of these
candidates, these signals make the prospect of discovering the particle
nature not only exciting, but possible.

A great effort has been undertaken to do this, especially for the broad ``WIMP"
and WIMP-like particles, with masses in the $1- 1000$ \gev\, range, and with
interaction strengths characterized by the weak scale. The standard set of
searches - nuclear recoil, missing energy, cosmic ray - have shown a diverse
set of anomalies\cite{Angloher:2011uu,Aalseth:2010vx,Knodlseder:2005yq,Adriani:2008zr,Accardo:2014lma,Bernabei:2010mq} which have been interpreted as various
DM candidates. For many of these anomalies, systematics have shown up
\cite{Angloher:2011uu, Aalseth:2010vx}, others have stayed, but with strong
alternative hypotheses \cite{Knodlseder:2005yq,Adriani:2008zr,Accardo:2014lma}, while others persist with neither clear resolution, nor viable alternatives \cite{Bernabei:2010mq}.

Of late, a particular candidate signal has been growing in significance - both
statistically and systematically. Originally argued by Hooper and Goodenough
\cite{Goodenough:2009gk}, a component of the gamma ray signal from the vicinity
of the Milky Way's center could be explained by DM. While the candidates have
varied somewhat (from a $\sim$ 7 \gev\, WIMP annihilating to
$\tau \bar \tau$ to a $\sim$ 35 \gev\, WIMP annihilating to $b \bar b$),
the signal has been relatively persistent, peaking in $E^2 dN/dE$ near 2 \gev
\cite{Hooper:2010mq,Hooper:2010im,Han:2012au, Abazajian:2012pn, Huang:2013pda,Gordon:2013vta, Abazajian:2014fta, Daylan:2014rsa, Zhou:2014lva,Calore:2014xka}.

Hooper et al \cite{Daylan:2014rsa} argue for and explanation of a
35 GeV WIMP annihilating to $b \bar b$, claiming that such a scenario is quite simple. Moving beyond this narrative to simplified models provides more information \cite{Boehm:2014hva, Huang:2013apa,Alves:2014yha,Berlin:2014tja}. However,  UV-complete models that respect the
low energy constraints from direct detection and colliders (e.g., \cite{Ipek:2014gua})
are often more complicated and constrained than these simple descriptions would suggest. Moreover, other indirect detection constraints should be considered here~\cite{Bringmann:2014lpa,Cirelli:2014lwa,Abdo:2010ex,GeringerSameth:2011iw,Ackermann:2013yva,Ng:2013xha}.
Recent studies of anti-proton constraints
\cite{Bringmann:2014lpa,Cirelli:2014lwa} would show that these hadronic models
are already under serious pressure by the data, although we note a conflicting interpretation of the anti-proton data \cite{Hooper:2014ysa}. This has prompted an explosion of models with a variety of features \cite{Anchordoqui:2013pta,Modak:2013jya, Guo:2014gra, Yu:2014pra, Cahill-Rowley:2014ora, Borah:2014ska, Banik:2014eda, Okada:2014usa, Cheung:2014lqa, Basak:2014sza, Berlin:2014pya, Ghosh:2014pwa,Ko:2014gha,Balazs:2014jla,Agrawal:2014una,Agrawal:2014aoa,Izaguirre:2014vva,Cerdeno:2014cda,Ipek:2014gua,Boehm:2014bia,
Wang:2014elb,Fields:2014pia,Arina:2014yna,Huang:2014cla,
Ko:2014loa,Cao:2014efa,Ghorbani:2014qpa,Heikinheimo:2014xza,Cheung:2014tha}. Recently, \cite{Agrawal:2014oha} have argued that the uncertainties also admit heavier models.

There is an exceedingly simple framework to explain the excess that
manifestly avoids a number of constraints \cite{Hooper:2012cw}, and helps us understand why the scale of these models may be low, and yet so far elusive.
The idea builds on the idea
of DM with cascade annihilations into a dark force carrier
\cite{Finkbeiner:2014sja,Pospelov:2007mp,ArkaniHamed:2008qn,Pospelov:2008jd,Mardon:2009rc,Gabrielli:2013jka}. In these scenarios DM is charged under a
``dark'' U(1) \cite{Holdom:1985ag,Boehm:2002yz,Boehm:2003hm}, which kinetically mixes with the SM, or if DM couples to a dark scalar, which mixes with the Higgs.
DM annihilates via $\chi \chi \rightarrow \phi_\mu \phi_\mu$ followed by
$\phi_\mu \rightarrow SM$, yielding significant cosmic ray signals are possible,
without immediate constraints from colliders. Instead, the terrestrial
constraints come from low energy, high luminosity experiments, such as
APEX \cite{Abrahamyan:2011gv}, MaMi \cite{Merkel:2011ze}, broad constraints from
BaBar \cite{Lees:2014xha}, CLEO \cite{Love:2008aa}, and future experiments
\cite{hpsweb, Kahn:2013nma}.

In this paper, we will revisit this scenario, focusing on the ``light'' mediator
window (i.e., $m_\phi \lesssim 1.5 \gev$) proposed in \cite{Hooper:2012cw}, which
is less constrained than the case with heavier mediators, which has also been
explored elsewhere \cite{Martin:2014sxa,Abdullah:2014lla,Berlin:2014pya,Cline:2014dwa}. In this
window, gamma rays from the Galactic Center can come either from ``prompt''
photons (from $\pi^{0}$'s in the decay of the $\phi$) or radiatively
(from final state radiation or internal bremsstrahlung in e.g.,
$\phi \rightarrow e^+e^-$), or from subsequent interactions (such as ICS, Inverse Compton Scattering).

In section \ref{sec:model}, we will restate the model. In section \ref{sec:fitting}, we discuss the parameter space where the dark mediator can explain the Galactic Center excess.
In section \ref{sec:constraint}, we discuss connections to other experiments and in section \ref{sec:conc}, we conclude.

%% file: Model.tex
\section{A new Dark Force}
\label{sec:model}

The class of models we consider in this article consists of a DM particle $\chi$ and a
dark force $\phi$ with a mass $\mathrm{MeV}  <m_\phi< \mathrm{GeV}$, which is lighter than the DM mass, $m_\chi$.
The DM has a dominant annihilation process, $\chi + \chi \rightarrow \phi + \phi$, followed by
cascade decays of the dark force to the Standard Model particles.
We consider the dark force to be either a gauge field $\phi_\mu$ or a scalar field $\phi_0$.
Generically, we will use $\phi$ to denote mediator without regard to its spin.

With a  $U(1)_D$ gauge field as a dark force, the models are quite simple. With a dark photon field strength strength $\phi_{\mu \nu}$, we have kinetic mixing with Standard Model hypercharge
$Y_{\mu \nu}$,
\begin{equation}
   - \frac{\tilde{\epsilon}}{2}  \phi_{\mu \nu }  Y^{\mu \nu}.
\end{equation}
At low energy, the mixing occurs with the EM field strength, and the cascade decay is triggered by the coupling of dark force and the Standard Model currents
\begin{equation}
   \mathcal{L}_{int} \simeq - \tilde{\epsilon} \cos \theta_w \phi_\mu J_{em}^\mu = - \epsilon \phi_\mu 
      J_{em}^\mu
\end{equation}
where $\epsilon \equiv \tilde{\epsilon} \cos \theta_w$ to simplify the notation.


For a detectable signal, we must have a present day annihilation rate of $\langle\sigma v\rangle\sim 10^{-26}{\rm cm^3 s^{-1}}$.
For a vector dark force, we take the DM to be a Dirac fermion.\footnote{Alternatively, we can consider a pseudo-Dirac fermion, in which case the ``thermal'' cross section is naturally a factor of two larger $\langle\sigma v\rangle \approx 6 \times 10^{-26}{\rm cm^3 s^{-1}}$. See the discussion in \cite{Schutz:2014nka}.}
The cross section for DM-DM annihilation is s-wave,
\begin{equation}
   \sigma v_{ \chi\chi \rightarrow \phi \phi} \simeq \frac{g_X^4} { 16 \pi m_\chi^2}
      \frac{\left( 1- x\right)^{3/2} } { \left( 1 - \frac{x}{2} \right)^2 }  \ ,
\end{equation}
where $g_X$ is the gauge coupling of the dark force, and $x=  m_\phi^2  / m_\chi^2 $.

In the case of a scalar dark force, we can take a real scalar to be the dark force ($\phi$) and a complex scalar as DM ($\chi$). The potential for the scalar dark force is

\begin{align}
 \mathcal{V}_{int} & =  g_{X_1 } \phi \chi ^* \chi  + \frac{g_{X_2 }}{2} \phi ^2 \chi ^* \chi  + \kappa _1 \phi \left| H \right|^2  + \kappa_2 \phi ^2 \left| H \right|^2 \\  \nonumber
 &  +\frac{m_\phi^2}{2} \phi ^2  + \frac{{\lambda _\phi  }}{2}\phi ^4  - \mu ^2 \left| H \right|^2  + \frac{\lambda }{2}\left| H \right|^4
\end{align}

We neglect the Higgs portal term $\chi \chi^* |H|^2$, which can affect the relic abundance and direct detection signals, but could be absent if the theory arises from a SUSY theory at a higher scale, or if the sectors are sequestered, such as via an extra dimension. We assume that DM carries some quantum number (e.g. a $Z_2$ charge, or hidden global charge). The singlet will acquire a mixing term via the trilinear when the Higgs gets a vev. \footnote{
The singlet could also acquire a vev spontaneously, and mix without a trilinear term. We will not pursue this possibility here, because of the possibility of domain walls and the subsequent cosmological issues. For our purposes the phenomenology is the same.} We assume the mixing is small, so as to avoid a sizable direct detection cross section.

%
%


The DM annihilation to $\phi \phi$ is s-wave with the following form,
\begin{align}
\sigma v_{ \chi\chi \rightarrow \phi \phi} \simeq
\frac{{\sqrt {m_\chi ^2  - m_\phi ^2 } }}{{64\pi m_\chi ^3 }}\left( {g_{X_2 }  + \frac{{2g_{X_1 }^2 }}{{2m_\chi ^2  - m_\phi ^2 }}} \right)^2.
\end{align}

While we have considered the scalar DM case, one can also consider a fermionic scenario. The principle obstacles to this is that for a fermion the annihilation of $\chi \overline \chi$ to $\phi \phi$ is p-wave suppressed. This can be evaded if the annihilation is into a complex scalar. In this case, either the pseudoscalar would be massless (and thus would be an additional relativistic degree of freedom), or could mix with the Higgs via a CP-violating mixing term $e^{iQ} \phi \left| H \right|^2 + {\rm h.c.}$. Our points below do not depend crucially on these details, however.

%% file: fitting.tex
\section{Fitting the data}
\label{sec:fitting}
The branching ratios and photon spectra are complicated, but straightforward. We refer the reader to the appendices for details. In the appendix, we calculate the branching ratio of the dark mediator decay in section \ref{sec:BR}. In section \ref{sec:photonspc}, we show how to calculate photon spectrum in lab frame, with the assumption that the spectra from each daughter particle are known. In section \ref{App::photonS}, we briefly interpret how we calculate the photon spectra from each channels. In section \ref{sec:e-channels}, we introduce how we calculate the electron spectra in a same way as for photon spectra.

\subsection{The role of prompt photons}
With the BR information and photon spectrum from each decay channel, we can calculate the prompt photon flux as below.

\begin{align}
E_\gamma ^2 \frac{{d\Phi^{Prompt}_\gamma  }}{{dE_\gamma  }} = J_f  \cdot \left\langle {\sigma v} \right\rangle \cdot BF \cdot \frac{{R_{\odot} \rho _{\odot}^2 }}{{8\pi m_{DM}^2 }} E_\gamma ^2 \sum\limits_i {BR_i } \frac{{dN_i }}{{dE_\gamma  }},
\end{align}
where  ${R_{\odot} }$ is $8.5$~kpc, the distance to the GC; the ${\rho _{\odot} }$ is the local DM density, $0.4 \rm GeV \rm cm^{-3}$; and the $\left\langle {\sigma v} \right\rangle$ is the annihilation cross-section taken as $3 \times 10^{-26}\rm cm^{3} s^{-1} $.  BF stands for the boost factor of the cross section, and  $J_f$ is the standard dimensionless factor for the \textit{l.o.s.} integration
with the following expression,

\begin{align}
J_f (\Omega ) = \frac{1}{{R_{\odot} \rho _{\odot}^2 }}\int\limits_{los} {dr} \rho _{DM}^2 (r,\Omega )
\end{align}

\begin{figure*}
\includegraphics[width=0.45\textwidth]{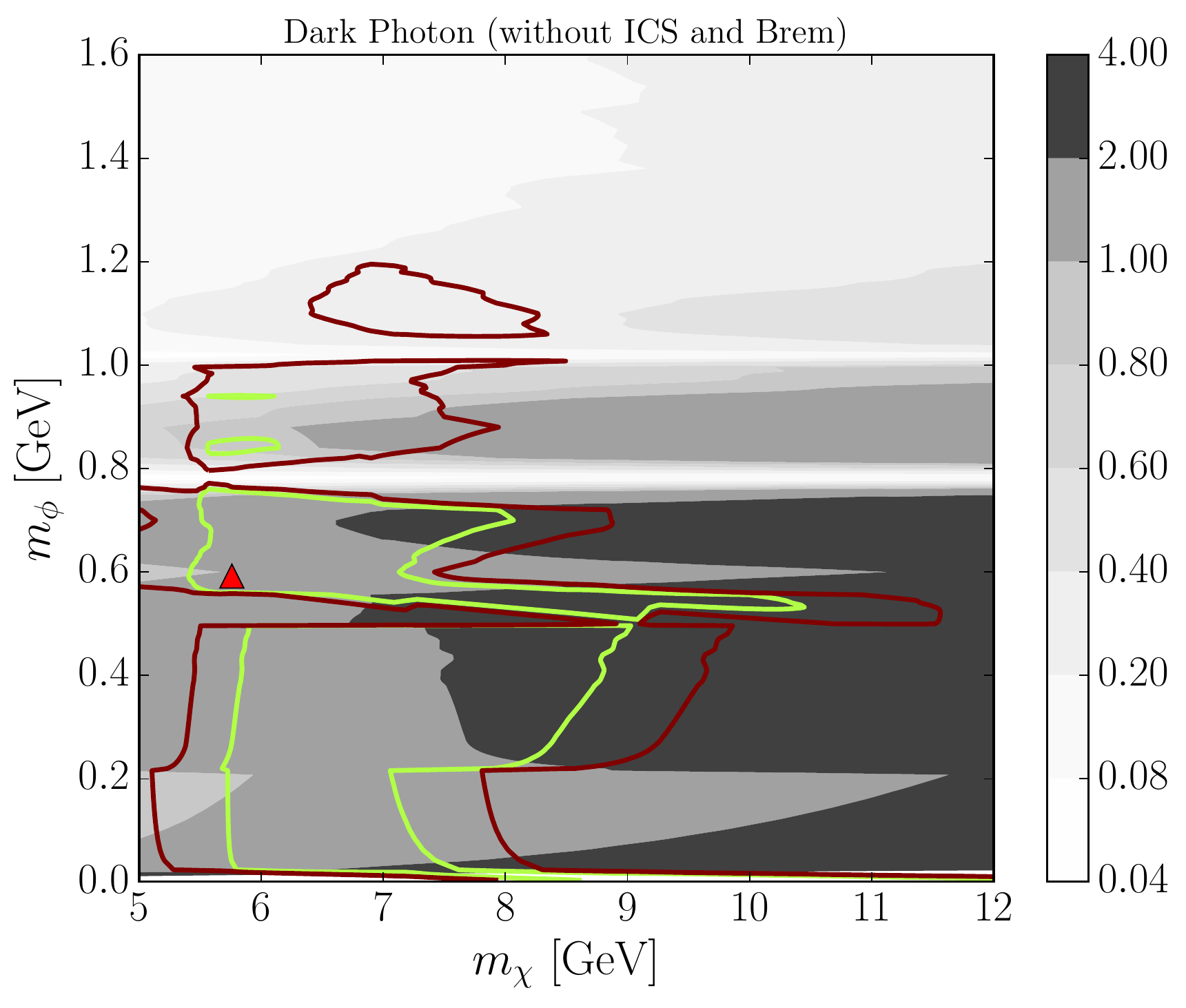} %
\includegraphics[width=0.45\textwidth]{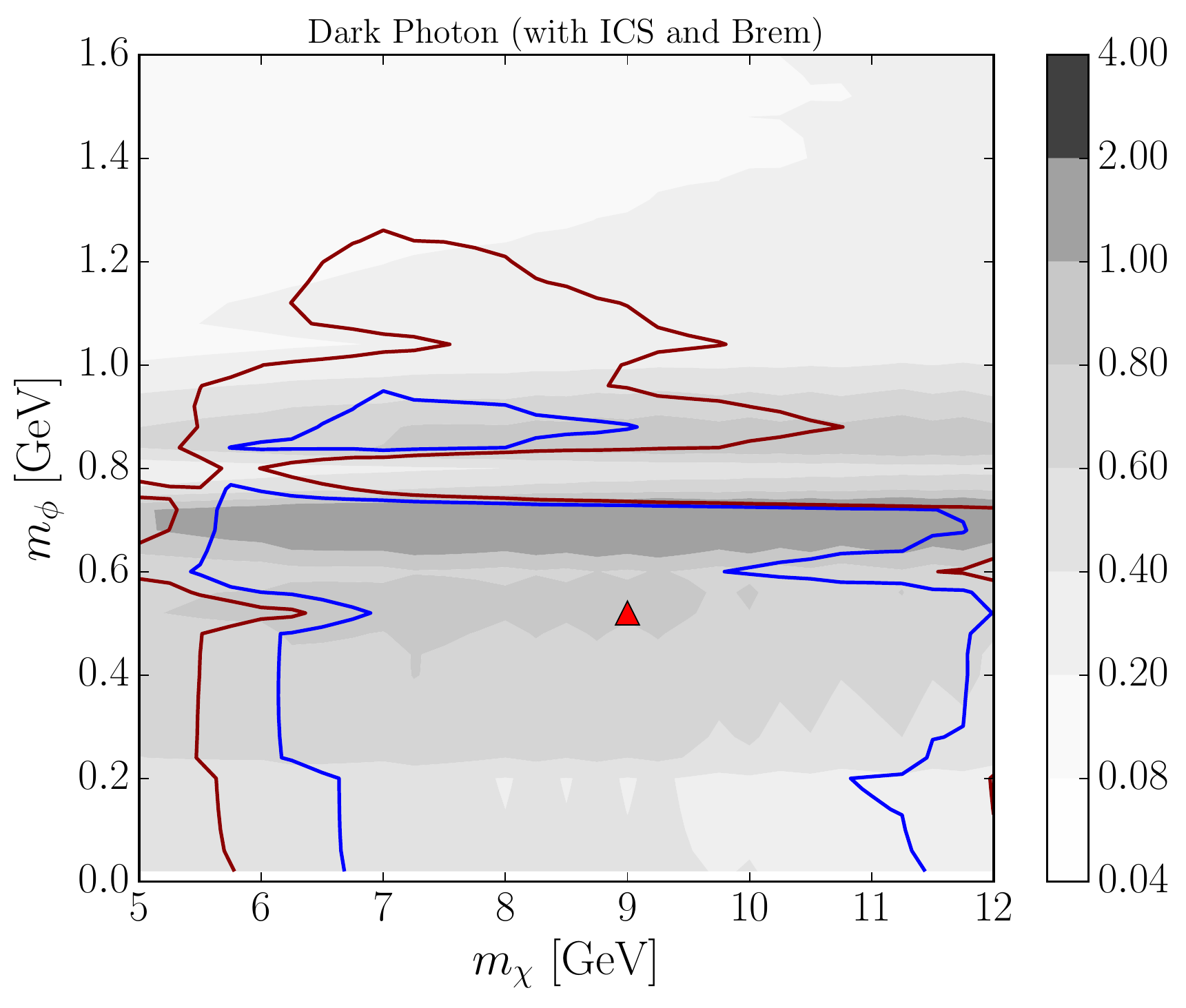} %
\caption{The $2\sigma$ and $ 3\sigma$ fitting contours for the dark photon by prompt only (\textit{left panel}) and including ICS and Bremsstrahlung (\textit{right panel}). The red triangle is the best fit point for the model. The gray scale indicates the BF from the $\chi^2$ fitting. We use twice the error-bar of  \cite{Daylan:2014rsa}.  \label{fig:DFcontour}}
\end{figure*}

  $J_f$ is calculated by taking a $5^\circ$ cone from GC, to match the data from \cite{Daylan:2014rsa},
which is taken as 268.7 for the generalized NFW profile ($\gamma$ = 1.26).
For each parameter point $\{ m_{DM} ,m_\phi  \}$, the BR for each channel and photon spectrum $\frac{{dN_i }}{{dE_\gamma  }}$ are fixed. We scan over BF to minimize the $\chi ^2$ for each point. The fitting resulting from a consideration only prompt photons for annihilations into dark photons are shown in the left panel of Figure ~\ref{fig:DFcontour}. The gray scale indicates the BF from the $\chi^2$ fitting. To count the uncertainty in the error estimation, we show the contour plot with double error-bar of the \cite{Daylan:2014rsa}. For the moment, we focus only on prompt photons from the decays of the $\phi$, and do not include additional contributions from ICS and bremmstrahlung.

\begin{figure*}
\includegraphics[width=0.43\textwidth]{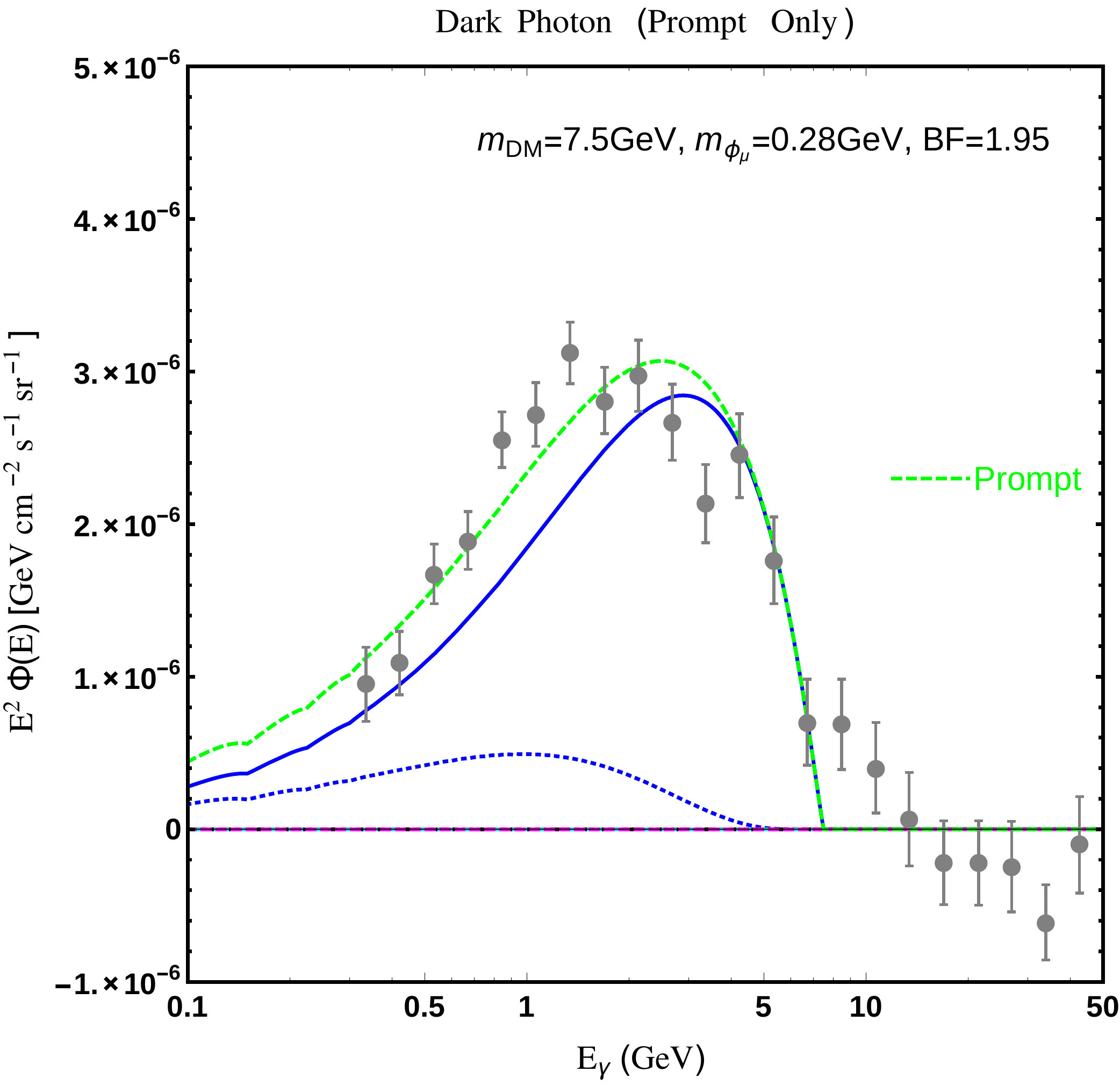} %
\includegraphics[width=0.43\textwidth]{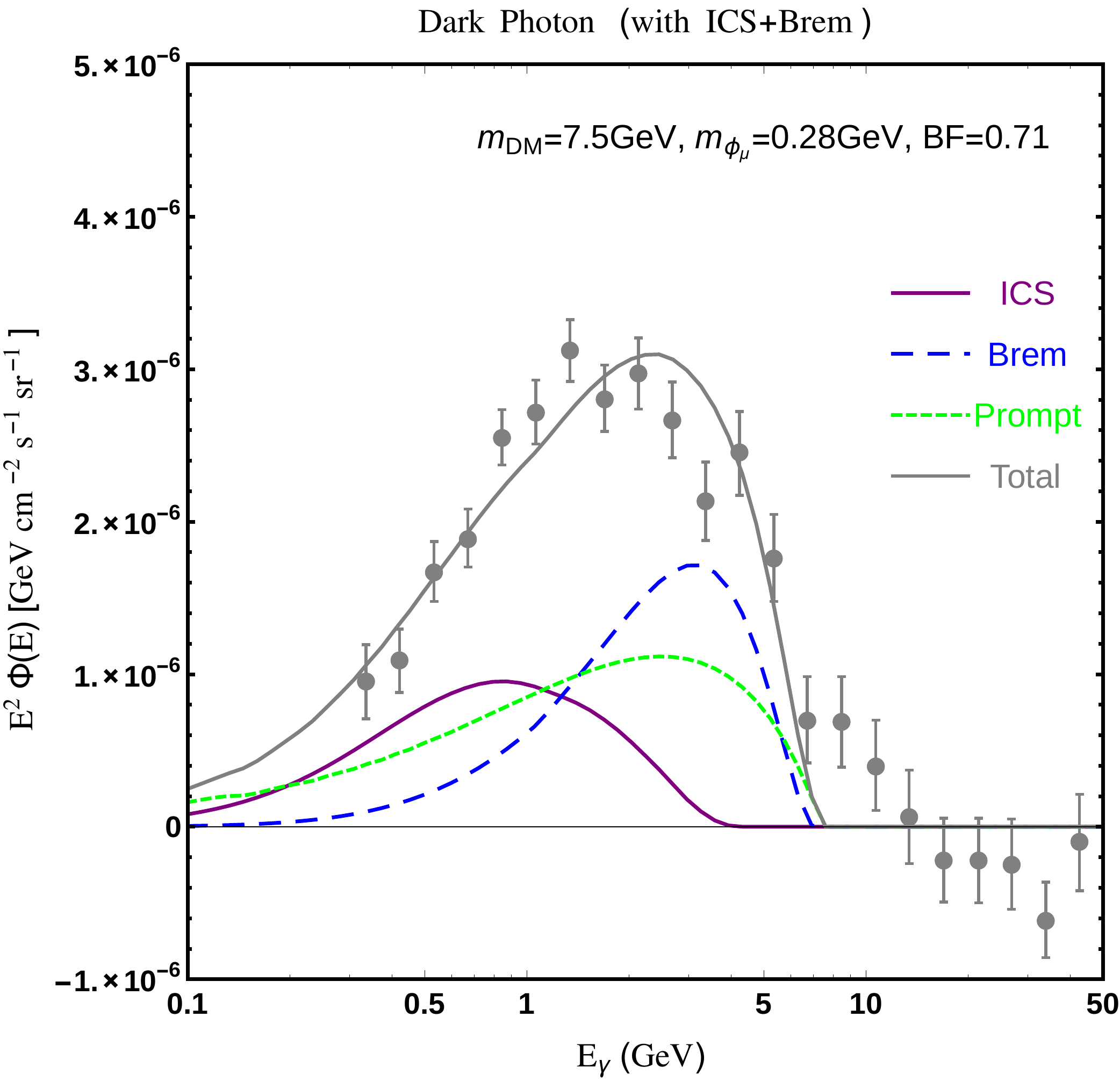} %
\\
\includegraphics[width=0.43\textwidth]{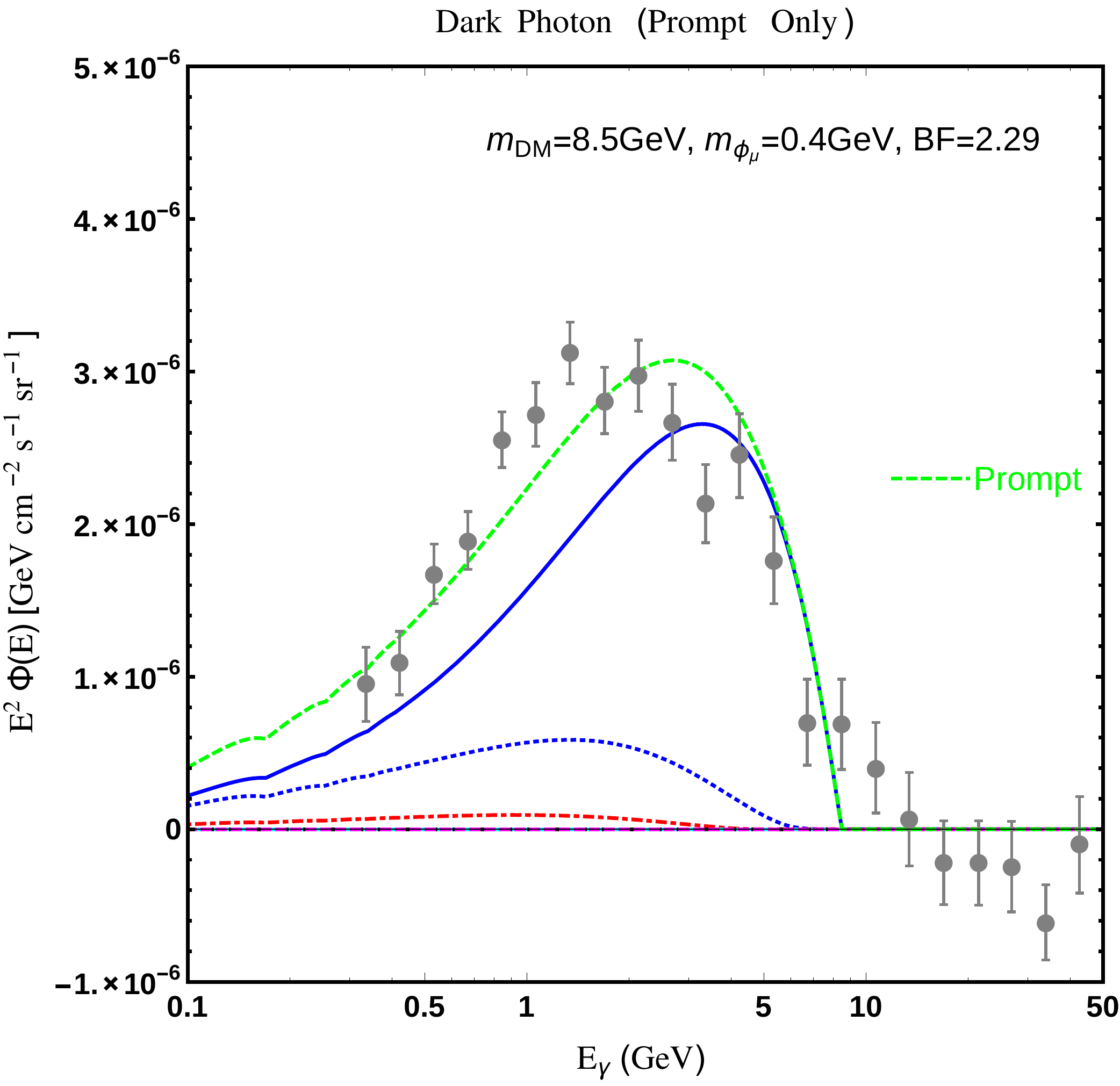} %
\includegraphics[width=0.43\textwidth]{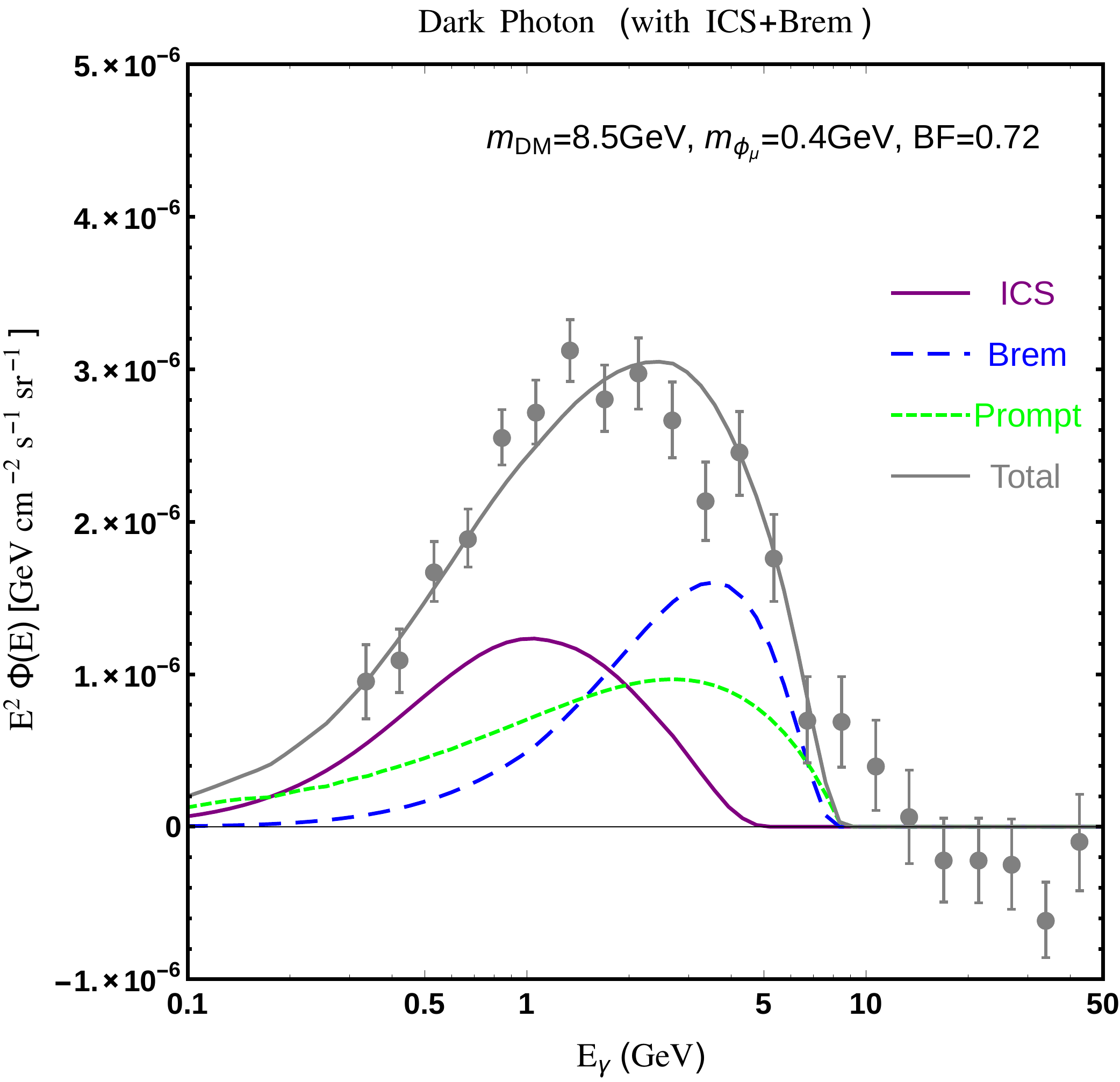} %
\\
\includegraphics[width=0.43\textwidth]{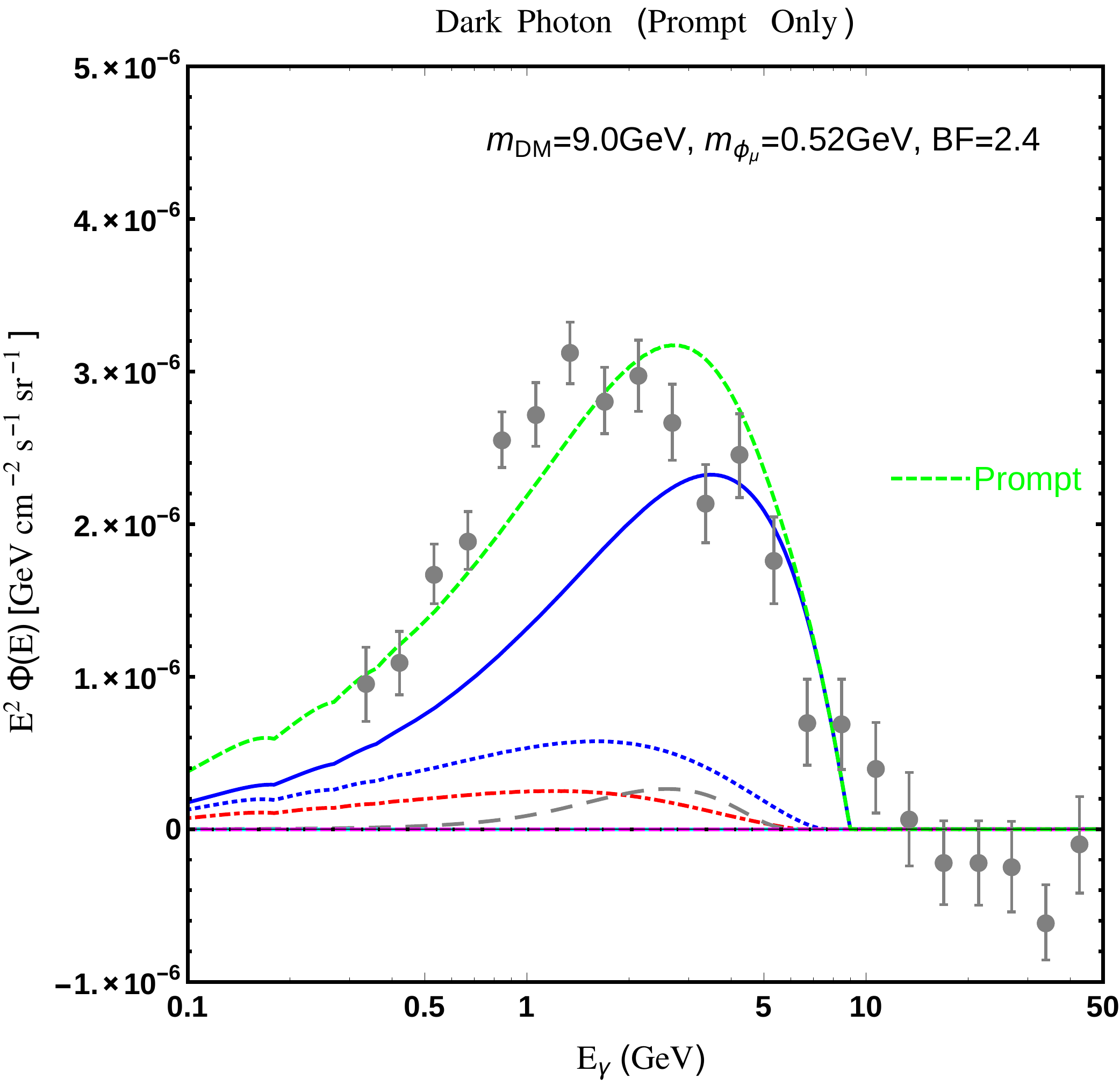} %
\includegraphics[width=0.43\textwidth]{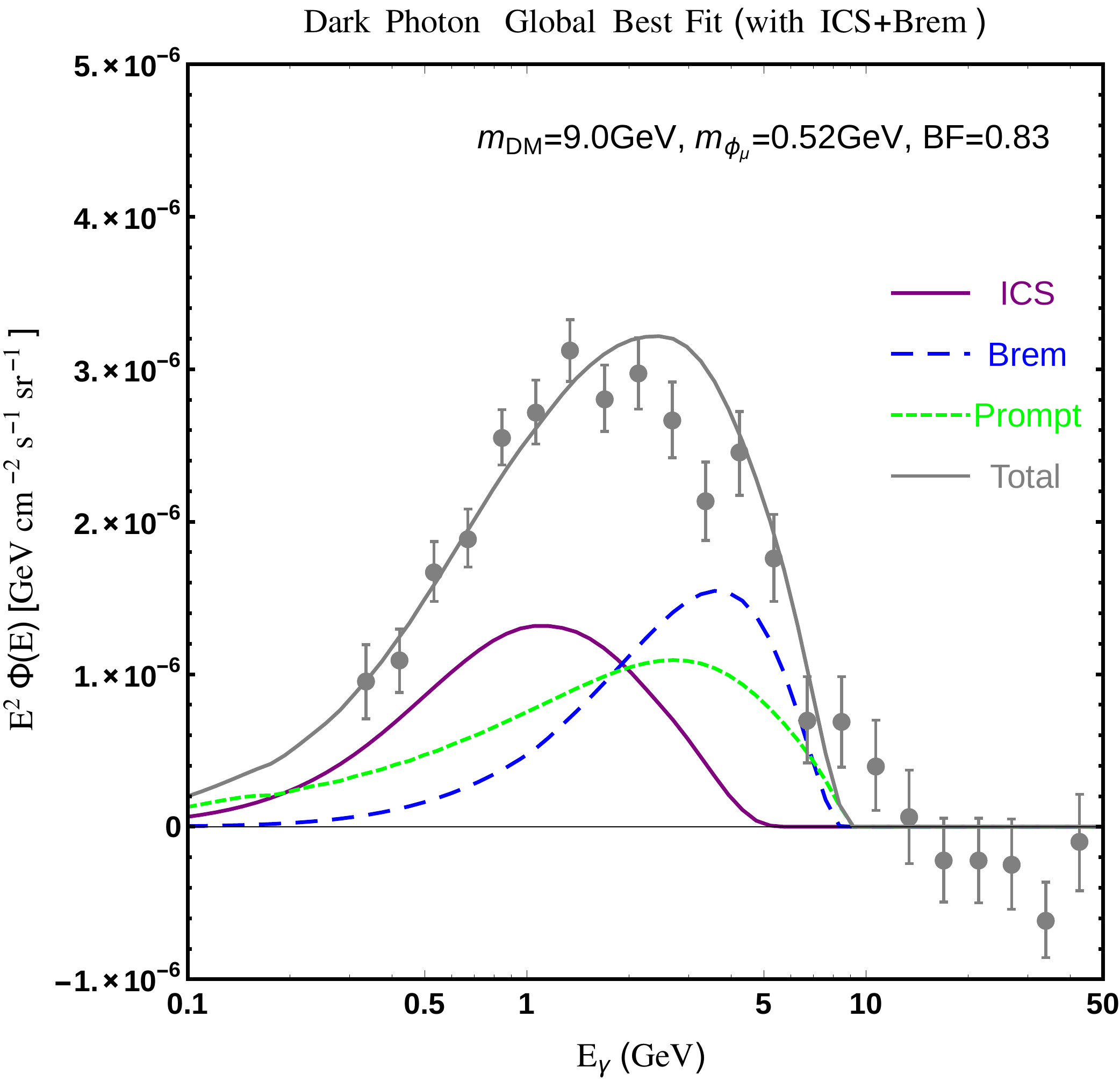} %
\caption{\textit{Left Panel}: the prompt photon spectra from FSR and IB for the dark photon scenario with different DM mass and mediator mass. The ICS and regular
Bremsstrahlung are assumed to be negligible.
The dashed green is the total prompt photon spectrum, while the other color lines correspond to decay channels for dark photon in the Figure ~\ref{fig:BRdarkphoton}. \textit{Right Panel}: the  photon spectra including the ICS and regular Bremsstrahlung. (see text) \label{fig:DFspectrum1}}
\end{figure*}

\begin{figure*}
\includegraphics[width=0.45\textwidth]{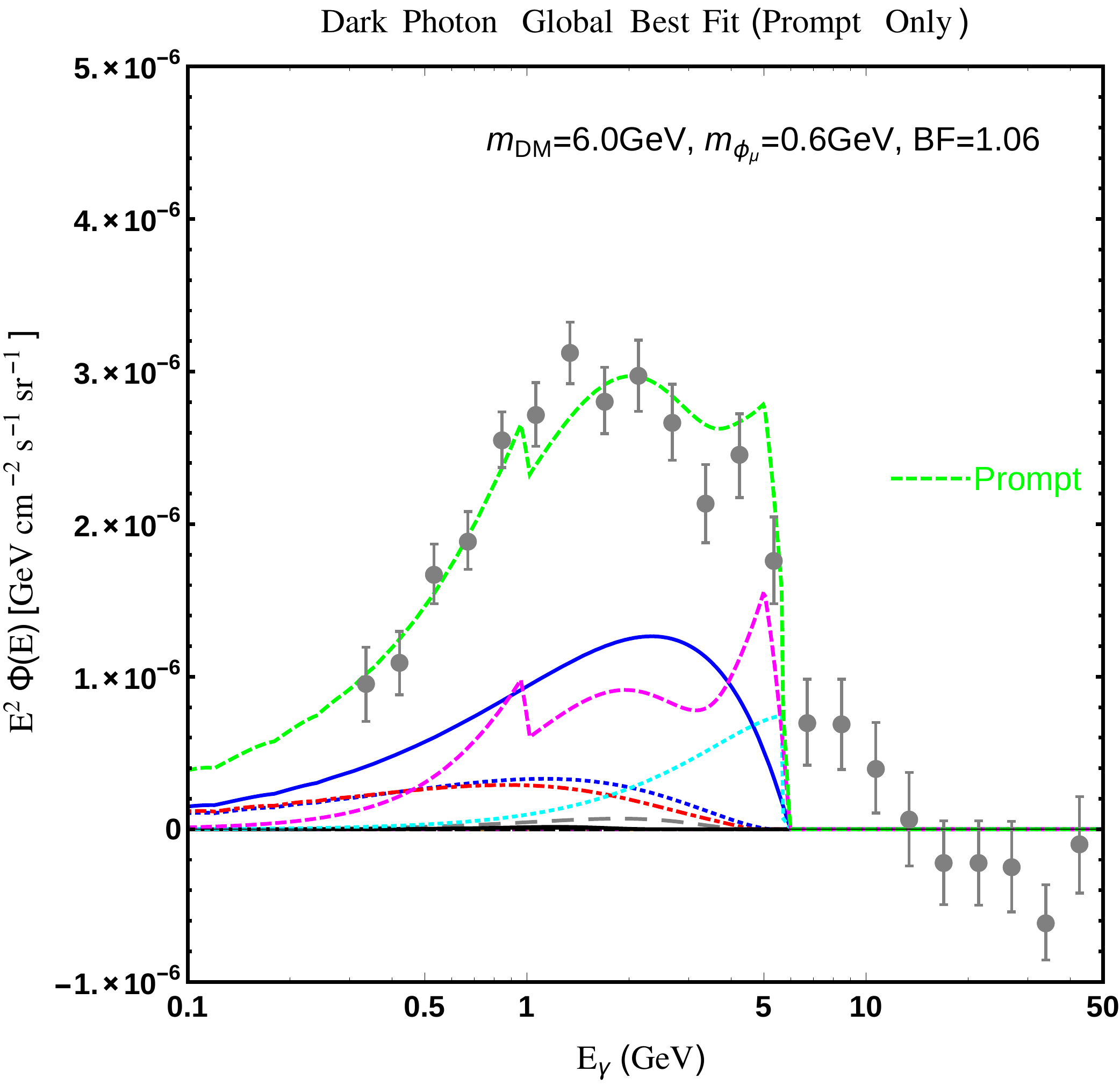} %
\includegraphics[width=0.45\textwidth]{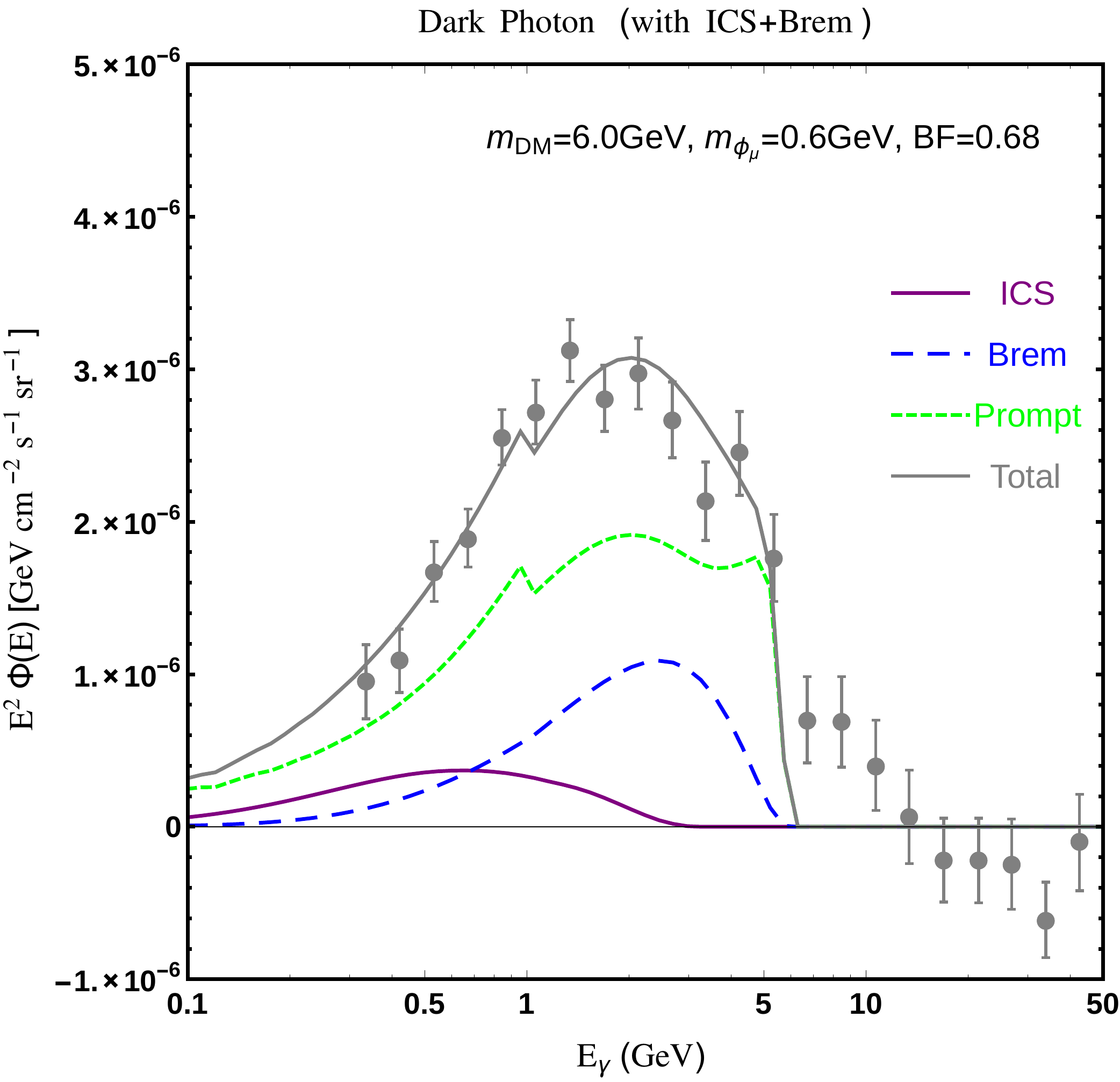} %
\\
\includegraphics[width=0.45\textwidth]{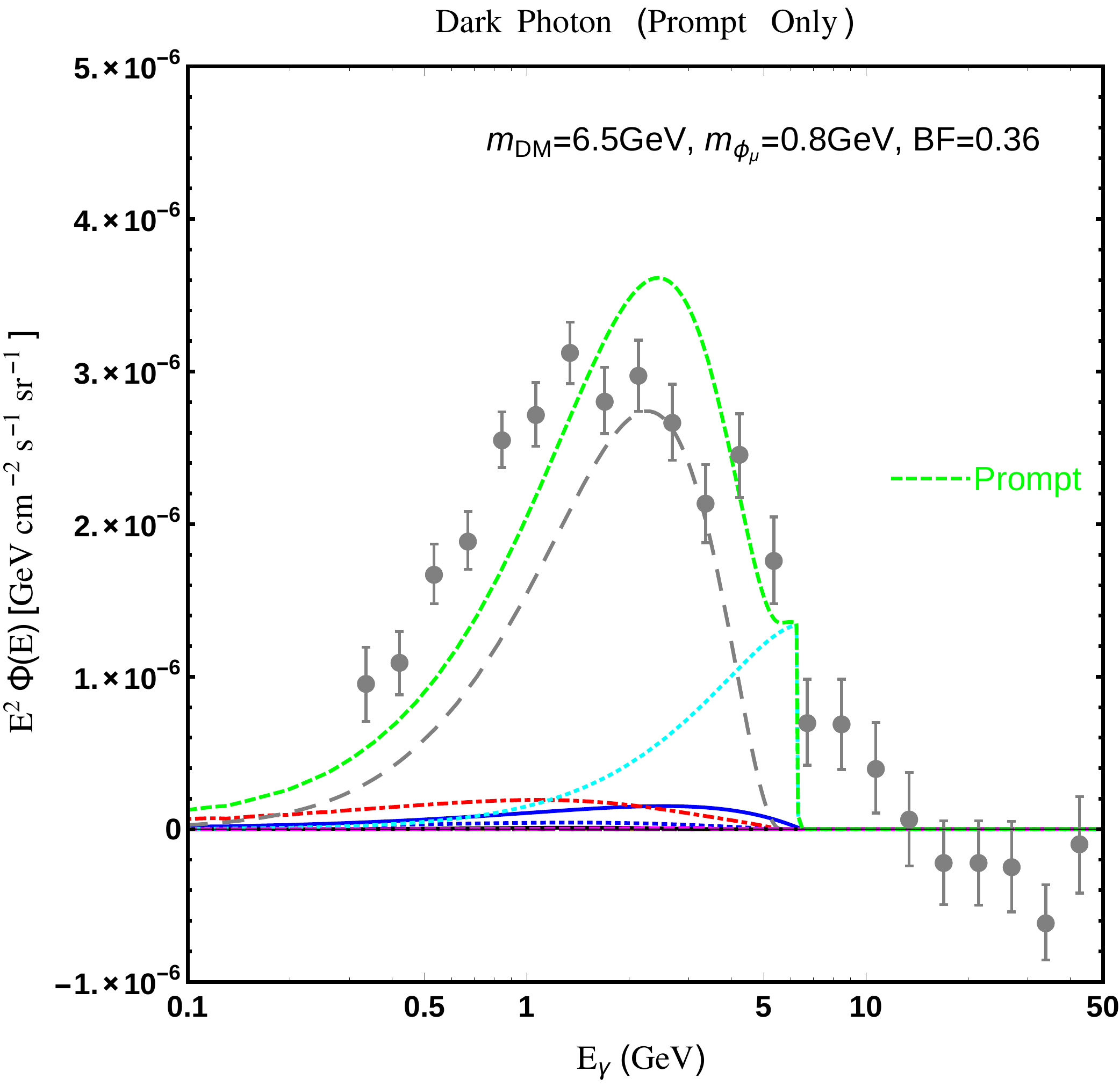} %
\includegraphics[width=0.45\textwidth]{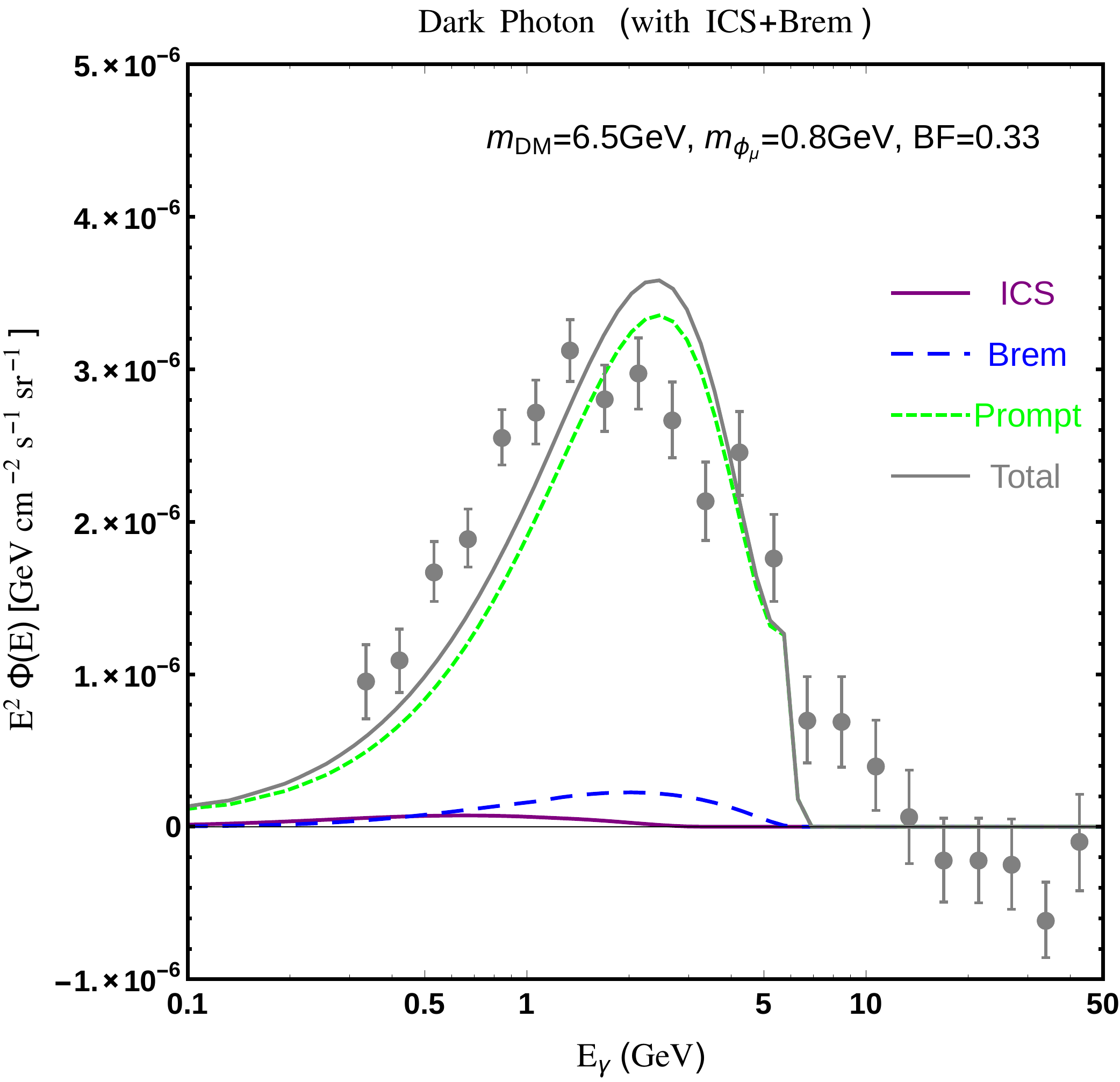} %
\\
\includegraphics[width=0.45\textwidth]{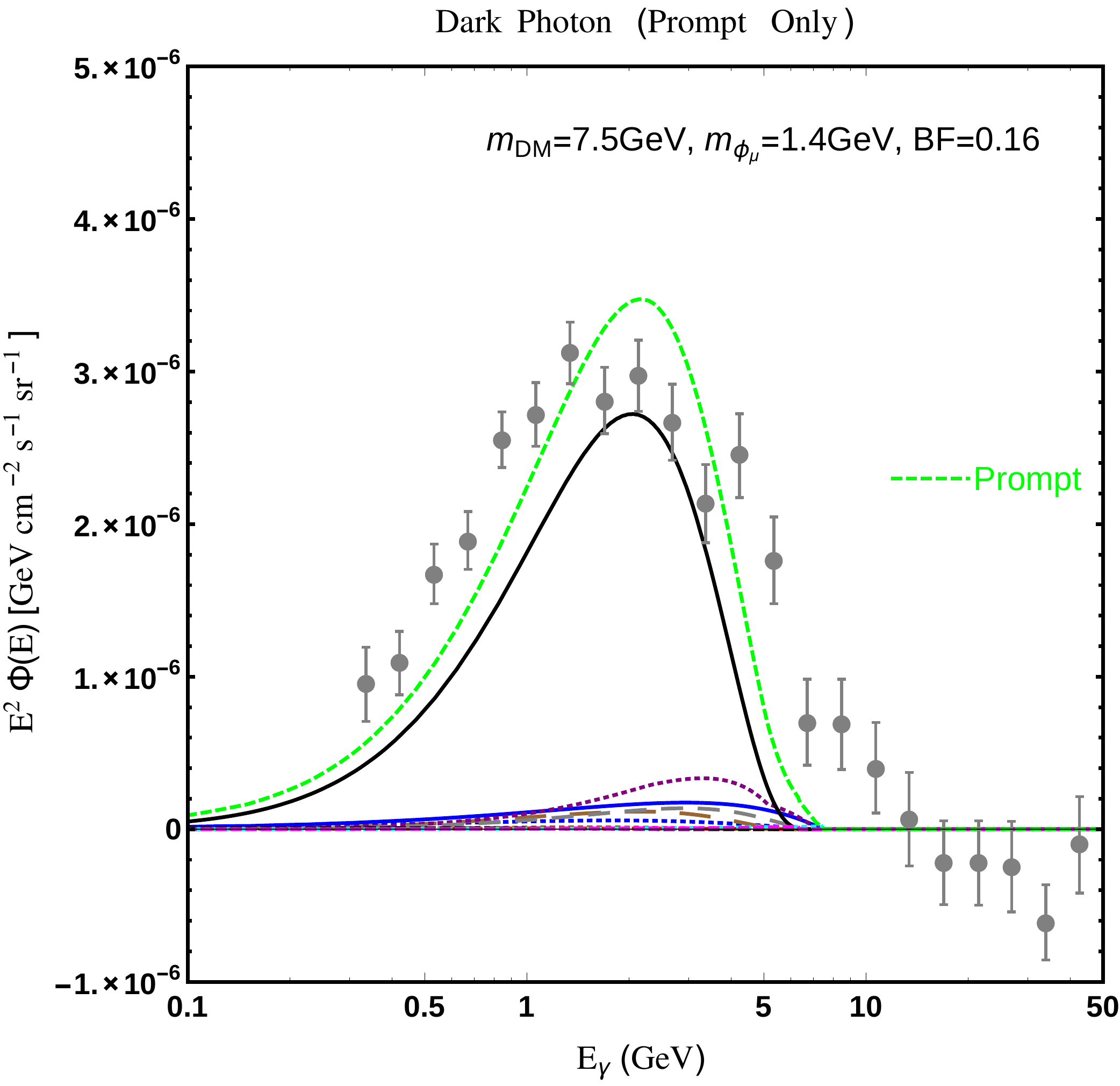} %
\includegraphics[width=0.45\textwidth]{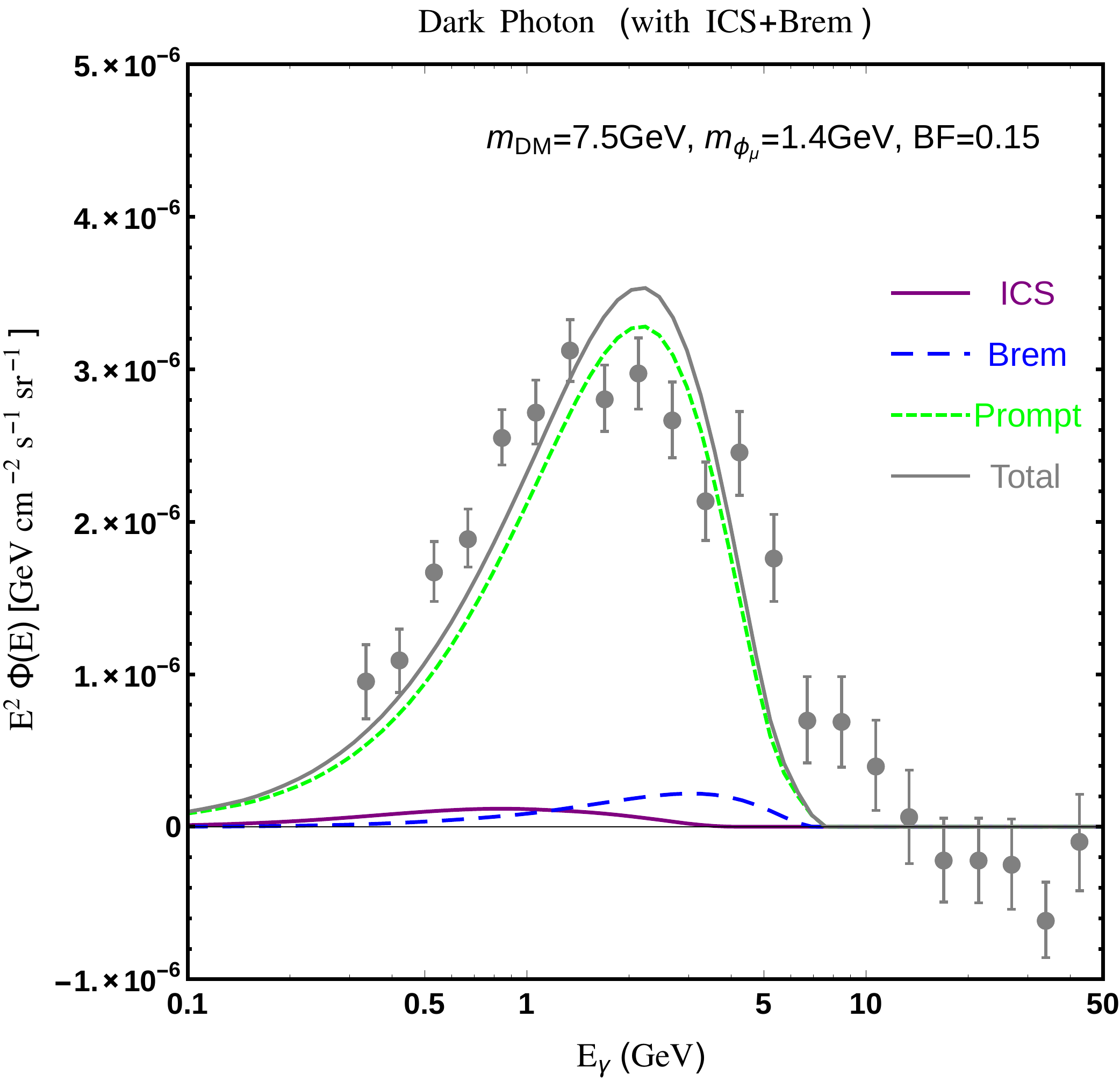} %
\caption{Same as Figure ~\ref{fig:DFspectrum1}, but for $m_\phi$ values that produce $\pi^0$ contributions, when prompt photon signals are dominant. \label{fig:DFspectrum2}}
\end{figure*}

 The best prompt photon fit for dark photon is $\{ 5.7\gev, 0.59\gev \}$ for DM mass and mediator mass respectively, shown as red triangle in the plot. We plot $2\sigma$ and $3\sigma$ contours for the parameter space. The color bar shows the BF for each point, after minimizing the $\chi ^2$. We can see the best regions are around $5.5 \sim 9$GeV for DM mass and $0.2 \sim 0.8$GeV for mediator mass. In these regions, the BR of $e ^ +  e ^ -$, $\mu ^ +  \mu ^ -$ and $\pi ^ +  \pi ^ -$ channels dominate in the decay. We plot the prompt photon spectra for each channels with different mediator mass in the left panel of Figure ~\ref{fig:DFspectrum1} and Figure ~\ref{fig:DFspectrum2}. Interestingly, the best fit for prompt photon spectra are dominated by $e ^ +  e ^ -$, $\pi^0 \gamma $ and $\eta \gamma$. The latter two have small BR but high photon yield, because the number of hard photons in $e ^ +  e ^ -$ goes as $\alpha/\pi$, while the $\pi^0 \gamma $ and $\eta \gamma$ channels have $O(1)$ number of photons. For mediator mass smaller than $0.4$GeV, the photon spectrum is dominated by radiative processes arising from $e^+e^-$. However, for heavier mediator around $1$GeV, the contribution comes from meson channels like $KK$, $\pi^0 \gamma$, $\pi^+ \pi^- \pi^0 $ and $\pi^+ \pi^- \pi^0 \pi^0 $. It shows that including meson channels is quite important in the light mediator analysis. The BF in these regions are around ${\rm O}(1)$, which means the fitting is quite reasonable.

\begin{figure*}
\includegraphics[width=0.45\textwidth]{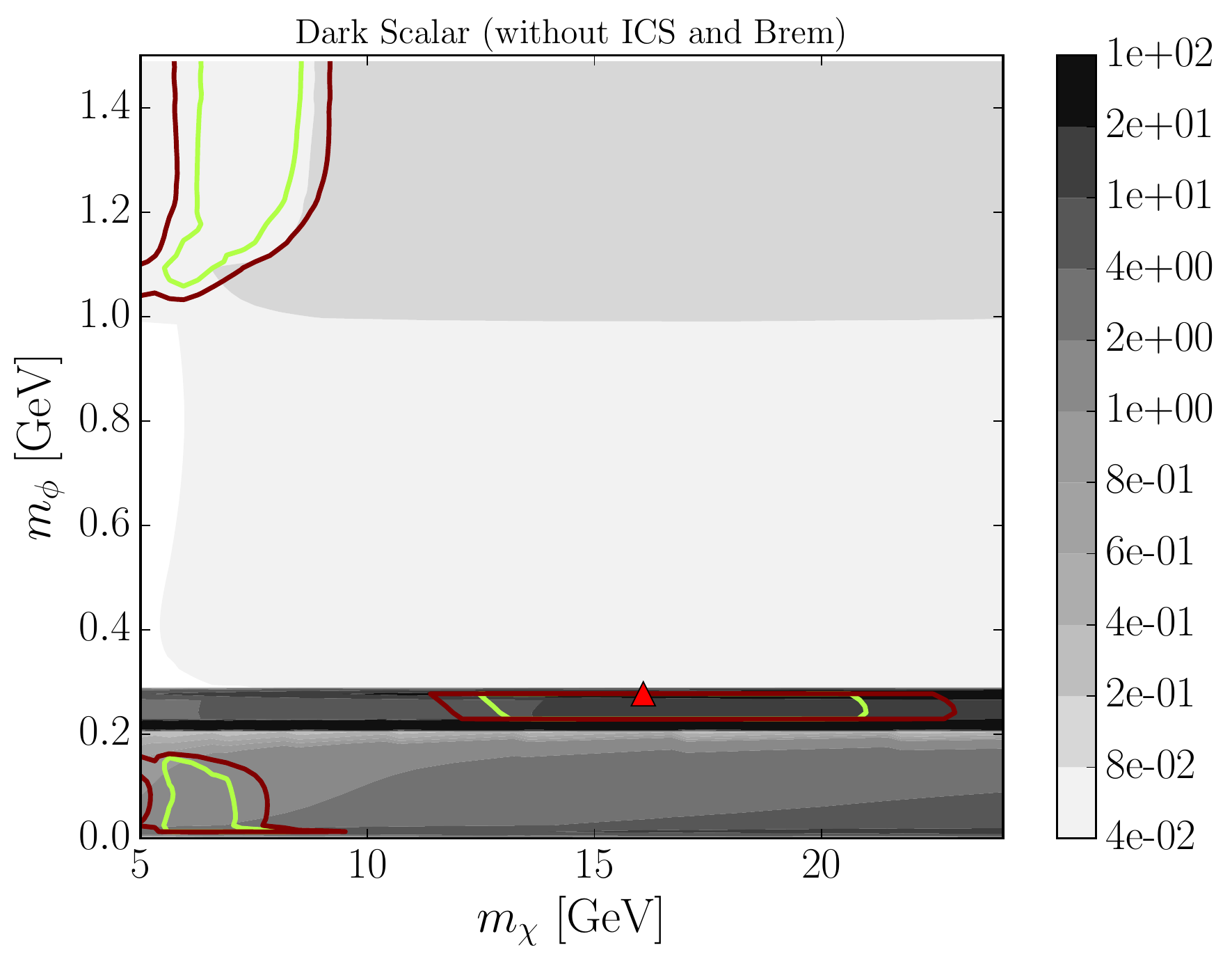} %
\includegraphics[width=0.43\textwidth]{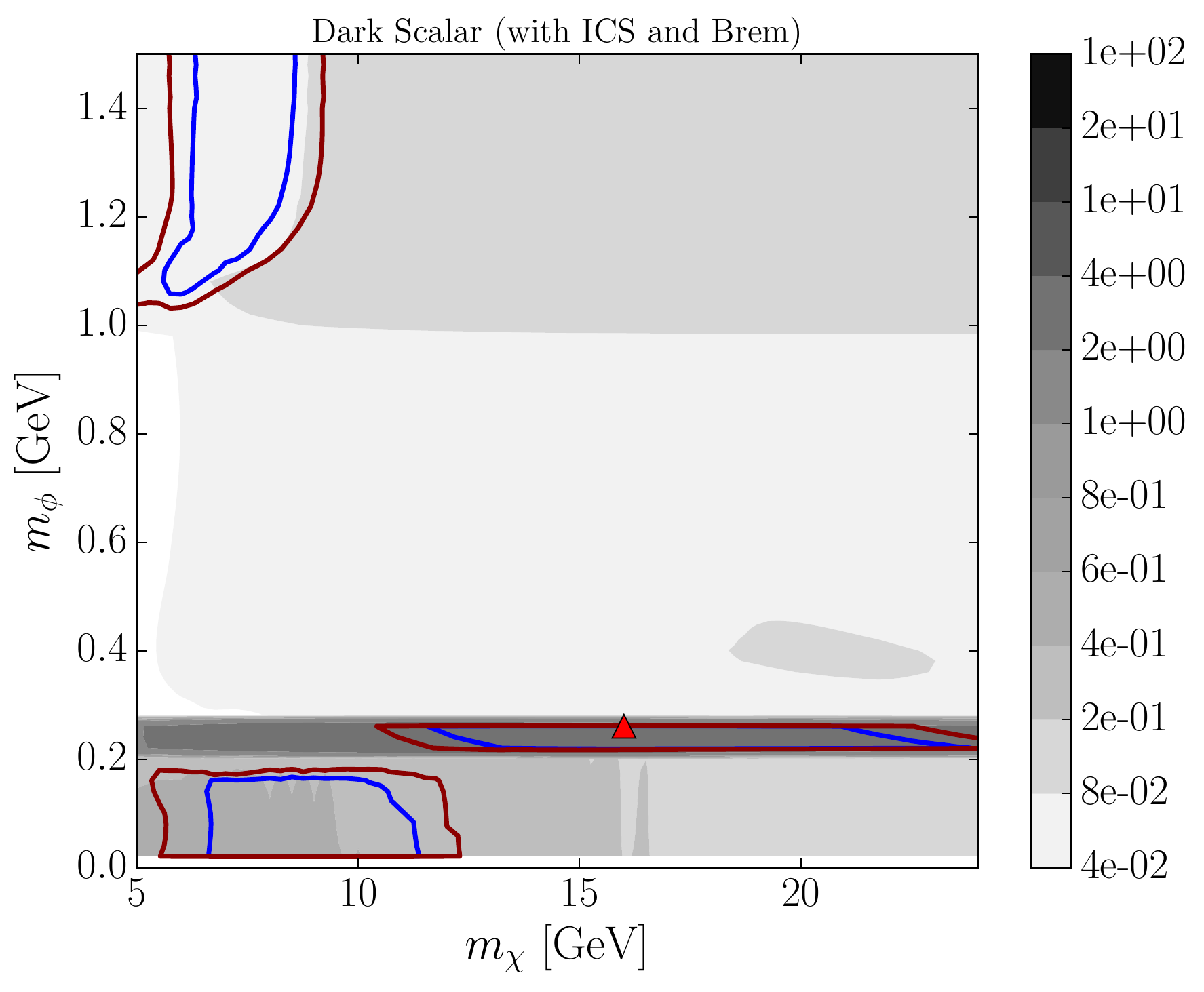} %
\caption{The $2\sigma$ and $ 3\sigma$ contour plot for the dark scalar with prompt photon only. The red triangle is the best fit point for the model. We use twice the error-bar of the \cite{Daylan:2014rsa}.  \label{fig:DScontour}}
\end{figure*}

\begin{figure*}
\includegraphics[width=0.43\textwidth]{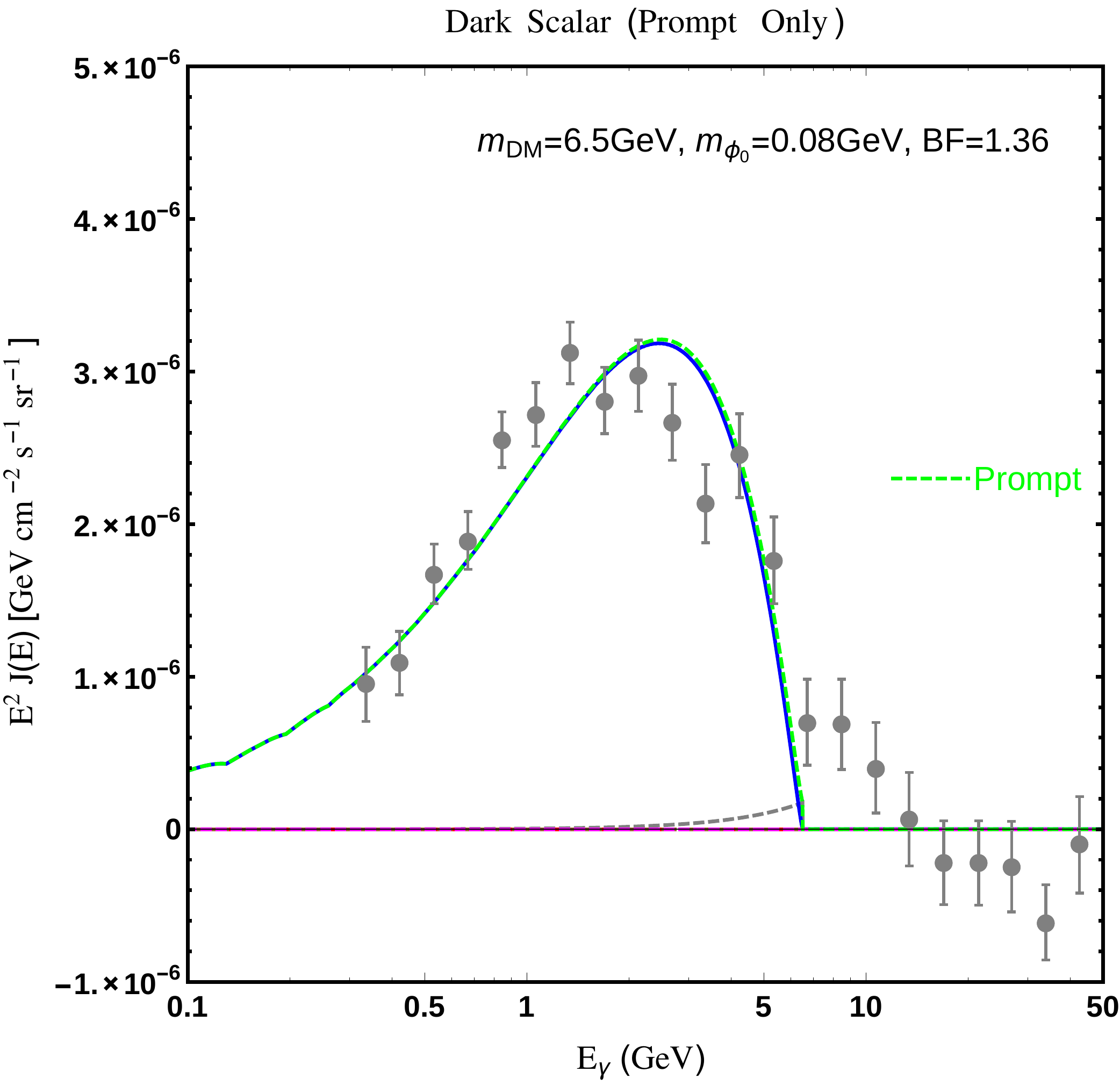} %
\includegraphics[width=0.43\textwidth]{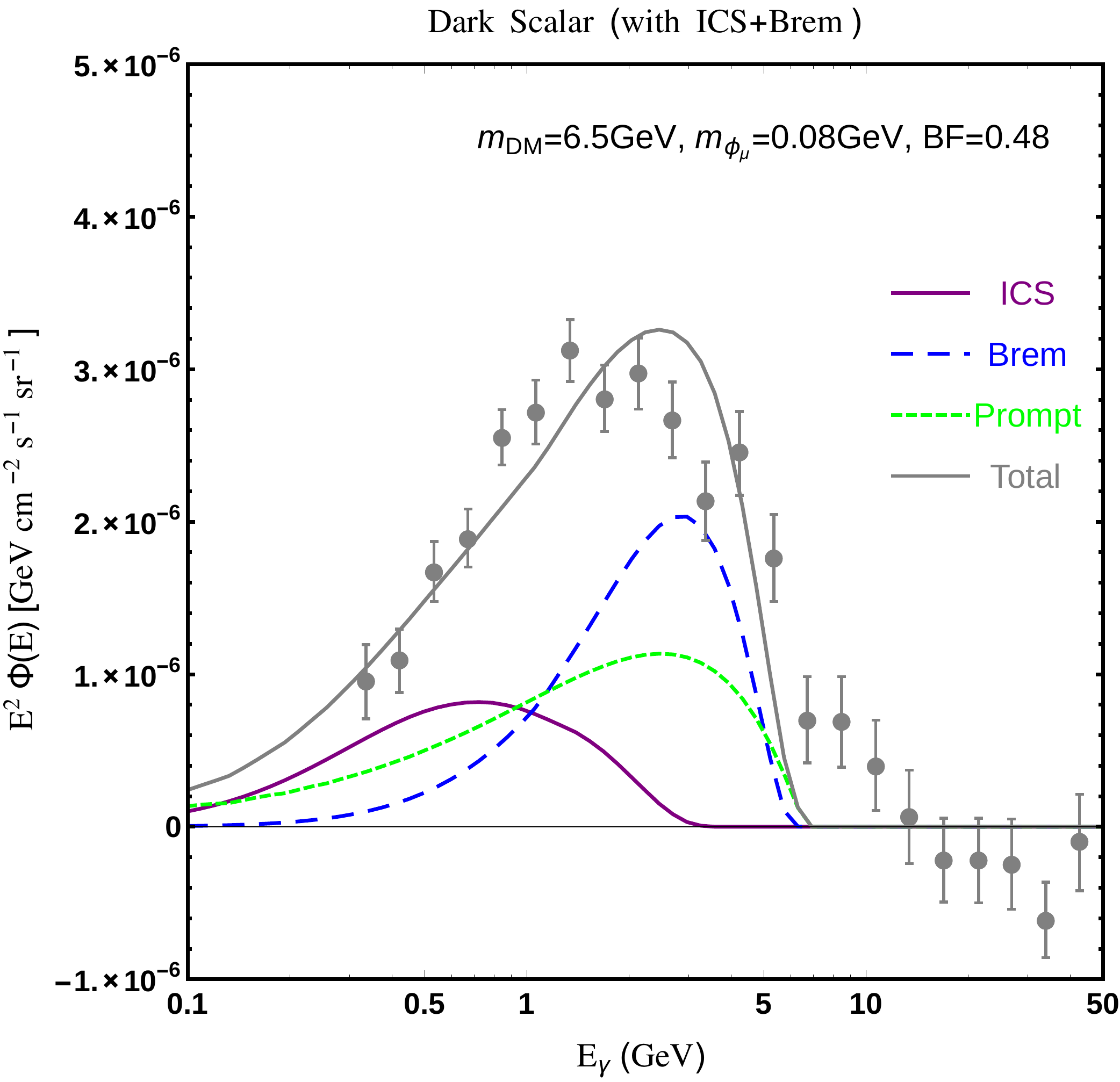} %
\\
\includegraphics[width=0.43\textwidth]{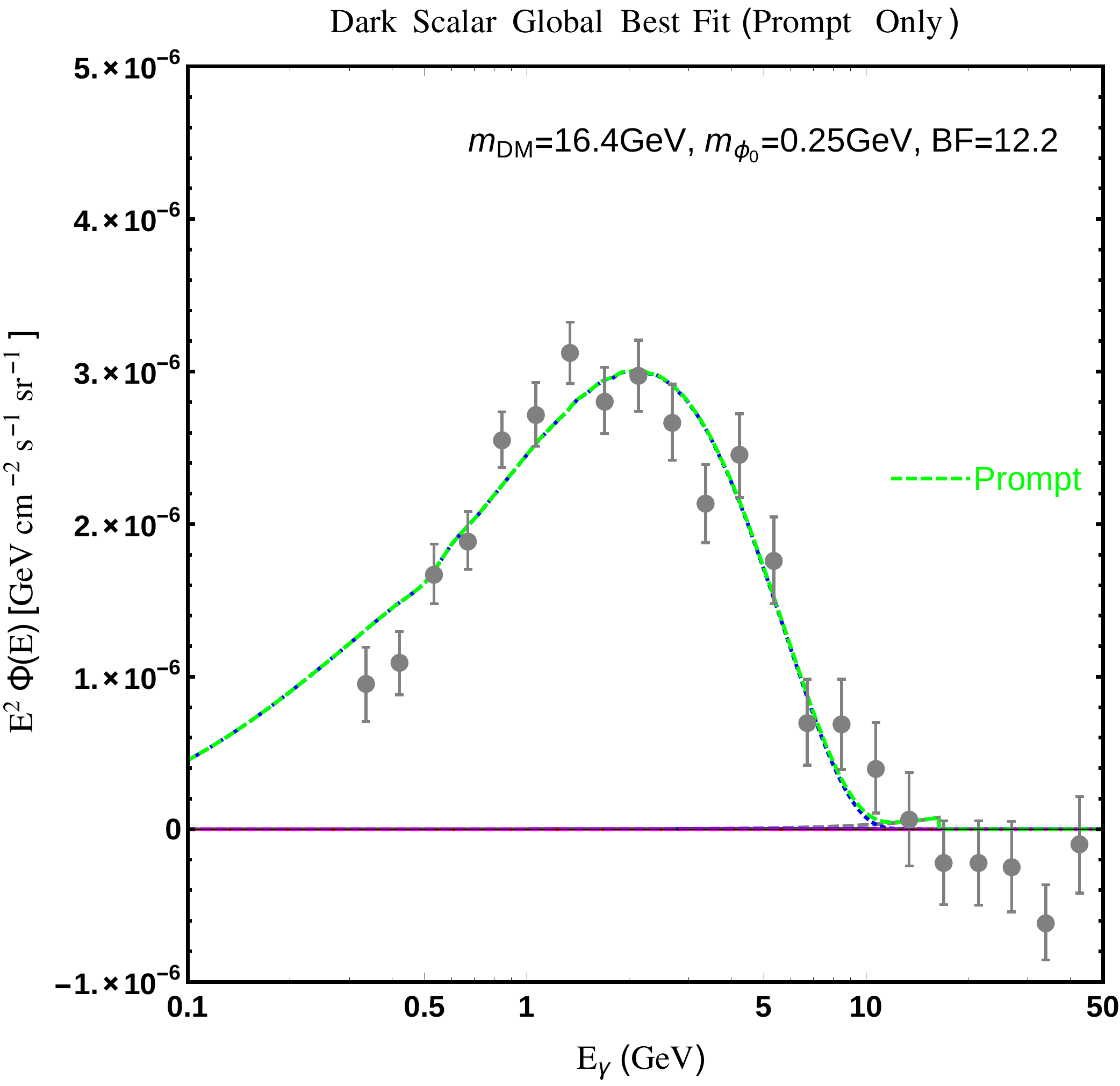} %
\includegraphics[width=0.43\textwidth]{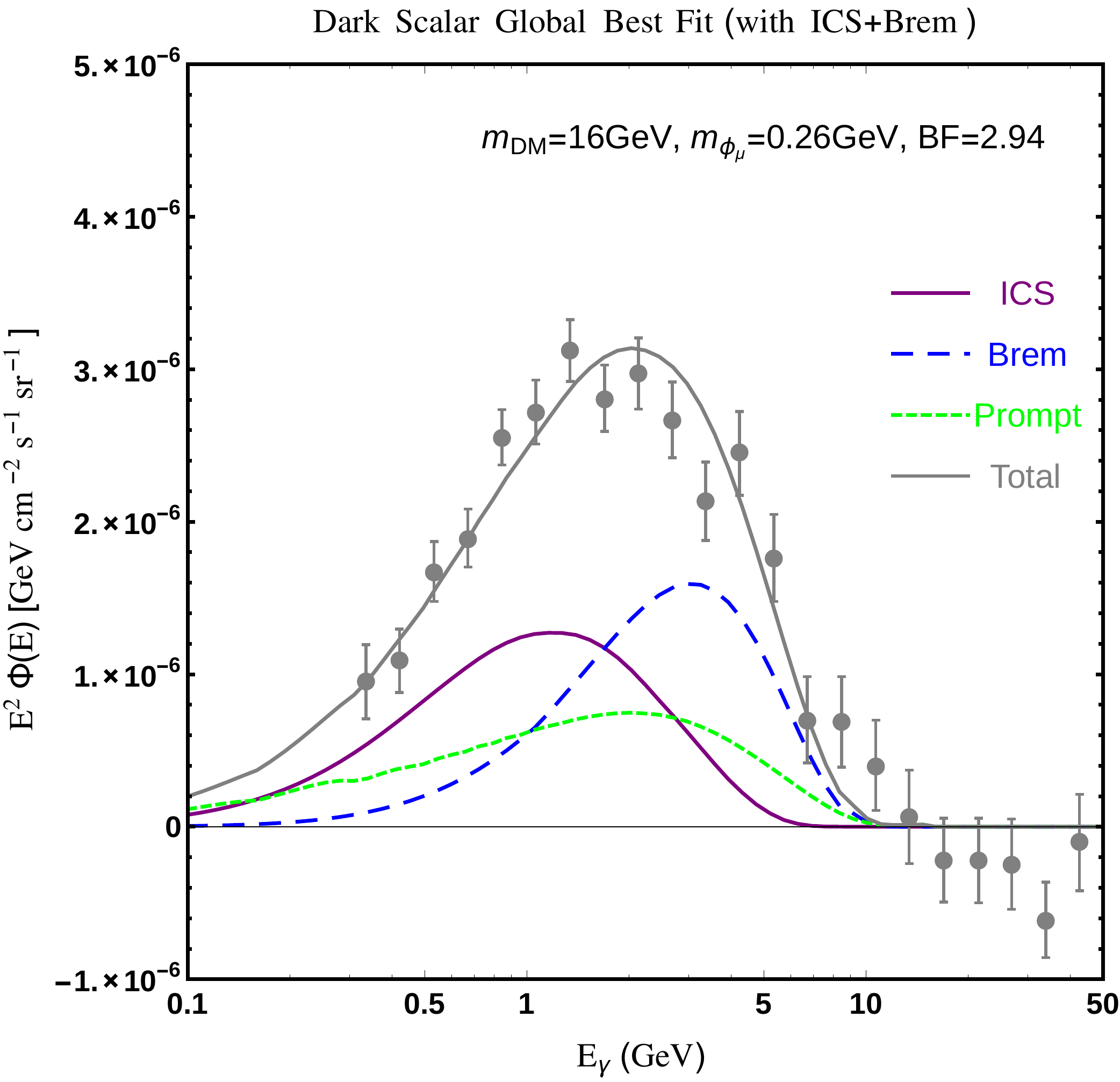} %
\\
\includegraphics[width=0.43\textwidth]{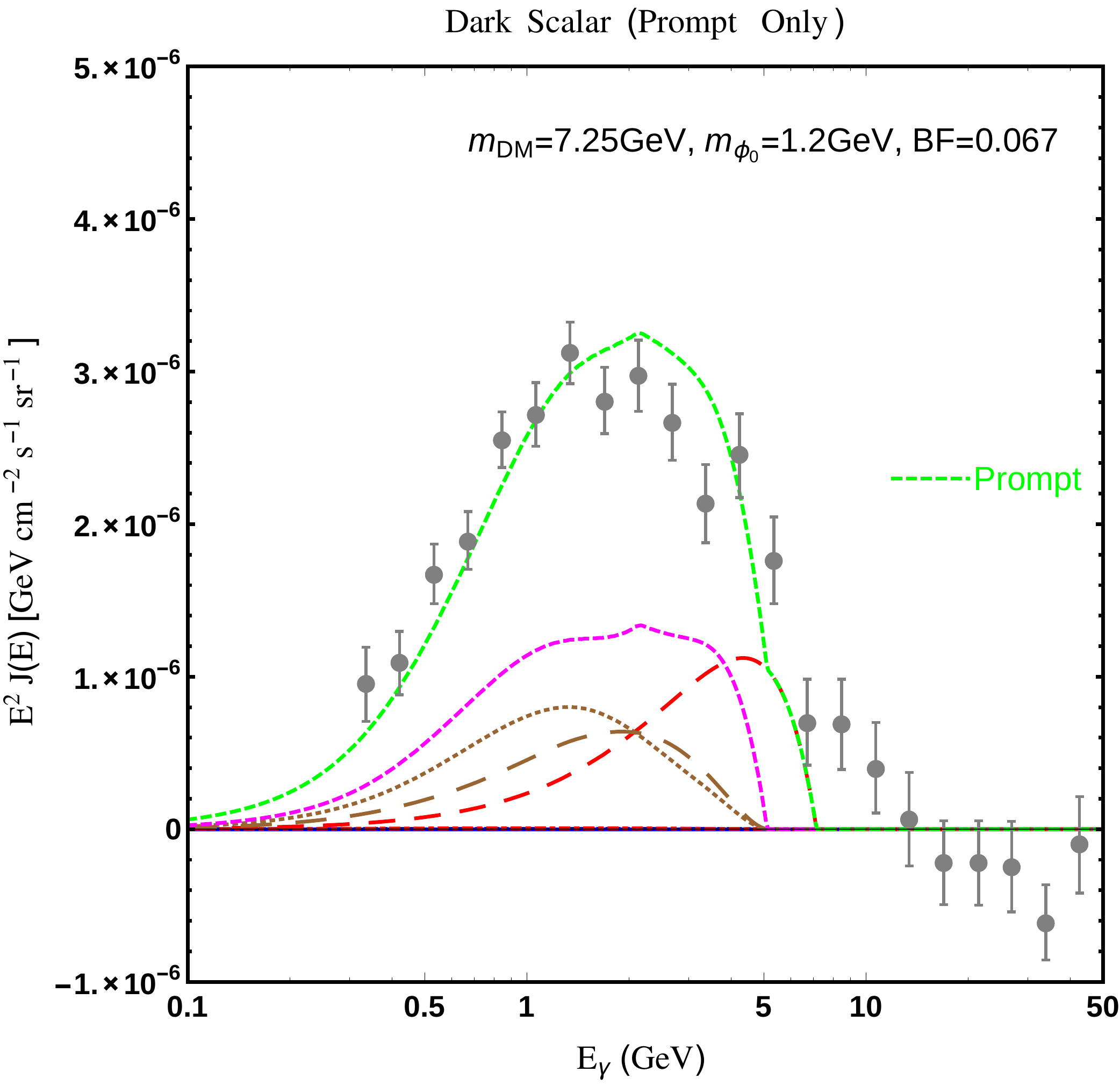} %
\includegraphics[width=0.43\textwidth]{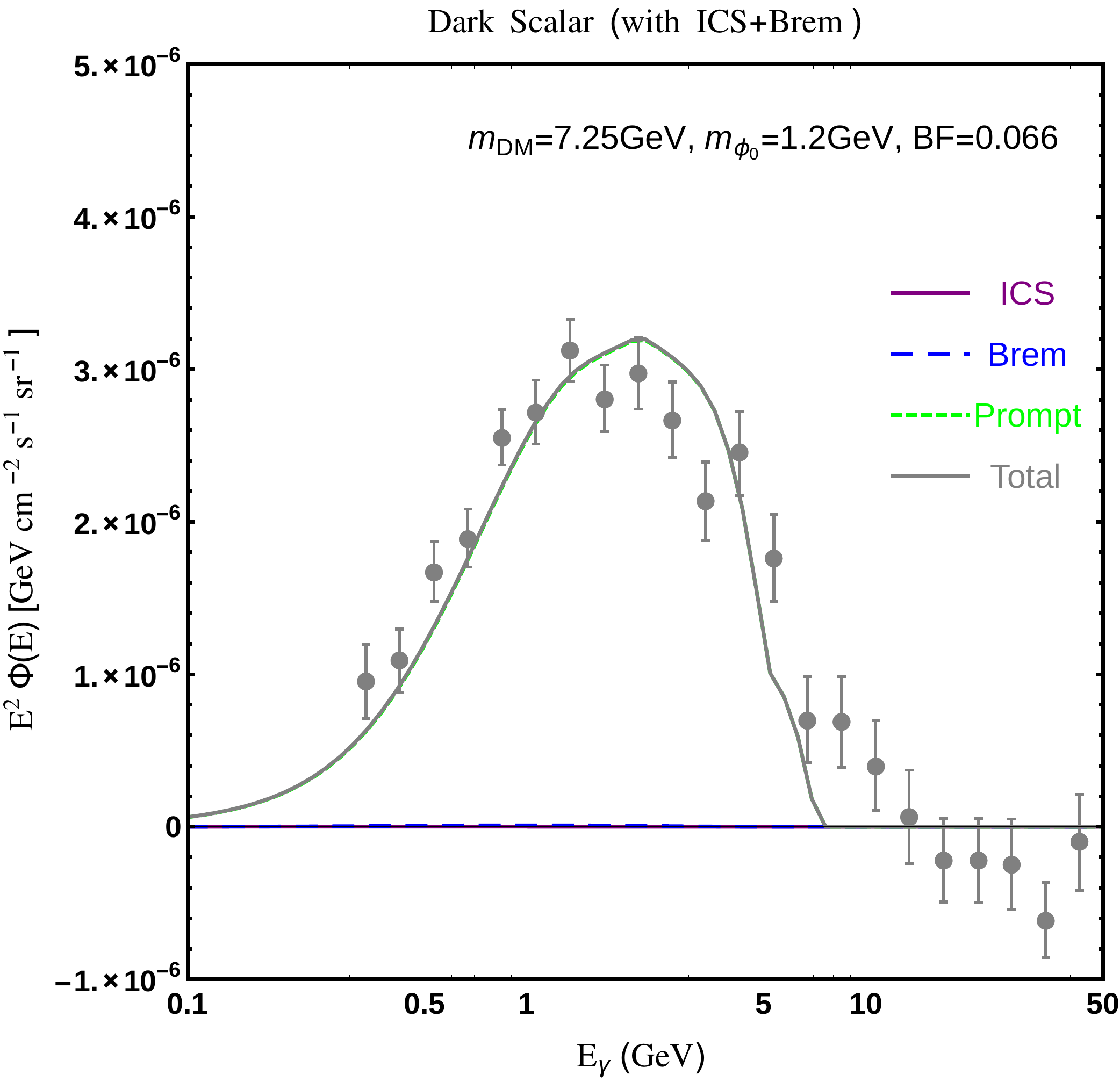} %
\caption{\textit{Left Panel}: the prompt photon spectra from FSR and IB for the dark scalar scenario with different DM mass and mediator mass. 
The ICS and regular Bremsstrahlung are assumed to be negligible.
The dashed green is the total prompt photon spectrum, while the other color lines correspond to decay channels for dark scalar in the Figure ~\ref{fig:BRdarkscalar}. \textit{Right Panel}: the photon spectra including the ICS and regular Bremsstrahlung. (see text).  \label{fig:DSspectrum}}
\end{figure*}

For the dark scalar, we show the fitting by prompt photon in the Figure ~\ref{fig:DScontour}. The best fit point for dark scalar is $\{ 16.4\gev, \sim 0.25\gev \}$ for DM mass and mediator mass respectively. The best regions are separated as three regions. The first region is around $5.5 \sim 7.5$GeV for DM mass and $0.0 \sim 0.2$GeV for mediator mass, where $e ^ +  e ^ -$ channel dominates. The next region is around $12.5 \sim 22$GeV for DM mass and $\sim 2m_{\mu}$ for mediator mass, where $\mu ^ +  \mu ^ -$ channel dominates due to mediator mass opens for $\mu ^ +  \mu ^ -$ channel but not for pions. The best fit is also in this region, and we plot the prompt photon spectra for each channels in left panel of Figure ~\ref{fig:DSspectrum}. One can see the best fit is dominated by mediators where the photons arise from radiative processes involving $\mu ^ +  \mu ^ -$. For these points, however, the BF is quite large, about $\sim 10$, due to the small number of photons from these radiative processes. For mediator mass between $ 2m_{\pi} \sim 1\gev$, there is no good fit because the $\pi ^ 0\pi ^ 0$ provides too many hard photons. The third region is $5.3 \sim 8.5$GeV for DM mass and $1.1 \sim 1.5$GeV for mediator mass, where Kaon channels dominates over the pion channels. Although the Kaon decays to $\pi^0$, yielding copious photons, since it is cascade decay the photon spectra are generally softer than $\pi ^ 0\pi ^ 0$ channel. We plot the prompt photon spectra for the $1.2$GeV scalar mediator in the Figure ~\ref{fig:DSspectrum} as an example. It is interesting that although the BR of $\pi ^ 0\pi ^ 0$ channel and $\eta \eta$ are smaller than Kaon channels, but they still dominate in the photon spectrum. The BF is quite small here, around 0.1, due to the high photon yields from those meson channels.

For these points in parameter space, we can ask about alternative indirect constraints. Gamma rays from dwarf galaxies are a natural constraint \cite{Abdo:2010ex,GeringerSameth:2011iw,Ackermann:2013yva}. In scenarios where the ICS (Inverse Compton Scattering) component is negligible , we expect the dwarf constraints are similar to those for comparable models (such as $\tau \tau$ annihilation). In scenarios where the ICS is significant, it will be weaker, with no starlight or confining magnetic fields to trap the electrons near the dwarfs to produce a comparable signal. Searches for $\bar p$ are clearly not relevant, as they are kinematically forbidden, and are an important distinguishing feature of these models. CMB constraints
from WMAP is not sensitive to our scenario currently, but the updated Planck
constraints may put new limits on the dark photon model \cite{Slatyer:2009yq,Finkbeiner:2011dx,Madhavacheril:2013cna}. We will return to the AMS constraints on positrons shortly.

\subsection{The role of ICS and Bremsstrahlung}
Models that produce copious $e^+e^-$ pairs can produce secondary photons from interactions with the surrounding medium (gas, starlight, cosmic rays). These components can contribute to the total signal~\cite{Cirelli:2013mqa,Lacroix:2014eea,Abazajian:2014hsa}. In particular, we find that for very light dark mediators $m_\phi \lesssim 0.5 \gev$, these can be the dominant component in the central region. For heavier mediators, it can be an O(1) change to the spectral shape at low energies, while for the heaviest mediators $m_\phi \gtrsim 1 \gev$, which have $\pi^0$'s, it is a small effect.

Bremsstrahlung is perhaps the hardest to model, because it has a profile that is tightly correlated to the gas, and thus to the disk. However, not all of this will be absorbed into the disk model. To account for this, we calculate the contributions from bremsstrahlung by masking out the disk region $-1^\circ <b<1^\circ$.
These plots should be understood to be the contributions to the signal in the inner galaxy region, where
$1^\circ <|b|< 20^\circ$ and $|l| < 20^\circ$.

We see in right panel of Figures \ref{fig:DFspectrum1} and \ref{fig:DFspectrum2} that for light mediators, where the dominant contribution is IB (Internal Bremsstrahlung) and FSR (Final State Radiation), that the ICS and Bremsstrahlung signals contribute at a sizable level, while for heavier mediators, the effect can be merely to add additional soft gamma, or to have a marginal effect. Interestingly, once taking into account the effects of these secondary photons, no point in parameter space requires a boost factor much larger than 1. In the Figure \ref{fig:DSspectrum}, the dark scalar also has similar story.

Furthermore, this raises the prospect, however, if at some point we have an accurate map of this signal, to look for deviations in the spectral shape as we move from the inner region to the outer, where these secondary gammas are less prevalent. Indeed, this may lead to a more rapid falloff in the size of the signal that would have been expected from the DM profile alone, simply because these secondary photons become less significant in the outer region.

%% file: Constraints.tex
\section{Constraints}
\label{sec:constraint}
Constraints on this scenario can be grouped into constraints on the signals of the DM, itself, or on the dark mediator.

\subsection{ Constraints on $\epsilon$}
\begin{figure*}
   \includegraphics[width=0.7\textwidth] {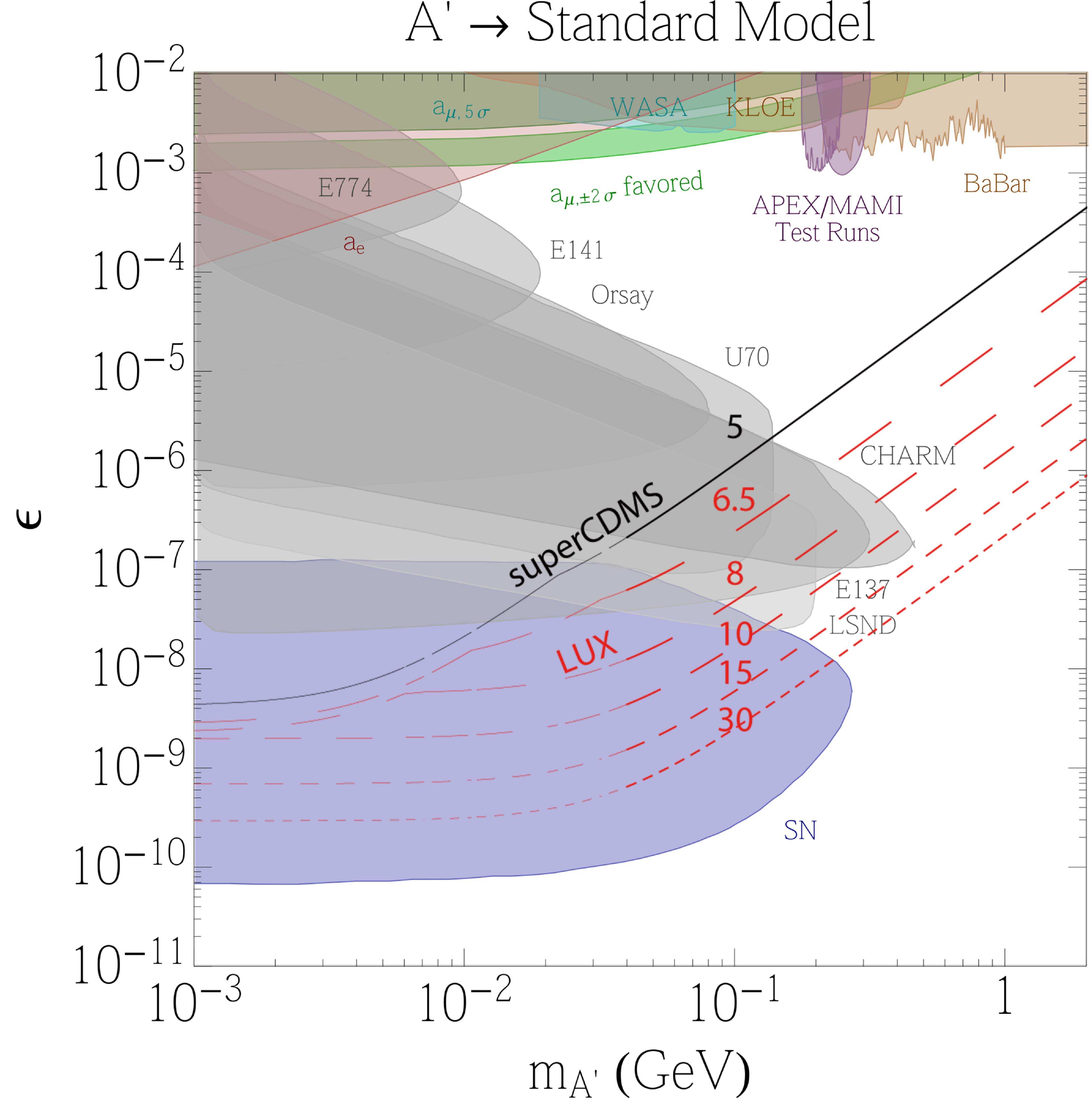}
   \caption{ Parameter space for Dark Photon \label{fig:PS-darkphoton}.
Diagonal lines : contours of spin-independent direct detection constraints
for different DM mass from LUX and superCDMS.
Backgrounds shows current dark photon
constraints from other dark photon search~\cite{Essig:2013lka}.  These limits do not apply to the scalar mediator, or pseudo-Dirac DM case.}
\end{figure*}

The constraints on the mediator are strongest when it is a dark photon, and come
mainly related to its mixing parameter with Standard Model, $\epsilon$. These limits are
derived from searches in beam-dump experiments, fixed target experiments, and
e-e collisions, among others. For a given DM mass $m_\chi$ and
DM coupling to the dark photon, $g_X$, a constraint can also be derived from
DM direct detection searches. These constraints are summarized in
Figure ~\ref{fig:PS-darkphoton}.

Figure ~\ref{fig:PS-darkphoton} shows the parameter space for dark photon.
The beam dump experiments, such as E141~\cite{Riordan:1987aw},
E137~\cite{Bjorken:1988as}, E774~\cite{Bross:1989mp}, etc used the displaced
decay vertex covering the lower left corner of the parameter space.
The fixed target experiments, the anomalous magnetic moment measurement and
$e^+ e^-$ and hadronic collisions give the constraints on the upper part of the
space. Much of the high mass range has been explored by the BaBar experiment  \cite{Lees:2014xha}. There is much parameter space left for the dark photon search in the
dark photon mass from $10 ~\mev$ to a few $\gev$, although this is now being probed by MaMi \cite{Merkel:2011ze}, APEX \cite{Abrahamyan:2011gv}, HPS \cite{hpsweb}, and DarkLight \cite{Freytsis:2009bh,Balewski:2013oza}, among others.

We display the constraints from direct detection on this plot as well. DM-nucleus scattering arises via dark photon exchange. The DM-proton scattering cross section is
\begin{equation}
   \sigma_p \simeq   \frac{\epsilon^2 ~ g_X^2 ~ e^2}{\pi}   \frac{\mu_{\chi p}^2 }{     \left( Q^2 + m_{A'}^2 \right)^2}
            \simeq  1 \times 10^{-43} cm^2
         \left( \frac{ g_X}{0.1} \right)^2
         \left( \frac{ \epsilon}{1 \times 10^{-8}} \right)^2
         \left( \frac{ 0.1 \GeV}{ m_{A'} }\right)^4
   \label{eq:sigma_p}
\end{equation}
where $\mu_{\chi p}$ is the DM and proton reduced mass; and $Q$ is the monmentum
transfer $Q = \sqrt{2  m_N E_r}$, which is related to the nuclei mass $m_N$ and the
recoil energy $E_r$.

 In the second equality of (\ref{eq:sigma_p}), we assume the dark photon mass is
larger than the t-channel momentum transfer
of the scattering process. The dark photon mass should be larger than
$\mathcal{O}(10)\mev$ for this assumption to be valid. For smaller dark photon masses, this breaks down and the t-channel momentum transfer becomes important.
To clarify this effect and the limits of validity of our curves, we have inserted a momentum transfer Q into the propagator, in which $Q = 35~\mev$ for LUX, $ Q = 5~\mev$ for CDMSlite and $ Q= 17~\mev$ for superCDMS.
This changes the behavior of the limits in Figure ~\ref{fig:PS-darkphoton}, and we have changed color into lighter ones in this regions where it occurs, in which case these limits are only approximate.

With the DM mass given, we can fix $g_X$ through the relic density constraint, (e.g. for $m_\chi =10\gev$, $g_X = 0.06$). In Figure ~\ref{fig:PS-darkphoton},
superCDMS~\cite{Agnese:2014aze} and LUX~\cite{Akerib:2013tjd} are considered,
which are currently the best constraints of
spin-independent cross section in the DM mass range of 5~$\GeV$ - 30~$\GeV$.

Importantly, is that these limits are {\em only} present if the dark matter is a Dirac fermion. If the DM is split into a pseudo-Dirac state after U(1) breaking, then the scattering is inelastic and can be kinematically suppressed \cite{TuckerSmith:2001hy}, leaving no appreciable constraint on these models.

Finally, these constraints are on the dark photon model. For the dark scalar, with its weaker interaction with ordinary matter, both the production and direct detection constraints are weaker.

%% file: AMS02.tex
\subsection{Constraint from AMS-02}


AMS-02 precisely measured the smooth electron, positron spectrum and
the positron ratio. We can turn these smooth data into a constraint on
light DM~\cite{Bergstrom:2013jra,Hooper:2012gq,Ibarra:2013zia}. If the light DM annihilates to
electrons and positrons and this
cross section is large enough, after the transportation of the electrons and positrons, a bump
feature would expect to be seen in the AMS-02
positron ratio data. Since we have not seen this bump yet,
the current measurement is able to put stringent constraints on light DM models.

We revisit the study of \cite{Bergstrom:2013jra} on the limit of DM annihilation from AMS-02,
and consider more channels and the systematic uncertainties from solar modulation and magnetic fields.
Our limits are not as stringiest as those in \cite{Bergstrom:2013jra}, and so we list the major differences here:
\begin{itemize}
   \item we use 2 parameters ($m_\chi$ and $\langle \sigma v \rangle $) to compute the relevant regions for $\delta \chi^2$, while \cite{Bergstrom:2013jra} use 1 parameter to do so. Furthermore, we plot a 3 sigma contour, and $\Delta \chi^2 = 11.83$, while \cite{Bergstrom:2013jra} plots
         $90\%$CL.
   \item we consider the uncertainties of solar modulation, while \cite{Bergstrom:2013jra} consider
         specific values of solar modulation parameters
   \item we choose one plain diffusion model, but test the uncertainties from the
         parameters in the cosmic ray diffusion.
         It turns out that the variation of the magnetic field or the effect of the energy loss
         influence the AMS-02 constraints most.
   \item we set $\rho_{\odot} = 0.4 \GeV / cm^3$ to be consistent with our Galactic Center analysis,
         while in \cite{Bergstrom:2013jra}, the minimum density is $\rho_{\odot} = 0.25 \GeV / cm^3$
\end{itemize}

First of all, instead of simulating the astrophysical background,
we apply polynomial functions to fit the AMS-02 electron spectrum
and positron ratio separately from 1 \GeV. After obtaining the two functions, we
derive the positron spectrum, and recheck the fit to AMS-02 positron data.
Secondly, we compute the positron or electron flux from
DM annihilation propagating in our galaxy, by using a public cosmic ray code
DRAGON~\cite{Evoli:2008dv}.

Before propagation, the positron spectrum is delta function for the process of
$\chi + \chi \rightarrow e^+ + e^-$, $\frac{\d N_e} {\d x} ( 2e )= \delta ( 1 -x )$ by
neglecting fragmentation.
For this process with one step cascade decay, $\chi + \chi \rightarrow \phi + \phi$ and
$\phi \rightarrow e^+ + e^-$, the spectrum is a box-like function
$\frac{\d N_e} {\d x} ( 4e )=  2 \theta ( 1- x ) $. After propagation,
the diffusion and energy loss make the positron flux softer.

\begin{figure*}
   \includegraphics[width=0.45\textwidth] {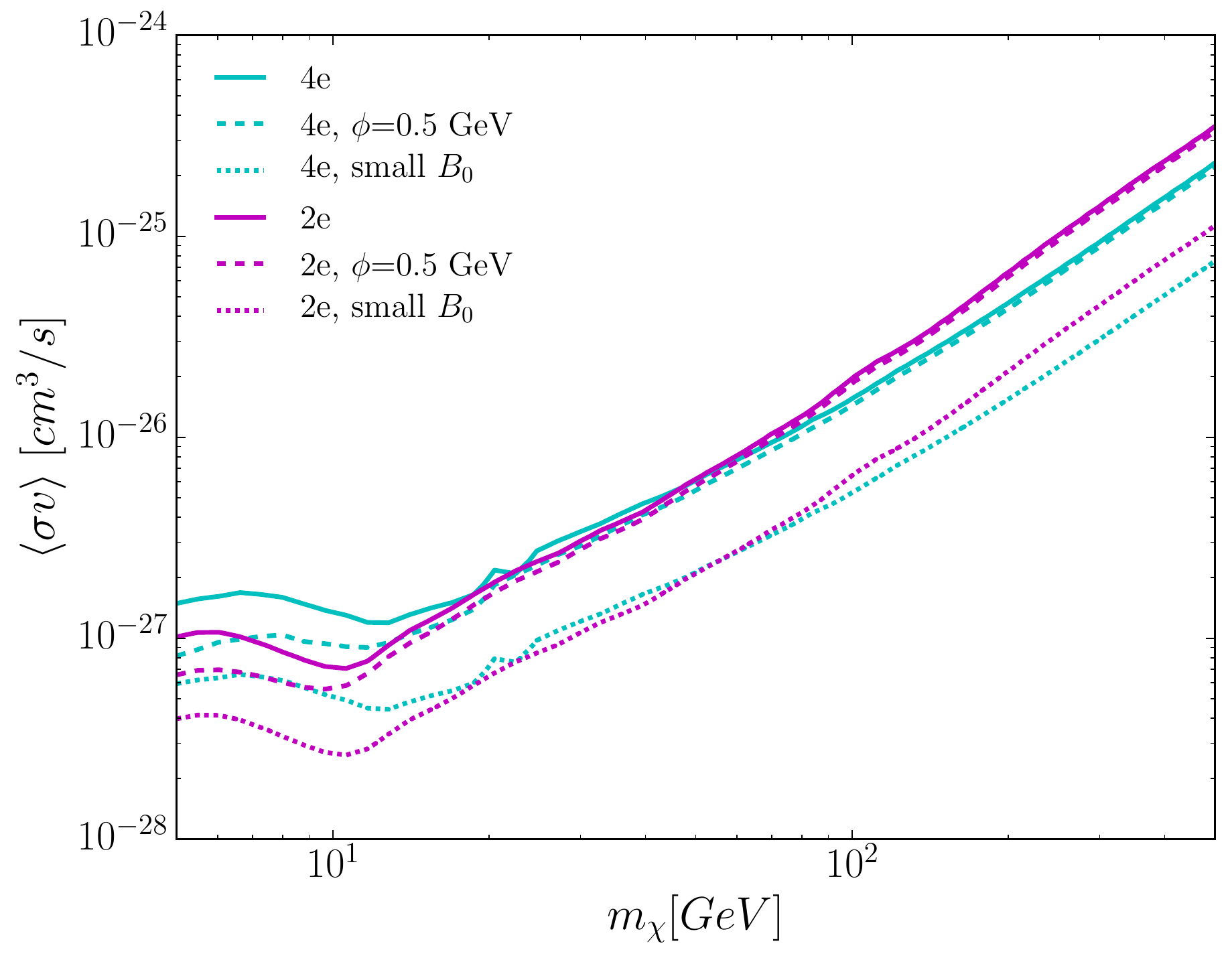}
   \includegraphics[width=0.45\textwidth] {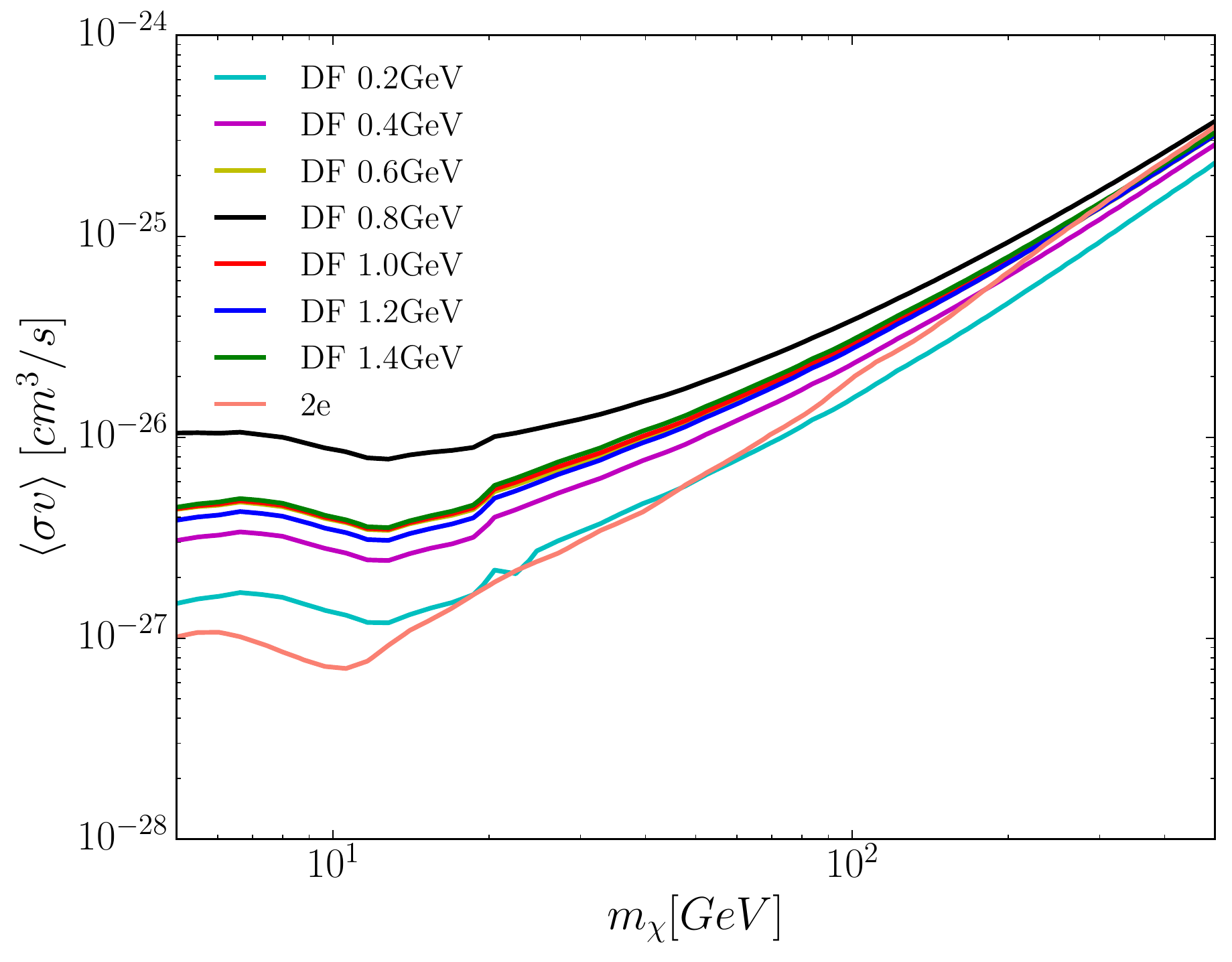}
   \caption{ Exclusion curves for different DM models and for different assumptions of cosmic ray
      propagation. In the left panel, the process of $\chi + \chi \rightarrow 2 e$ and $ \chi + \chi \rightarrow 4e $
      are considered. The solid lines take into account of the uncertainties from solar modulation, and choose
      large magnetic fields $B_\odot = 15 \mu G$. The dashed lines choose the solar modulation $\phi = 0.5$ \GeV,
      and the dotted
      lines consider a smaller magnetic fields $B_\odot = 7.5 \mu G$. In the right panel, the exclusion limit of
      various dark force mass assuming a dark photon model are included.  }
   \label{fig:amscontoursC}
\end{figure*}

We compare the cross section limits by choosing different magnetic fields and considering
the variation of the solar modulation or not in Figure ~\ref{fig:amscontoursC}.
The magnetic field is modeled as two main components, regular one and the turbulent one~\cite{Pshirkov:2011um,DiBernardo:2012zu},
but little is known for the size of magnetic field.
The total magnetic field we choose at Sun is $B_\odot = 15 \mu G$.
In the left panel of Figure ~\ref{fig:amscontoursC}, the solid line is $B_{\odot} = 15 \mu G$, while the
dotted lines corresponds to $B_{\odot} = 7.5 \mu G$.
In addition, the solid line considers the variation of
the solar modulation, while the dashed line fixes the solar modulation potential by $\phi = 0.5$ \GeV.
The limits differ by a factor of 2 for DM mass smaller than $10 \GeV$.
In the right panel of Figure ~\ref{fig:amscontoursC}, we plot the exclusion limit for different mass of
dark force mediator.

The implication for result is that for $\sim 10$ \GeV~ DM, if the branching ratio of $ \chi + \chi \rightarrow e^+ + e^-$
or $\chi + \chi \rightarrow 2 e^+ + 2 e^-$ is larger than $\sim 5\%$ and the cross section is the thermal cross
section $3 \times 10^{-26} cm^3/s$, the model has tension with AMS-02. In other words, if the branching ratio is
$100\%$ to $2 e$ and $4 e$, the cross section should be smaller than $\sim 1 - 2\times 10^{-27} cm^3/s$.
For the dark photon models, the branching ratio to $4e$ is generally about $30\%$, except in the resonance region. In the resonance region (e.g. $m_{\phi _\mu  }  \sim 0.8\gev$), $4e$ channel is suppressed and AMS constraint could be satisfied. In the non-resonance region, one needs either a small BF by large $\pi^0$ production in heavy dark photon region or a large ICS and Bremsstrahlung contribution in the light dark photon region, to alleviate the AMS constraint.

We see that most of the light dark photon mediator models would appear to be constrained. For instance, for the light mediators, we require a cross section $\sim 2\times 10^{-26}\rm cm^3 s^{-1}$, while the limits are $2\sim 3  \times 10^{-27}\rm cm^3 s^{-1}$. However, for heavier mediators, this is less of a problem. For a 1.4 GeV mediator, for instance, we need a cross section $\sim 4.5\times 10^{-27}\rm cm^3 s^{-1}$, while the limit is $\sim 5\times 10^{-27}\rm cm^3 s^{-1}$, comparable to the cross section we need. For a 0.8 GeV mediator, the limits are around $10^{-26}\rm cm^3 s^{-1}$, again comparable to the cross section we need. For dark scalar models, the constraint is generally much weaker, because $e^+e^-$ channel has much smaller BR than dark photon by Yukawa coupling. We note that since the Fermi signal arises from the central galaxy, while AMS is from more local annihilation, a somewhat steeper profile than what we take here could lead to alleviations in the remaining tensions.

%% file: CMB.tex
\subsection{Constraint from CMB}

\begin{figure*}[t]
   \includegraphics[width=0.45\textwidth] {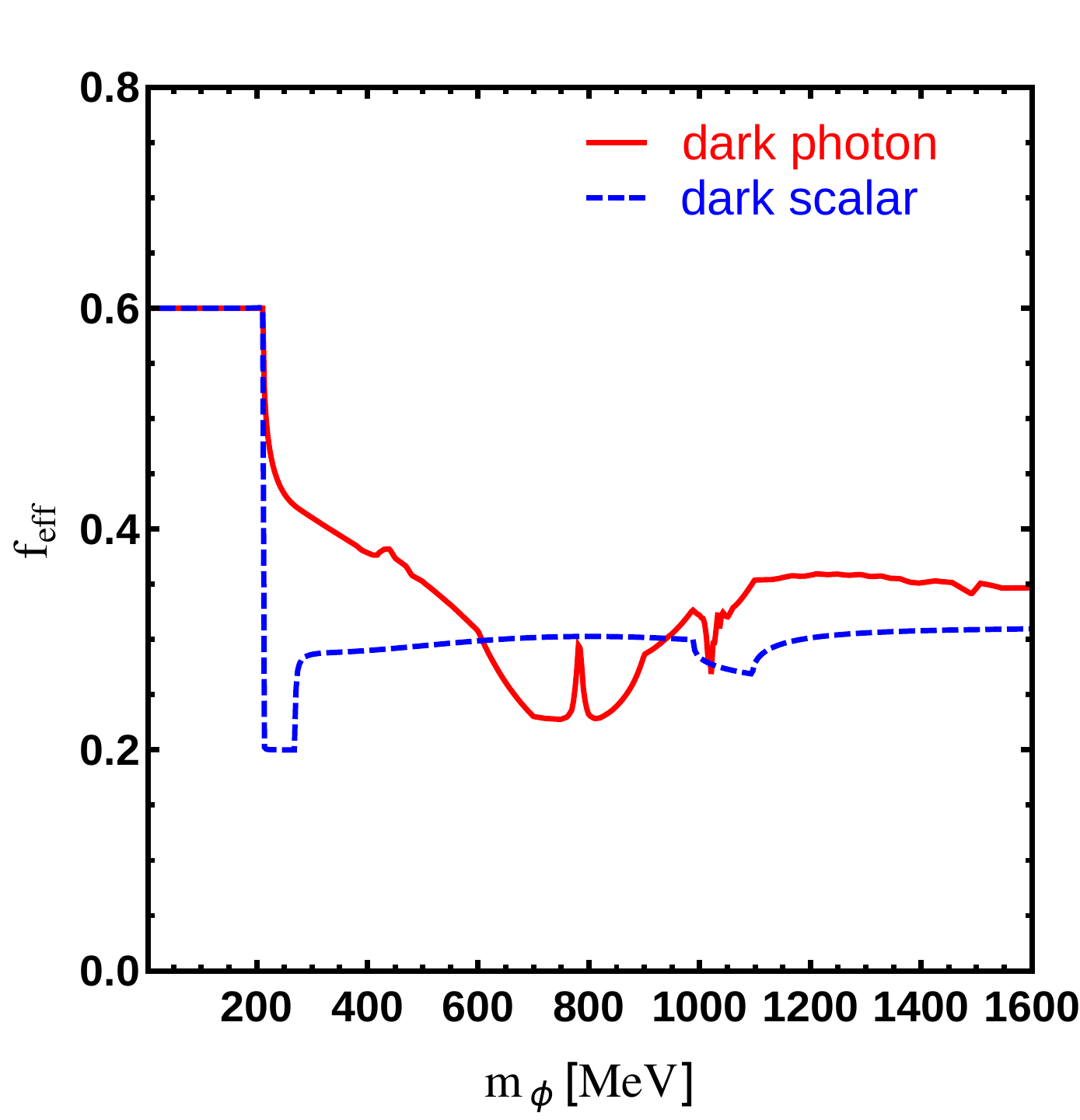}
   \caption{  $f_{eff}$ for the dark photon and dark scalar.}
   \label{fig:feff}
\end{figure*}

DM annihilation can inject energy into the CMB, which distort its temperature and polarization power spectra \cite{Chen:2003gz, Padmanabhan:2005es}. The anisotropy of CMB can constrain the DM annihilation \cite{Galli:2009zc,Slatyer:2009yq,Finkbeiner:2011dx}. In 2015 Planck data \cite{Planck:2015xua}, it shows very strong constraint on low mass DM annihilation. To calculate the constraint the annihilation to dark mediators, we start with the efficiency factor $f_{eff}$, which describes the fraction of the energy injected into the gaseous background. Following the data in ref.~\cite{Madhavacheril:2013cna, Cline:2013fm}, we assume $f_{eff}$ are $0.6$, $0.2$, $0.16$ and $0.62$, for dark mediator decay channels $e^+e^-$, $\mu^+\mu^-$, $\pi^+\pi^-$ and $\gamma \gamma$ respectively. $f_{eff}$ has some mild dependence on the dark matter mass $m_\chi$, but since we consider a small range of $m_\chi$ around $10$~GeV, we neglect it. For other particles, we can build up their $f_{eff}$ through decay branching ratio and decay products. For $\pi^0$,  we assume its $f_{eff}$ is the same with $\gamma$. After some calculation, $f_{eff}$ for $K^{\pm}$, $K_L^{0}$, $K_S^{0}$ and $\eta$ are $0.18$, $0.37$, $0.42$ and $0.54$ respectively. We calculate $f_{eff}$ for the dark photon and dark scalar, according to their decay branching ratios, in Fig.~\ref{fig:feff}.

\begin{figure*}[t]
   \includegraphics[width=0.45\textwidth] {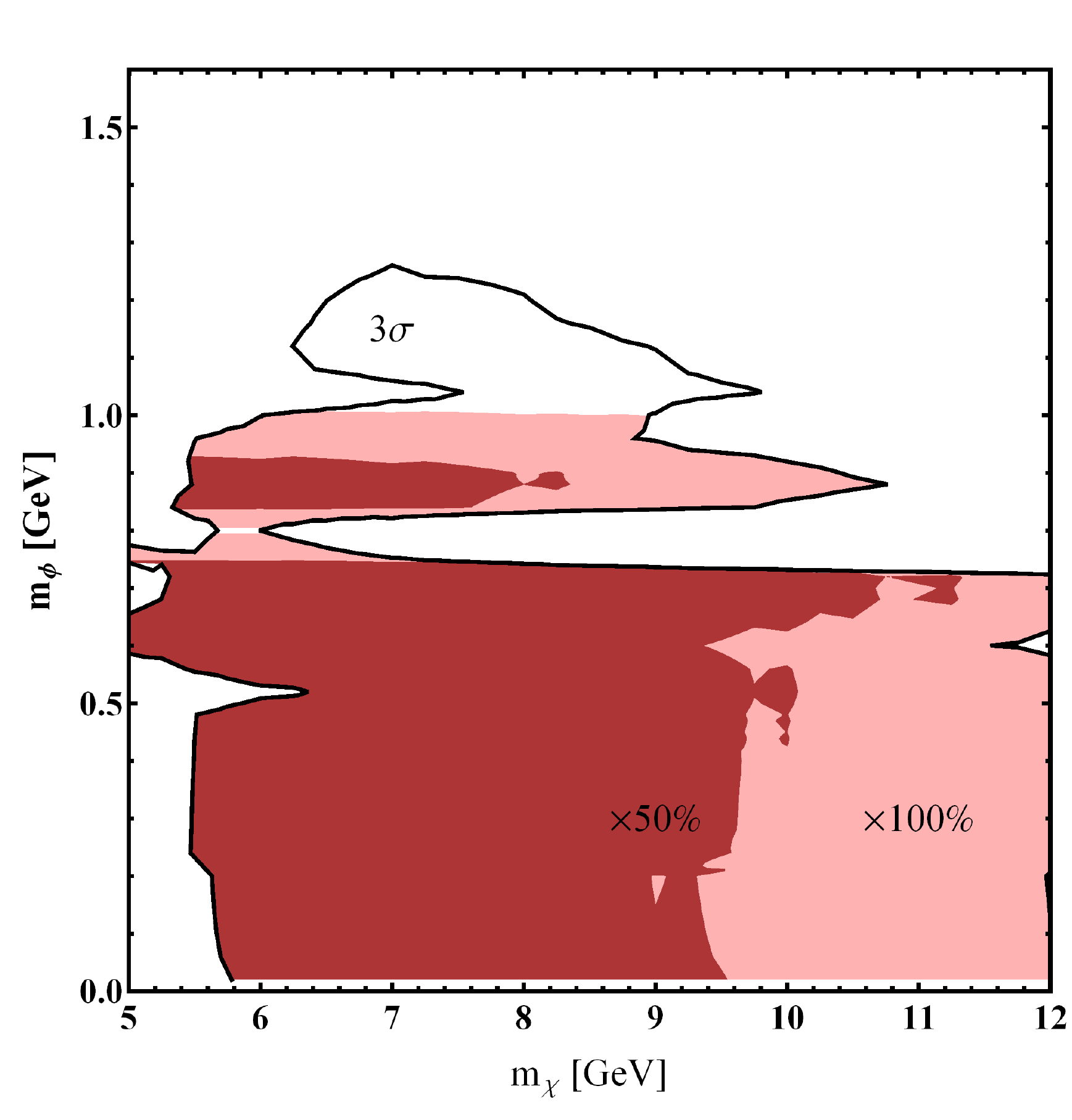}
   \includegraphics[width=0.45\textwidth] {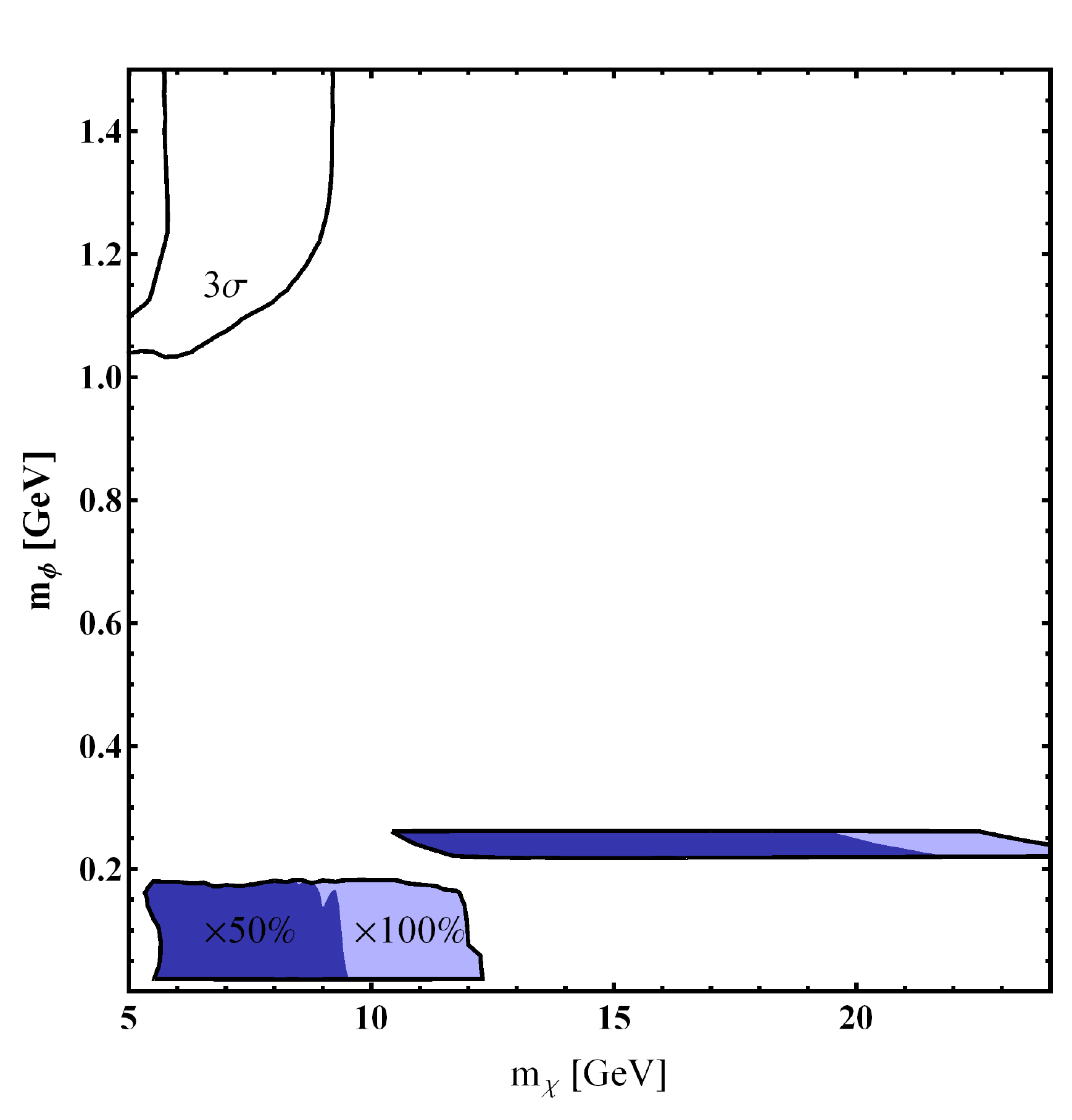}
   \caption{The CMB constraints on the DM annihilation in $m_\chi - m_\phi$ plane for dark photon (\textit{left panel}) and dark scalar (\textit{right panel}). Inside the solid black contours are the $3 \sigma$ best fit region for dark photon and dark scalar in Fig.~\ref{fig:DFcontour} and \ref{fig:DScontour}. The light (dark) color shaded regions are excluded by CMB, assuming best fit cross-section for GCE times $\times 100\%$ ($\times 50\%$).}
   \label{fig:planck}
\end{figure*}

Planck can constrain the annihilation cross-section $\left\langle {\sigma v} \right\rangle _{rb}$ at recombination times the efficiency parameter $f_{eff}$ \cite{Planck:2015xua}. We assume the boost factor for annihilation at recombination is the same as today. To derive the constraints on the light dark force scenario, we apply the annihilation cross-section from the $\chi^2$ fit, which is the thermal cross-section times the BF from right panel of Fig.~\ref{fig:DFcontour} and Fig.~\ref{fig:DScontour}.

We plot the constraints on the light dark scenario in Fig.~\ref{fig:planck} in $m_\chi - m_\phi$ plane. Inside the black contour, it is the $3 \sigma$ best fit region for dark photon and dark scalar. We can see that most of the best fit region for GCE are excluded, as indicated by light red shaded region for dark photon and light blue shaded region for dark scalar, except when dark mediator is heavier than $1$~GeV. Those region survive because their needed cross-section are quite small due to direct photon contribution from meson decay. This is true for both dark photon and dark scalar. Moreover, if we weaken our signal by a factor of $50\%$, significant parameter space opens for $m_\phi < 1$~GeV. It means if we allow a partial fit to GCE, more parameter space could survive. In summary, the GCE excess from dark mediator interpretation can still survive significant parameter space, e.g. $m_\phi >1$~GeV or if we allow a partial interpretation for GCE. We also plot the contours of excluded annihilation cross-section at freeze-out from Plank as a function of $m_\chi$ and $ m_\phi$ in Fig.~\ref{fig:CMBthermal1} in Appendix~\ref{sec:CMBthermal}. We assume BF at freeze-out and recombination are the same. It shows DM with thermal cross-section $3 \times 10^{-26} cm^3/s$ in the dark mediator models should be larger than $\sim 20$~GeV.

%% file: conclusion.tex
\section{Summary and Conclusions}
\label{sec:conc}

While the nature of dark matter has remained elusive, tremendous progress has been made in constraining its nature. The recent evidence of a $\gev$ excess from the Galactic Center, arising from analysis of data from the FGST
Galactic Center \cite{Daylan:2014rsa,Calore:2014xka} invites interpretations as being of a DM origin.

We have revisited the proposal of DM annihilating into a light mediator as an explanation for these signals.
We have carefully studied the decay branching ratios of the light
mediator and the various meson channels which produces the gamma-rays. We have scanned
the best fit region for the dark force scenario, both with dark photons and dark scalars. The result shows that for
mediator masses $ \lesssim 1.5 \gev$ and DM mass  $\lesssim 10 \gev$,
lepton final states or combination with meson final state could give a very
good fit for the GeV excess, which is in agreement with \cite{Hooper:2012cw}.

We note that what we have discussed here should be considered simplified models for this scenario. Annihilations $\chi \chi \rightarrow \phi_\mu h$, where $h$ is the Higgs field for the dark photon can occur at a parametrically similar rate for the Dirac DM case. There may be multiple dark photons (i.e., as in \cite{ArkaniHamed:2008qn}), leading to more complicated cascade spectra. And, if there are additional scalars in the dark sector, there could be an intermediate step in the cascade as well. Thus, the spectral shape may vary as these complications are present, which may lead to changes in interpretation. Much of these can be considered as combinations of the dark photon and dark scalar spectra presented.

While the prompt photon spectrum is typically dominant, the contributions from Bremsstrahlung and ICS can change the picture. For light mediators, it can be an O(1) component of the total signal in the GC, while for heavier mediators it becomes less important. As the lightest mediator models are more tightly constrained by AMS, it is unlikely that these secondaries are the dominant sources of the gamma rays we observe if DM is in the mass range we consider. However, it still may be important and lead to spectral changes going from the GC to the inner Galaxy regions.

Since we lack understanding about the detailed nature of the diffusion of cosmic rays near the GC, there are important systematic uncertainties in calculating the ICS contribution to the gamma ray signal.
Still, it is clear that the ICS from DM-induced electrons and
positrons gives contributions to the gamma-ray spectrum, especially at slightly lower energy than the prompt photons.
Interestingly, in some diffusion models, the morphology for ICS is similar to
the one of the GeV excess, while in other models it is different. Finally, while these uncertainties are present, it is
essential to understand its effects on GeV gamma-ray excess, both in the change of spectrum and of morphology, especially to do detailed comparisons of models and data.

Ultimately, while the nature of the gamma ray excess remains unclear, we do see here that annihilations into dark sector cascades provide a good explanation of the data. Upcoming searches, both terrestrial and astrophysical, may shed light on whether such a weakly coupled light sector exists in nature.

\vskip 0.5in
{\bf Note added:} As this work was being completed, \cite{Calore:2014nla} appeared, which considers the ICS signals from somewhat heavier DM candidates. Our results are in good agreement on the consequences of ICS for these signals.

%% file: branchingratios.tex
\section{Branching ratios}
\label{sec:BR}

In both dark photon and dark scalar scenarios,  $\phi$ will decay to leptons and mesons.
In order to obtain the photon spectrum from the decays,
we will first derive the branching ratios of their decaying
channels. For the dark photon, a data driven method is employed, and
for dark scalar, a theoretical analysis is provided.

\begin{figure*}
\includegraphics[width=0.8\textwidth]{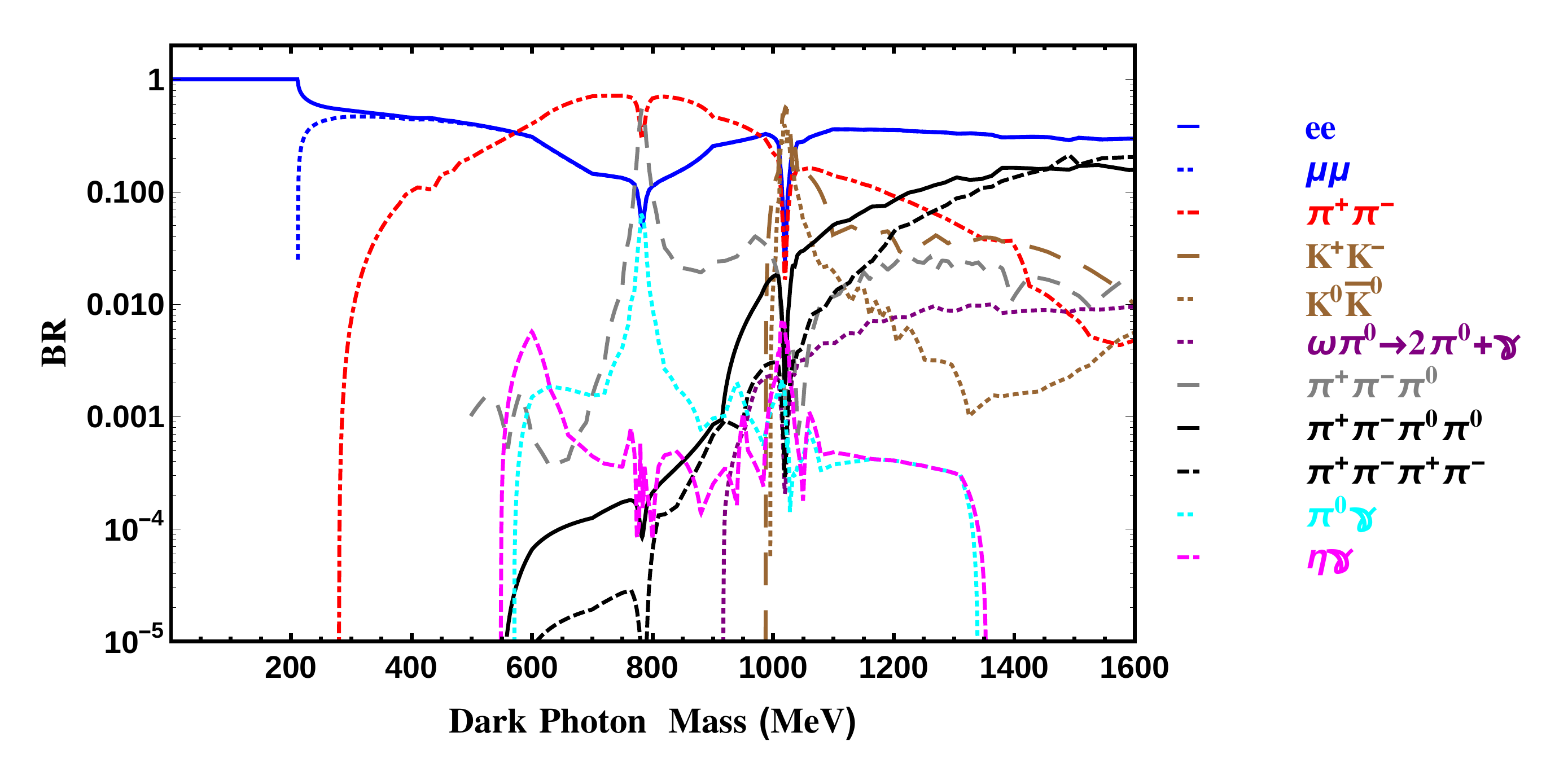}
\caption{The decay branching ratios for dark photon. \label{fig:BRdarkphoton}}
\end{figure*}

In the dark photon scenario, DM annihilation to dark photons is followed by
decay of the on-shell dark photons to SM particles.
Since the kinetic mixing between dark photon and
photon, the dark photon decay can be analyzed using the measurements of
$e^+ e^- \rightarrow \mathrm{hadrons}$ at different Center of Mass~(C.M.) energies.
Suppose the dark photon mass is the same as the C.M. energy of the $e^{+}e^{-}$ collision,
the ratio of the cross-section of the different final states reveals
the branching ratio of the dark photon decay products.
When the mass of dark photon is above $\sim 2 \mathrm{GeV}$, the perturbative
QCD is valid from the observation that the energy dependence of $R(s) = \frac{ \sigma \left(
e^+ e^- \rightarrow \mathrm{hadrons} \right) } { \sigma \left(  e^+ e^- \rightarrow \mu^+ \mu^-
\right) }$
matches with the QCD prediction~\cite{Beringer:1900zz}; hence the underlying processes are $\phi \rightarrow
q \bar{q}$ and $\phi \rightarrow l \bar{l}$. At the C.M. energy below
$\sim 2 \mathrm{GeV}$,
there are rich structure of resonance, such as $\rho, \omega$, and  $\phi$, and different
exclusive channels are measured separately. We obtain the branching ratio of the channels 
from the exclusive cross-sections at different C.M. energies \cite{0954-3899-29-12A-R01, hepdata}.
We have included all the two body final states shown in
Figure~\ref{fig:BRdarkphoton}. For multiple particle final states, we
only include three pion and four pion final states and neglect
others like $K^+K^-\pi^0$, as well as five pion and six pion states, because
these have subdominant contribution to the photon yield.
\footnote{Only for ${\pi ^ + }{\pi ^ - }{\pi ^ + }{\pi ^ - }$
channel, there is measurement at $3$GeV, while for other channels
the highest measurement is around $2.4$GeV.} As a caveat, in the $\omega {\pi ^0}$ channel, we only
include the final states when $\omega$ decays into ${\pi ^0}\gamma
$. The $\omega$ dominantly decays into three pions, but it is
already considered in the three and four pion final states. However,
in the ${K ^ + }{K^ - }$ and ${K ^ 0 }{K^ 0 }$ channel, their
cascade decays includes four pion final states, which are not
included in the four pion channel in Figure~\ref{fig:BRdarkphoton}.
Thus we calculate the spectrum of $KK$ and $4\pi$ states separately.

In the dark scalar mediator scenario, the DM annihilates into a pair of dark scalars, which, through their mixing, subsequently decay into SM fermions.
The dark scalar's coupling to SM fermions is proportional to the fermion mass, and suppressed by the mixing
term $\epsilon$, while the heavy fermions (c,b,t) are decoupled and will
influence the low energy hadronic process by coupling to gluons. Hence we are able to write down
the effective Lagrangian in the following form
\begin{equation}
   \mathcal{L}_{eff} =  \epsilon \frac{\phi } {v} \left( -\sum_{q = u,d,s} m_q \bar{q} q
   + \frac{ \alpha_s N_H} { 12 \pi} G_{\mu \nu}^a G^{\mu\nu a} \right) \ ,
   \label{eq:lint1}
\end{equation}
where $v$ is the Higgs vev, and $N_H = 3 $ is the number of the heavy quarks. Introducing the
trace of energy momentum tensor $\theta_\mu^\mu$ can relate the quark level interaction to
the hadronic process. First, $\theta_\mu^\mu$ illustrates the anomaly of the
conformal symmetry, which contains the terms proportional to QCD beta function $\beta$ and
the terms proportional to the mass of the light quarks,
\begin{equation}
   \theta_\mu^\mu = - \frac{ \beta } { 2 g_s }  G_{\mu \nu}^a G^{\mu\nu a}
      + \sum_{q = u,d,s} m_q \bar{q} q  \ .
\end{equation}
On the other hand, $\theta_\mu^\mu$ is related to the hadronic process, 
and at the leading order,
\begin{equation}
   \left<   \pi^+ \pi^-  |   \theta_\mu^\mu | 0 \right> = s + 2 m_\pi^2 + \mathcal{O} ( p^4) .
   \label{eq:matrix_tensor}
\end{equation}
From the first order of the chiral Lagrangian, we are able to derive the other hadronic matrix element,
\begin{equation}
   <   \pi^+ \pi^-  |   \sum_{q = u,d,s} m_q \bar{q} q | 0 > \simeq  m_\pi^2 \ .
   \label{eq:mq}
\end{equation}
After replacing the $G_{\mu \nu}^a G^{\mu\nu a}$ term by $\theta_\mu^\mu$ and
$\sum_{q} m_q \bar{q} q$ in the effective Lagrangian eq.~(\ref{eq:lint1}),
the decay width of the dark scalar is computed by combining the two matrix elements in eq.~(\ref{eq:matrix_tensor},
\ref{eq:mq}),
\begin{align}
 \Gamma (\phi  \to {\pi ^ + }{\pi ^ - } ) =  \frac{\epsilon^2 {m_\phi ^3}}{{324\pi {v^2}}}
      {\left( {1 - \frac{{4m_\pi ^2}}{{m_\phi ^2}}} \right)^{1/2}}
      {\left( {1 + \frac{{11m_\pi ^2}}{{2m_\phi ^2}}} \right)^2} \ .
\end{align}
Due to the isospin symmetry, the ratio of charged states (e.g. ${\pi ^ + }{\pi ^ - }$) to neutral states (e.g. ${\pi ^0}{\pi ^0}$) is just $2:1$.
The decay width to $K\bar K$ and $\eta \eta $ are similar with pion by adding a statistical factor of $4/3$ and $1/3$ respectively \cite{Gunion:1989we}
and substituting the pion mass by Kaon mass and Eta mass.\footnote{For the light mass Higgs, there are debates about the ratio
$BR({\mu ^ + }{\mu ^ - })/BR(\pi \pi )$ (see \cite{Clarke:2013aya} and references therein).
Our result are insensitive to such debate, because the photon spectrum from muon pair final states is similar to charged
pion pair final states. }
We also list the decay width to leptons here.

\begin{align}
\Gamma (\phi  \to \ell ^ +  \ell ^ -  ) = \frac{{\varepsilon ^2 m_\ell ^2 }}{{8\pi v^2 }}m_\phi  (1 - \frac{{4m_\ell ^2 }}{{m_\phi ^2 }})^{3/2}
\end{align}

The decay width to two photons are the same as the Standard Model Higgs, except the mixing factor. We explicitly list the width formula for photons in the following,

\begin{align}
\Gamma (\phi  \to \gamma \gamma ) = \frac{{\varepsilon ^2 \alpha _{EM}^2 }}
         {{256\pi ^3 }}\frac{{m_\phi ^3 }}{{v^2 }}
      \left| {\sum\limits_i {Q_C^i Q^i F_{1/2} (\frac{{4m_i^2 }}{{m_\phi ^2 }})
       + F_1 (\frac{{4m_W^2 }}{{m_\phi ^2 }})} } \right|^2  \ ,
\end{align}
where the $v$ is the Higgs vev. $i$ runs over all the fermions in the SM.
$Q_C$ is the color factor and $Q$ is the charge of the fermion.
$F_1$ and $F_{1/2}$ are the well known functions,
\begin{align}
 F_1 (x) & = 2 + 3x + 3x \left( 2 - x \right) f(x) \\ \nonumber
 F_{1/2} (x) & =  - 2x \left[1 + \left( 1-x \right) f(x)\right] \ .
\end{align}

The function $f(x)$ is the following,

\begin{eqnarray}
f(x) = \left\{
       \begin{array}{lr}
   \left(\sin^{-1}\sqrt {1/x} \right)^2 , & \quad x \ge 1
   \\
    - \frac{1}{4}\left[\ln \left(\frac{{1 + \sqrt {1 - x} }}
      {{1 - \sqrt {1 - x} }}\right) - i \pi \right]^2 , & \quad x < 1
\end{array}
      \right.
\end{eqnarray}

\begin{figure*}
\includegraphics[width=0.7\textwidth]{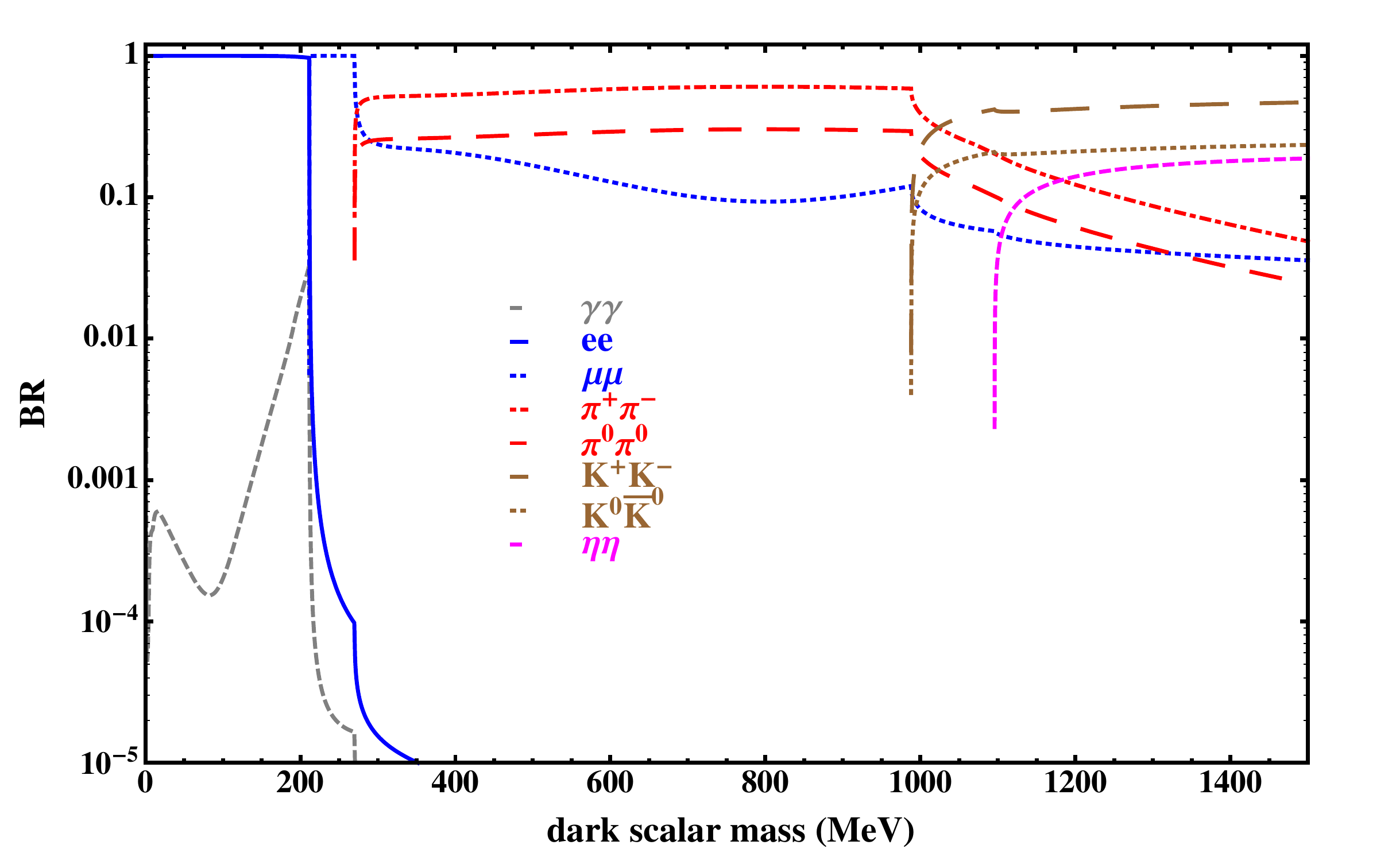}
\caption{The decay branching ratios for dark scalar mediator. \label{fig:BRdarkscalar}}
\end{figure*}

Since we are dealing with very light scalar mass, the quark mass will have significant influence on the width. Here we take the current quark mass. We plot the decay branching ratios for dark scalar mediator in the Figure~\ref{fig:BRdarkscalar}.

%% file: photon-spectrum.tex
\section{Photon spectrum in the lab frame}
\label{sec:photonspc}

We present how we calculate the photon spectrum in the cascade decays. We generally follow the notation and procedure in the \cite{Mardon:2009rc}. The difference is we take into account the finite mass of the mother particle and daughter particles, however, in \cite{Mardon:2009rc} the daughter particles are treated as massless to simplify the calculation. In our case, since we want to scan for dark photon and dark scalar mass, there are regions where their mass are close to the threshold of the daughter particle, thus taking account the finite mass into boost calculation makes the photon spectrum more accurate. We take into account the dark photon and scalar mass, and also the various meson mass in their cascade decays.

To show the boost calculation quantitatively, we assume a process where mother particle $A$ decays to daughter particles $B_i$, where the $i$ is the $i$th daughter particle.

\begin{align}
A \to \sum\limits_i {{B_i}}
\end{align}

The number density distribution of photons from particle $B_i$ in the $B_i$ center frame is denoted as $d{N_{{B_i}}}/d{x_{{B_i}}}$. The distribution from FSR and radiative decay are described in detail in section \ref{subsec:FSR} and \ref{subsec:radiativedecay}. The ${x_{{B_i}}}$ is dimensionless quantity defined as

\begin{align}
{x_{{B_i}}} \equiv \frac{{2{E_i}}}{{{m_{{B_i}}}}}
\end{align}

 ,where $E_i$ is the energy of photon from particle $B_i$ in the $B_i$ center frame and ${{m_{{B_i}}}}$ is the mass of particle $B_i$. If the $B_i$ decays directly to photons, for example ${\pi ^0}$, then the total number of hard photons ${N_{{B_i}}}$ in the $B_i$ center frame is about $O(1)$. However, if the photons from $B_i$ are from initial and final state radiation, then ${N_{{B_i}}}$ is about $O({\alpha _{EM}})$. This means once $B_i$ decays directly to photons, then the spectrum $d{N_{{B_i}}}/d{x_{{B_i}}}$ are usually determined by the direct photons. The mesons ${\pi ^0}$, $\omega $ and $\eta $ can directly decay to photons, which are quite important. The Kaon mesons also makes $O(1)$ number of photons, because their decay usually contains ${\pi ^0}$. There are various decay channels for those mesons, we only calculate the leading photon source in the cascade decay. To be concrete, take the $\eta $ decay to ${\pi ^0}{\pi ^ + }{\pi ^ - }$ as an example, we only account the photons from $\pi^0$. The photons from cascade decay in $\pi^\pm$ into muon and finally electron are subdominant. The only exception is when dark photon decays into four pion, two charged and two neutral pions, we account both photon from neutral and charge pions. A detailed description of leading contribution for each channel is in section \ref{subsec:channels}.

With the $d{N_{{B_i}}}/d{x_{{B_i}}}$ in hand, we want to know the photon distribution in the center frame of mother particle $A$, the $d{N_A}/d{x_A}$, where the $x_A$ is $\frac{{2E}}{{{m_A}}}$ and $E$ is the energy of photon in the $A$ center frame. Suppose the momentum of particle $B_i$ has an isotropic spherical distribution in the $A$ center frame and $B_i$ has energy $E_{B_i}$ in $A$ center frame, then the connection between the two distribution is,

\begin{align}
d{N_A}/d{x_A} = \int_{{x_A} \cdot \frac{{{m_A}}}{{{m_{{B_i}}}}} \cdot \frac{{{\varepsilon _{{B_i}}}}}{{1 + \sqrt {1 - \varepsilon _{{B_i}}^2} }}}^{Min[1,{x_A} \cdot \frac{{{m_A}}}{{{m_{{B_i}}}}} \cdot \frac{{{\varepsilon _{{B_i}}}}}{{1 - \sqrt {1 - \varepsilon _{{B_i}}^2} }}]} {d{x_{{B_i}}}} \frac{{d{N_{{B_i}}}}}{{d{x_{{B_i}}}}}\frac{1}{{2{x_{{B_i}}}}}\frac{{{m_A}}}{{{m_{{B_i}}}}}\frac{{{\varepsilon _{{B_i}}}}}{{\sqrt {1 - \varepsilon _{{B_i}}^2} }} \label{eqn:fullboost}
\end{align}

, where ${\varepsilon _{{B_i}}} = \frac{{{m_{{B_i}}}}}{{{E_{{B_i}}}}}$. Sometimes, the number of daughter particles is larger than 2, so $E_{B_i}$ is not fixed by two body final state. In the multi-particles final state like three pion and four pion, the pions do not have a definite energy as in the two body decay. We assume those pions have isotropic spherical distribution in momentum direction, and their energy distribution satisfy the natural phase space distribution. The natural phase space distribution means the momentum satisfy the phase space constraints, assuming the matrix element is a constant. The calculation of momentum distribution is in section \ref{subs:nbody}. With the distribution in hand, we can average $d{N_A}/d{x_A}$ over $E_{B_i}$ with proper possibility function. We use this method to trace back the number distribution of photons level by level, until to the lab frame and take fully account the mass of all the daughter and mother particles. We only omit the daughter mass in the last step, when boosting the photon back into lab frame, $DM + DM \to \phi \phi $. The $\phi$ is dark photon or dark scalar. The last boost can be seen as a hypothetical particle with mass of twice DM mass and decay into two $\phi$. We assume the $\phi$ mass is negligible to this this hypothetical particle and set it to zero. In this case, the equation \ref{eqn:fullboost} can be simplified as

\begin{align}
d{N_A}/d{x_A} = \int_{{x_A}}^1 {d{x_{{B_i}}}} \frac{{d{N_{{B_i}}}}}{{d{x_{{B_i}}}}}\frac{1}{{{x_{{B_i}}}}}
\label{eqn:boost2lab}
\end{align}

, where $x_A = E/m_{DM}$ and $E$ is the photon energy in the lab frame. This simplification will not change the accuracy of the photon spectrum significantly, because in our region of interest, the dark photon has mass around $O(1)$ GeV, while twice DM mass is around $O(10)$ GeV.

\begin{figure*}
\includegraphics[width=0.45\textwidth]{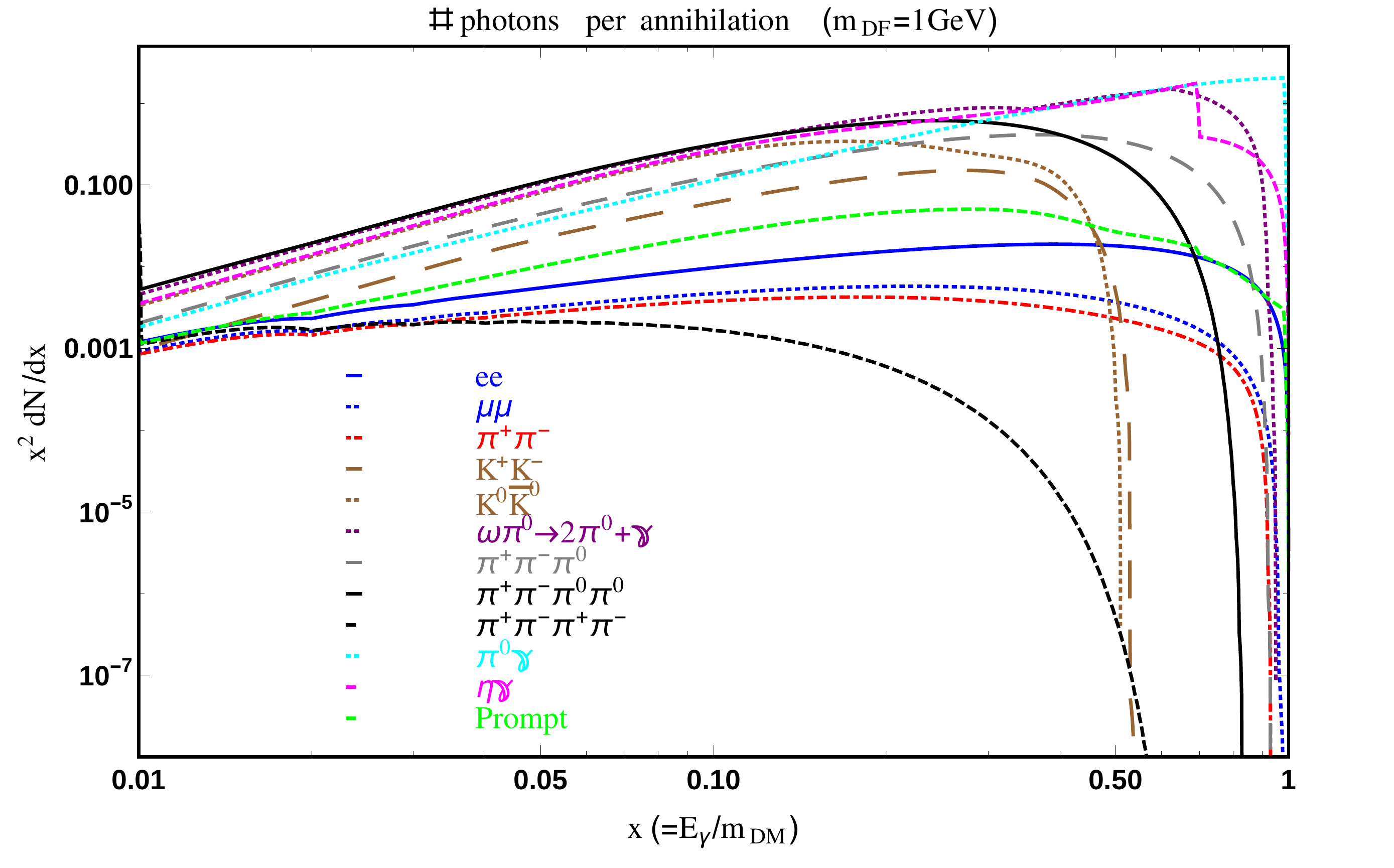} %
\includegraphics[width=0.45\textwidth]{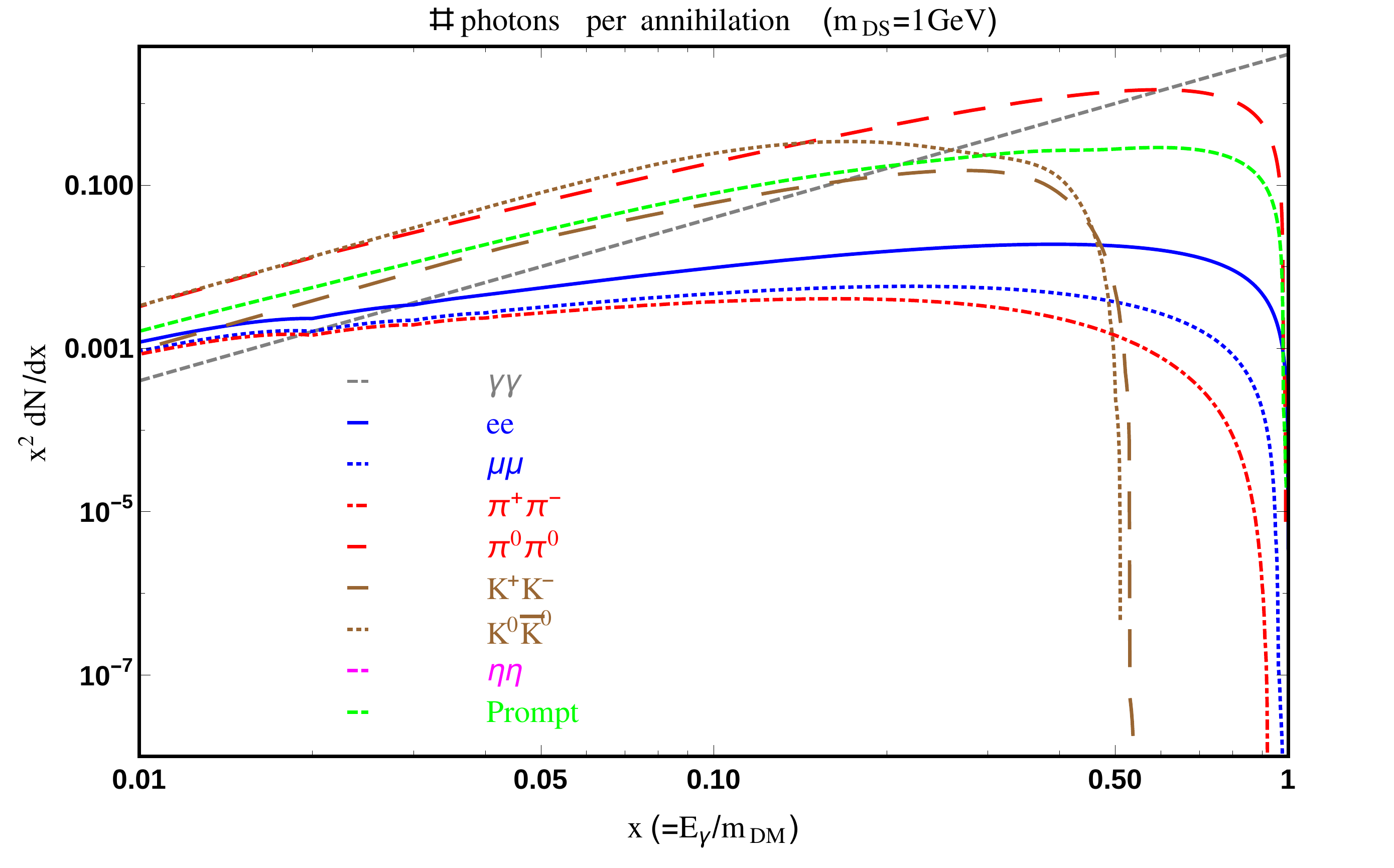} %
\\
\includegraphics[width=0.45\textwidth]{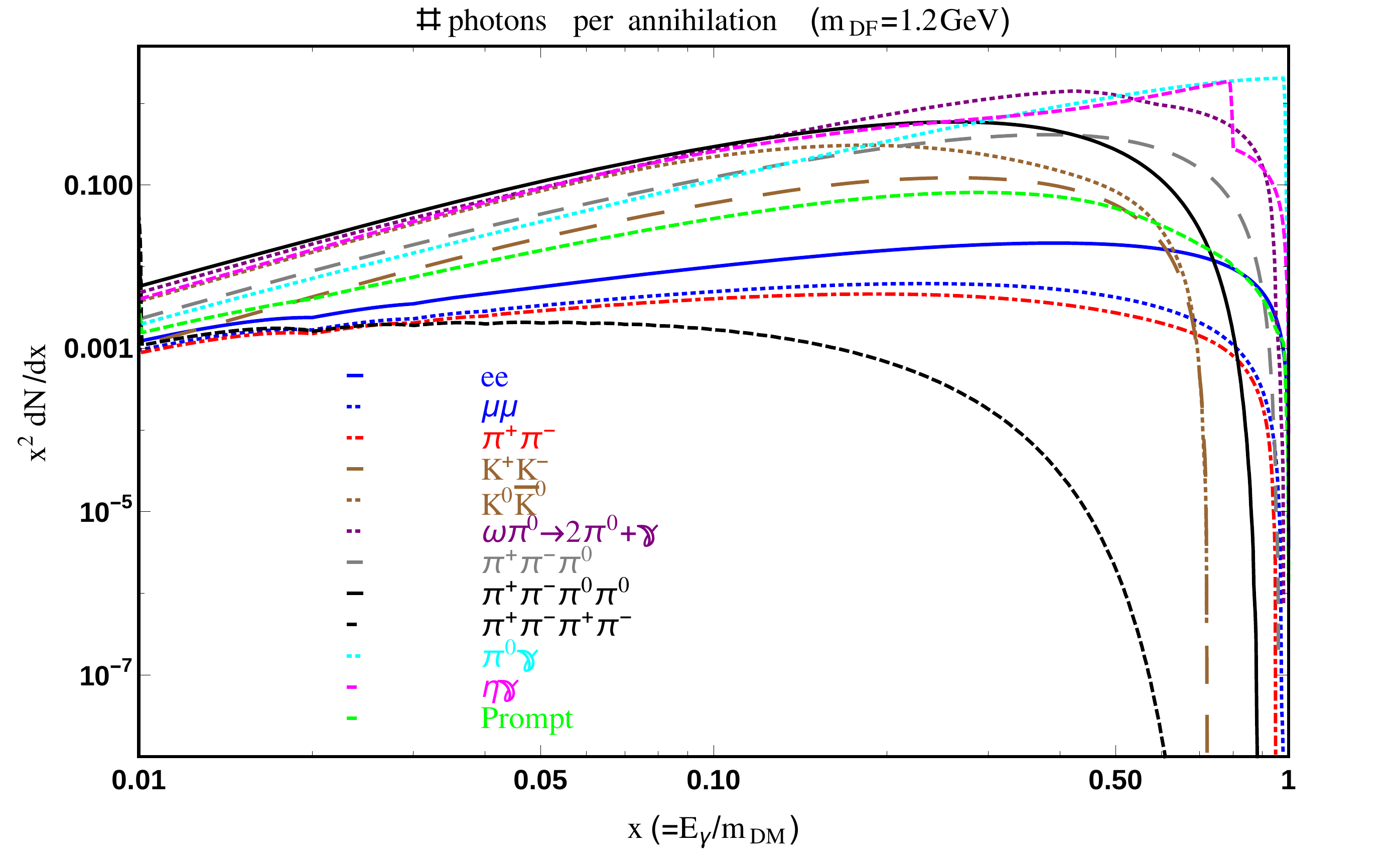} %
\includegraphics[width=0.45\textwidth]{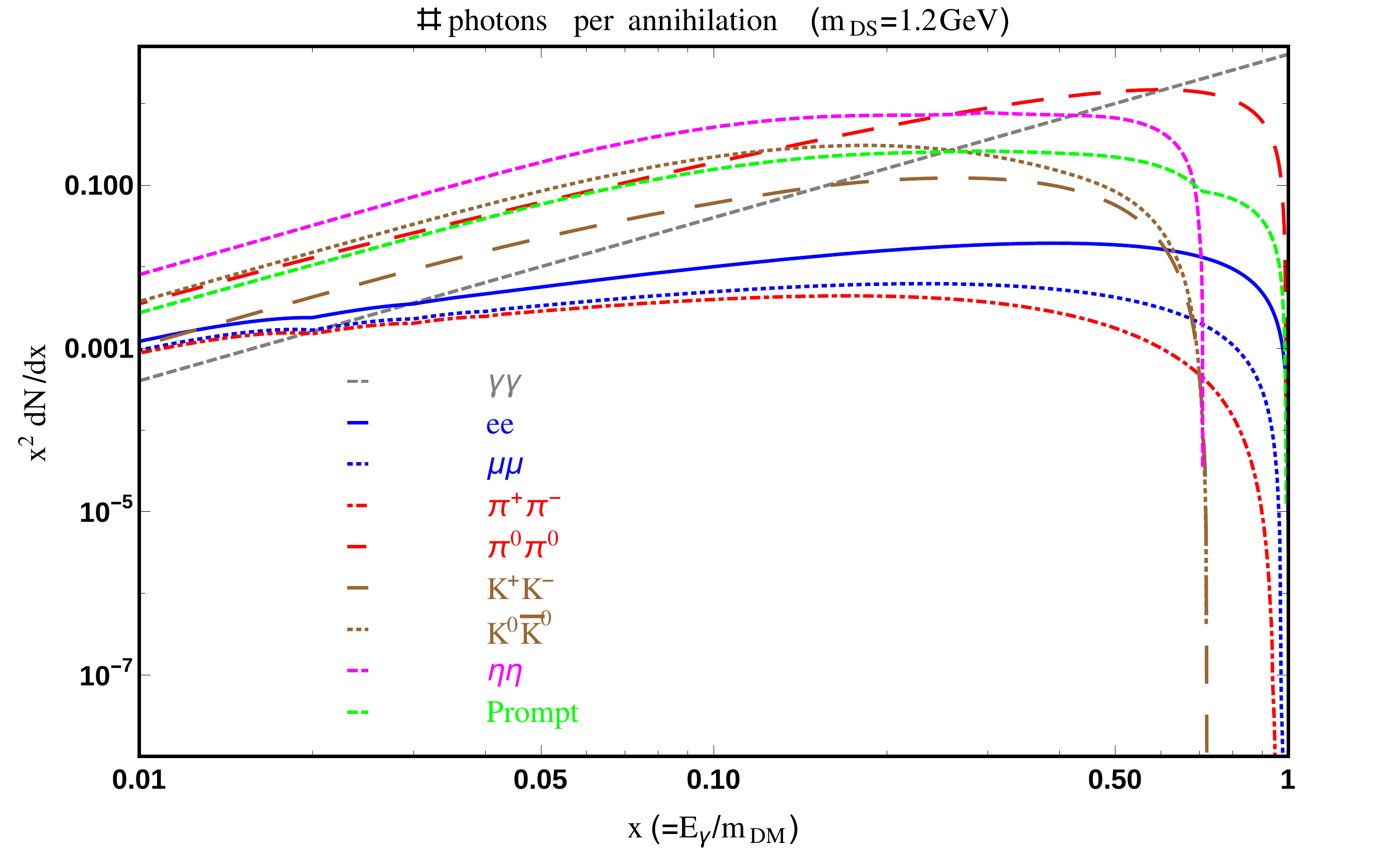} %
\caption{The photon distribution $x^2dN/dx$ for dark photon (\textit{left panel}) and dark scalar (\textit{right panel}) in the lab frame. The prompt photon means summing all the channels according to BR. \label{fig:x2dNdx}}
\end{figure*}

We plot the photon distribution $x^2dN/dx$ for dark photon and dark scalar in the lab frame in Figure \ref{fig:x2dNdx}. It is clear that those channels with direct photons are dominant. In the channel $\eta \gamma$ on the left panel, there is a kink structure from direct photon and continuous photon from $\eta$ decay. In the Kaon channel, one can see that the photon spectrum for $1\gev$ is different from $1.2\gev$, because two Kaon mass is close to $1\gev$ and have mass threshold effect in the equation \ref{eqn:fullboost}. The other channels like $\pi \pi$ are not affected by the mass difference.

%% file: spectrum2.tex
\section{ Photon Spectrum from different final states}
\label{App::photonS}

Here we will present the photon spectrum from Final State Radiation (FSR),
three-body, four-body final states, etc.

\subsection{ FSR}
\label{subsec:FSR}

FSR from charged fermionic pairs and charged bosonic pair should be treated
separately.
For $\phi$ decay to the bosonic field, as an example of $\pi^+ + \pi^-$, the composite
structure of $\pi^\pm$ brings the FSR computation some theoretical
uncertainties~\cite{Hoefer:2001mx,Gluza:2002ui}, which will be neglected here. Hence, the scalar
QED is employed to derive FSR spectrum.
To study fermionic fields or other bosons, the mass of the particles should replace the pion
mass $m_\pi$, and other changes needed is written below.

In general, the FSR spectrum is divided into three parts: the spectrum from
hard photon $\delta^H$, an exponential part taking into account the soft
mutli-photon emission $B_\pi x^{-1 + B_\pi}$ and the virtual photon correction
to the soft photon emission $\delta^{V+S}$.
The photon spectrum from boson FSR $\frac{\d N}{\d x_0}$ and fermion FSR
$\frac{\d N_f}{\d x_0}$ in the rest frame of $\phi$
are written as follows,
\begin{eqnarray}
   \frac{\d N} { d x_0} &=&    \delta^H ( x) +
      \left( 1 + \delta^{V+S} \right) x^{ B_\pi (x) -1} B_\pi (  x)
   \label{eq:FSR_b0}
   \\
   \frac{\d N_f} { d x_0} &=&    \delta^H_f ( x) +
      \left( 1 + \delta^{V+S}_f \right) x^{ B_\pi (x) -1} B_\pi (  x)
\end{eqnarray}
where
\begin{eqnarray}
   \delta^H ( x) &=&  \frac{\alpha}{\pi}  \frac{ 2 x \beta_\pi^\prime }
         { \beta_\pi^3  }
\\
   \delta^H_f ( x) &=& \frac{\alpha}{\pi} \frac{2 x}{ 3 - \beta_\pi^2} \frac{\beta_\pi^\prime }
         { \beta_\pi  } \left[ - 1 + \frac{1}{ \beta_\pi^\prime}
         \ln \left( \frac{ 1+ \beta_\pi^\prime }{ 1- \beta_\pi^\prime} \right)
         \right]
\\
   B_\pi (x) &=& \frac{\alpha}{\pi}\frac{ 2 ( 1-x) \beta_\pi^\prime} { \beta_\pi}
      \left[
      \frac{ 1+ {\beta_\pi^\prime}^2 }{ 2 \beta_\pi^\prime}
      \ln \left( \frac{1 + \beta_\pi^\prime} { 1- \beta_\pi^\prime} \right) -1
      \right]
\\
   \delta^{V+S}  &=& \frac{\alpha}{\pi} \left\{ \frac{  2 + \beta_\pi^2}{ \beta_\pi}
      \ln \left( \frac{ 1+ \beta_\pi} {1- \beta_\pi} \right)  -2
      - 2 \ln \left( \frac{ 1 - \beta_\pi^2 }{4} \right)
      -\frac{  1+ \beta_\pi^2}{ 2\beta_\pi}
      \left[
         \ln \left( \frac{ 1+ \beta_\pi }{ 1- \beta_\pi}\right)
           \right. \right.
\nonumber\\
      &&
      \left. \left.
            \ln \left( \frac{ (1+ \beta_\pi) \beta_\pi }{2} \right)
      +
       \ln \left( \frac{ 1+ \beta_\pi }{ 2 \beta_\pi}\right)
       \ln \left( \frac{ 1- \beta_\pi }{ 2 \beta_\pi}\right)
         + 2 \mathrm{Li}_2 \left( \frac{ 2 \beta_\pi }{ 1+ \beta_\pi}\right)
           \right. \right.
\nonumber\\
      &&
      +
      \left. \left.
          2 \mathrm{Li}_2 \left( -\frac{ 1- \beta_\pi }{ 2 \beta_\pi}\right)
         -\frac{ 2 }{3} \pi^2
      \right] \right\}
\\
   \delta^{V+S}_f &=& \delta^{V+S}  - \frac{\alpha}{\pi} \frac{1} {2 \beta_\pi} \ln \left(
      \frac{ 1+ \beta_\pi } { 1- \beta_\pi } \right)  \ ,
\end{eqnarray}
where $ \beta_\pi  = \sqrt{ 1 - 4 m_\pi^2 /s }$ is the pion velocity without photon radiation,
$ \beta_\pi^\prime  = \sqrt{ 1 - 4 m_\pi^2 /((1-x)s ) }$.
Notice that the soft-virtual part $\delta^{V+S}$ taking into account the
one-loop correction to $\phi \rightarrow \pi^+ + \pi^-$, does not depends
on $x$.

Boosting the spectrum $\frac{\d N_0}{\d x}$ at $\phi$ reference give the photon spectrum at the frame of DM. In the limit of $m_{\chi} \gg m_\phi$, the spectrum is
\begin{equation}
   \frac{\d N_1} { \d x} = \int_x^1 \frac{\d x_0}{x_0}\frac{ \d N_0} {\d x_0 }
\end{equation}

The above formula are derived from QED or scalar QED, which is fitted well to the analysis of
dark photon. In the case of dark scalar, the chiral perturbation theory complicates the situation,
but due to the other uncertainties, such as branching ratio, this is a good approximation as well.

\subsection{  $\pi^\pm$ and $\mu$ radiative decay }
\label{subsec:radiativedecay}

$\pi^\pm$ is close to $100 \%$ decaying to $\mu + \nu_\mu$; besides that, there is $0.2 \%$ possibility
that the radiative decay $\pi^\pm \rightarrow \mu^\pm + \nu_\mu + \gamma$ happens. Inner
Bremsstrahlung from the weak decays as the dominant process contributing to the radiative decay
are considered here, while the other
decay processes from virtual hadronic are neglected since they are subdominant~\cite{Beringer:1900zz}.
At the rest frame of $\pi^\pm$, the photon spectrum is
\begin{eqnarray}
  \frac{ \d N_{\gamma} } {\d x_{-1} } & =& \frac{ \alpha } { 2 \pi}
      \frac{1}{ \left( r-1 \right)^2 ( x-1 ) x }
      \bigg\{  -\left[ (-2 +x )^2 + 4 r ( x -1 )  \right] \left( r +x -1 \right)
      \nonumber \\
      &&+ ( x -1) \left( -2 r^2 + 2 rx + x^2 - 2 x +2 \right) \ln \frac{1-x}{r} \bigg\}
       \ ,  \quad  \quad 0 \leq x \leq (1-r)
\end{eqnarray}
where $x$ is in the range of $0 \leq x \leq 1-r$, and $ r = (m_\mu/m_\pi)^2$.
Since $m_\mu$ is not quite small relative to $m_\pi$, we cannot assume $r \simeq 0$ to boost the
spectrum. Under the assumption that $m_\phi \gg m_\pi$ and $m_\chi \gg m_\phi$, the spectra in
$\phi$ frame and DM frame have analytical solutions, and in any frame, the spectrum has
the same range $0 \leq x \leq 1-r$.
The photon spectrum in the dark photon frame, $\frac{\d N_\gamma }  { \d x_0}$, and in the
DM rest frame, $\frac{\d N_\gamma }  { \d x_1}$, from the process of $\phi\rightarrow \pi^+ + \pi^- + \gamma$
can be derived,
\begin{eqnarray}
\frac{\d N_\gamma }  { \d x_0} &= &
      \frac{\alpha} { 2 \pi ( -1 +r )^2 x } \bigg\{
        -2 \left(-2 + 2r - x\right) \left(-1 + r + x\right) + 4  r^2 x \left( \tanh^{-1}(1 - 2r)
     \right.
   \nonumber\\
      &&+
      \left.
       \tanh^{-1}(1 - 2x) \right)
       \left(2x - 4 r x \right) \ln{\frac{1 - r}{x}} + \left[-2 + 2 r^2 + x - r x + x^2
     \right.
   \nonumber\\
      &&
     \left.
         + 2 ( -1 + r ) x \ln{x}\right]
      \ln{\frac{r}{1 - x}} +
           2 (-1 + r) x \left[\mathrm{Li}_2(r) - \mathrm{Li}_2(1 - x)
         \right] \bigg\}
\\
\frac{\d N_\gamma }  { \d x_1} &= &
\frac{\alpha}{ 12 \pi ( - 1 + r)^2 x} \left\{ -24 ( -1+r )^2
   + \left[ -42 (-1+r) - \pi^2 ( -1+r + 2 r^2 ) \right]   x
   -18 x^2
   \right.
   \nonumber\\
   &&+  \left.
    24 r^2 x \tanh^{-1} ( 1- 2x)
   - 12 \ln \frac{r}{1-x} -24 r^2 x \tanh^{-1} ( 1- 2r) ( -1 + \ln x )
   \right.
   \nonumber\\
   &&  \left.
   + 2 \big[ 3 ( -1 + r )   x \ln^2 ( 1-r ) ( -1+ r + \ln r)
   + 3 \left( 2 r^2 + 3x + x^2 \right) \ln \frac{ r} { 1-x}
   \right.
   \nonumber\\
   &&  \left.
   + x \left( \pi^2 ( -1 + r) + 6 r + 3 ( -1+r ) \ln r \right) \ln x
   + 3 ( -1 +r ) ( -1 + r - \ln r ) x \ln^2 x
   \right.
   \nonumber\\
   &&  \left.
   + x \ln ( 1-r ) (- \pi^2 ( -1 +r ) - 6r + 6 ( -1 +2r ) \ln x ) \big]
   + 6 x \big[ ( -1 +r + 2 r^2
   \right.
   \nonumber\\
   &&  \left.
   + 2 ( -1+r ) \ln \frac{1-r}{x} ) \mathrm{Li}_2 (r)
   + ( -1+r + 2 r^2) \mathrm{Li}_2 (x) +
   2 ( -1+r ) \left( \mathrm{Li}_3 ( 1-r)
   \right. \right.
   \nonumber\\
   &&  \left. \left.
   - \mathrm{Li}_3 (x) \right)
   \big]
   \right\}
\end{eqnarray}
The pion radiative decay formula can apply to Kaon directly, but its gamma ray
spectrum from radiative decay is negligible due to $\pi^0$ from Kaon decay.

If the final states are $\mu^+ + \mu^-$, the Branching ratio of
$\mu \rightarrow e^-  \bar{\nu_e} \nu_\mu \gamma$ is
$ ( 1.4 \pm 0.4 ) \%  $, which is one order magnitude larger than the branching ratio of
$\pi^\pm$ radiative decay. The photon spectrum in different frame are listed as follows,
\begin{eqnarray}
  \frac{ \d N } {\d x_{-1}  } &=& \frac{ \alpha} { 3 \pi } \frac{1-x}{x}
      \bigg\{     \left( 3 - 2 x + 4 x^2 - 2 x^3\right)
      \ln \frac{1}{r}  +
      \bigg[  -\frac{17}{2}  + \frac{23}{6}x  - \frac{101}{12} x^2
      + \frac{55}{12}x^3
      \nonumber\\
     &&
      + \left( 3 - 3x + 4 x^2 - 2 x^3 \right) \ln ( 1-x)
      \Big]
      \bigg\}
\\
   \frac{ \d N } {\d x_0} &=& \frac{\alpha} { 3 \pi} \frac{1}{x} \bigg\{
      \left( 3 + \frac{2}{3} x - 6 x^2 + 3 x^3 - \frac{2}{3} x^4 + 5 x \ln x \right)
      \ln \frac{1}{r}  +
      \Big[- \frac{17}{2} - \frac{3}{2} x + \frac{191}{12} x^2
   \nonumber\\
      &&
      - \frac{ 23}{3} x^3
      + \frac{7}{4} x^4
       + \left( 3 + \frac{2}{3} x - 6 x ^2 + 3 x^3 -
      \frac{2}{3} x^4 \right)  \ln ( 1-x)
       - \frac{ 28}{3} x \ln x
   \nonumber\\
     &&
         + 5 x \ln ( 1-x) \ln x + 5x \mathrm{Li}_2 (1-x)
         \Big]
      \bigg\}
\\
   \frac{ \d N } {\d x_1} &=& \frac{\alpha} { 3 \pi} \frac{1}{x} \bigg\{
    \left(3 - \frac{139 }{18}x  + 6 x^2 - \frac{3 }{2}x^3 + \frac{2 }{9} x^4- \frac{2 }{3} x \ln{x}
      - \frac{5 }{2} x {\ln^2{x}}\right) \ln \frac{1}{r}  +
         \Big[ - \frac{19}{2} +
   \nonumber\\
       &&
         \Big( \frac{2735}{108}  - \frac{\pi^2}{9}
         -5 \zeta(3) \Big) x
         - \frac{743}{36} x^2+ \frac{161}{36} x^3 - \frac{71}{108} x^4
        +\left( 3 - \frac{139}{18} x  + 6 x^2 - \frac{3}{2} x^3
      \right.
   \nonumber\\
      &&
      \left.
      +  \frac{2}{9} x^4 \right) \ln (1-x)
         +  \left(  \frac{9}{2} x -  \frac{5 \pi^2}{6} x \right) \ln{x}
         +   \frac{14}{3} x \ln^2{x} +
     -  \frac{2}{3} x \mathrm{Li}_2(x)+ 5 x \mathrm{Li}_3 (x) \Big]
      \bigg\}
\end{eqnarray}
where $r = \frac{m_e^2}{m_\mu^2}  \ll 1 $, and the range of $x$ is $(0,1)$ which does not depends on $r$
since $r$ is negligible.

\subsection{ n-body final states }
\label{subs:nbody}

Here we study the energy spectrum from the process of $\phi$ decay to $n$ particles. As $n = 2$, the photon spectrum is
a delta function, which is determined by kinematics. Whereas $n \geq 3$, the phase space integral and matrix elements
will influence the shape of spectrum.
The energy spectrum for $n$-body final states can be easily applied to $\phi \rightarrow \pi^+ \pi^- \pi^0 $,
$\phi \rightarrow \pi^+ \pi^- \pi^0 \pi^0 $ and $\phi \rightarrow \pi^+ \pi^- \pi^+  \pi^-$.

The n-body phase space integration $R_n ( s) $ is computed by a recursion relation~\cite{Byckling:1969sx,Kersevan:2004yh},
and assuming matrix element constant, the energy spectrum can be computed by the phase space integral,
The recursion relation of $R_n$ is written as,
\begin{eqnarray}
   R_n(s ) &=&  ( 4 \pi )^{n-1} \times  \int_{( m_1 + ...+ m_{n-1})^2}^{( \sqrt{s} - m_n)^2}  \d M_{n-1}^2
      \frac{ \sqrt{ \lambda ( s, M_{n-1}^2, m_n^2 ) }} { 8 s}
   \nonumber \\
   &&
   \times  \int_{( m_1 + ...+ m_{n-2})^2}^{( M_{n-1} - m_{n-1})^2}  \d M_{n-2}^2
      \frac{ \sqrt{ \lambda ( M_{n-1}^2, M_{n-2}^2,  m_{n-1}^2 ) }} { 8 M_{n-1}^2}
   \nonumber \\
   &&
   \times \dots \times  \int_{( m_1 + m_{2})^2}^{( M_{3} - m_{3})^2}  \d M_{2}^2
      \frac{ \sqrt{ \lambda ( M_{3}^2, M_{2}^2,  m_{3}^2 ) }} { 8 M_{3}^2}
      \frac{ \sqrt{ \lambda ( M_{2}^2, m_{1}^2,  m_{2}^2 ) }} { 8 M_{2}^2}
\end{eqnarray}
where the angular integration is equal to the prefactor $( 4 \pi )^{n-1}$ due to the assumption of constant
matrix amplitude, and the Lorentz invariant function $\lambda  (x, y, z) \equiv  x^2 + y^2 + z^2 - 2 x y - 2 yz -
2 z y$.
The energy spectrum of the $n$-th final states can be derived,
\begin{equation}
 \frac{\d N } { \d x } = \frac{1}{R_n} \frac{ \d R_n }{   \d x}  = \frac{ s }{ R_n} \frac{ \d R_n}{ \d M_{n-1}^2}
   \label{eq:dndxR}
\end{equation}
In the study of dark photon decaying to $3 \pi$ or $ 4 \pi$, we did not take the limit of $m_\pi$ to zero, since
the $\mathcal{O}(1)~\mathrm{GeV}$ dark photon mass is close to $pion$ mass, but if we set the masses of all the final states
to zero, eq.~(\ref{eq:dndxR}) has an analytical solution, \
\begin{equation}
   \frac{\d N} { \d x } = \left( n - 1 \right) \left( n- 2 \right) \left( 1- x \right)^{n-3} x    \ .
\end{equation}

\subsection{ photons from individual channels}
\label{subsec:channels}

The photon spectra are computed channel by channel. We will briefly mention the method
to obtain the spectrum for the different channels. With no explicit mention of the dark force $\phi$, we refer to  both
dark photon and dark scalar.

\begin{itemize}
   \item $\phi \rightarrow e \bar{e}  $, photon from electron FSR are considered

   \item $\phi \rightarrow \mu \bar{\mu}  $, photon from muon FSR and radiative decay

   \item $\phi_\mu \rightarrow \pi^+ \pi^- $, from pion FSR including hard photon spectrum $\delta^H(x)$
      in eq.~(\ref{eq:FSR_b0}) and $\pi^\pm$ radiative decay. In the radiative decay, the form factor are
      neglected.

   \item $\phi_0 \rightarrow \pi^+ \pi^- $, photon from pion FSR not including hard photon spectrum $\delta^H(x)$
      in eq.~(\ref{eq:FSR_b0}) and $\pi^\pm$ radiative decay. No including the hard photon spectrum is
      due to the fact that it mainly comes from the interaction term $A_\mu A^\mu \pi^+ \pi^-$, not for scalar mediator.

   \item $\phi_0 \rightarrow \pi^0 \pi^0$. $98.82~\%$ of pion cascade decays to $2 \gamma$. The
      photon spectrum of the $\pi^0$ decay in different frames are written as,
      \begin{eqnarray}
         \frac{\d N}{\d x_{-1} } &=&  2 \delta \left( 1- x \right)
         \\
         \frac{\d N}{\d x_{0} } &=&  \frac{2} { \sqrt{1- \epsilon_0^2} } \ , \quad \quad
         \left( \frac{1 - \sqrt{ 1- \epsilon_0^2}}{2} < x<  \frac{1 + \sqrt{ 1- \epsilon_0^2}}{2} \right)
         \\
         \frac{\d N}{\d x_{1} } &=&
               \left\{
               \begin{array}{lr}
                -\frac{2} { \sqrt{1- \epsilon_0^2} } \ln \frac{ 2 x}{1+ \sqrt{1- \epsilon_0^2} }  \ ,  & \quad \quad
         \left( \frac{1 - \sqrt{ 1- \epsilon_0^2}}{2} < x<  \frac{1 + \sqrt{ 1- \epsilon_0^2}}{2} \right)
                         \\
              \frac{2} { \sqrt{1- \epsilon_0^2} } \ln
                  \frac{ 1+ \sqrt{1- \epsilon_0^2} } {1- \sqrt{1- \epsilon_0^2} }  \ ,  & \quad \quad
                \left( 0  < x<  \frac{1 - \sqrt{ 1- \epsilon_0^2}}{2} \right)
               \end{array}
               \right.
      \end{eqnarray}
   where $\epsilon_0 = \frac{ 2 m_{\pi^0}}{m_\phi}$, and $\epsilon_1 = \frac{m_\phi}{m_\chi} \simeq 0$.

   \item $\phi \rightarrow K^+ K^-$. $20.66 ~ \%$ of kaon decaying to hadronic modes
      $K^+ \rightarrow \pi^+ + \pi^0 $ are major contribution. Due to the small branching ratio of
      $\phi \rightarrow K^+ K^-$, this process is the only one considered here.
      In the leptonic channel, $K^+ \rightarrow \pi^0 e^+ \nu_e $ and  $K^+ \rightarrow \pi^0 \mu^+ \nu_\mu $
      are suppressed by the smaller branching ratio and three-body phase space.

   \item $\phi \rightarrow K^0 \bar{K}^0 $, or we can think it as $\phi$ decays to CP even $K_S^0$ and CP odd $K_L^0$.
     For $K_S^0$, the photon yield originates from the modes of $\pi^0$. $K_S^0 \rightarrow \pi^0 \pi^0$ with the branching
      ratio $30.69~\%$. For $K_L^0$,
      $K_L^0 \rightarrow \pi^0 \pi^0 \pi^0 ,  19.52~\%$, $K_L^0 \rightarrow \pi^+ \pi^- \pi^0 ,  12.54~\%$.
      Photon from $\pi^\pm$ are not included here.

   \item $\phi \rightarrow \omega \pi^0 \rightarrow 2 \pi^0 + \gamma$. The second $\rightarrow$ means that we consider one
      modes of the $\omega$ decay. Due to some experimental reason, the mode of $\omega \rightarrow \pi^+ \pi^- \pi^0$, with $89.2 ~\%$ BR
      are included in the 4$\pi$ final states. Since these process are the process with two body final states, we
      can use kinematics to derive the photon spectrum.

   \item $\phi \rightarrow \pi^+ \pi^- \pi^0 $. Following \ref{subs:nbody}, We assume the scattering matrix element is
   constant and the photon from $\pi^0$ are considered.

   \item $\phi \rightarrow \pi^+ \pi^- \pi^0 \pi^0 $. Assume the scattering matrix element is
   constant and the photon from $\pi^0$ are considered.

   \item $\phi \rightarrow \pi^+ \pi^- \pi^+ \pi^- $. Assume the scattering matrix element is
   constant and the photon from $\pi^\pm$ radiative decays are considered.

   \item $\phi \rightarrow \pi^0 \gamma$. Two body final states.

   \item $\phi \rightarrow \eta \gamma $. $\eta \rightarrow \gamma \gamma, 39.31~\%$,
     $\eta \rightarrow \pi^0 \pi^0 \pi^0, 32.56~\%$,
     $\eta \rightarrow \pi^+ \pi^- \pi^0, 22.73~\%$. For the three body final states decay of $\eta$, constant
      matrix element are assumed, and photon from $\pi^\pm$ are neglected.

   \item $\phi \rightarrow \eta \eta $. The photon from $\eta$ decay is the same as the treatment in the process of
      $\phi \rightarrow \eta \gamma $. With the photon in the $\eta$ frame, we can boost it to the $\phi$ and DM frame.

\end{itemize}

\section{ Electron Spectrum Calculation}
\label{sec:e-channels}

The electron spectra are calculated channel by channel. We start with the electron spectrum for muon at rest. In SM, the unpolarized muon has the following electron spectrum in muon rest frame,

\begin{align}
dN_{e ^ \pm  } /dx = 2 x^2 (3 - 2x)
\end{align}

where $x \equiv 2E_e /m_\mu $. We have neglect the electron mass in the spectrum. As long as we know the electron spectrum in daughter particle frame, we do boost accordingly to get the spectrum in the lab frame, similar as in photon spectrum. For example, the dark matter annihilating directly into a pair of muon, the electron spectrum in lab frame is

\begin{align}
dN^{lab:2\mu}_{e ^ \pm  } /dx_2  = \frac{1}{3}\left( {4x_2^3  - 9x_2^2  + 5} \right)
\end{align}

, where $x_2  \equiv E_e^{lab} /m_{DM}$. The calculation uses the boost formula in equation \ref{eqn:boost2lab}. If we neglect the daughter particle mass at each step, we can have analytic expression for the cascade decay to four muon.

\begin{align}
dN_{e^ \pm  }^{lab:4\mu } /dx_2  = \frac{1}{9}\left( { - 8x_2^3  + 27x_2^2  - 30Log(x_2) - 19} \right)
\end{align}

\begin{figure*}[t]
\includegraphics[width=0.45\textwidth]{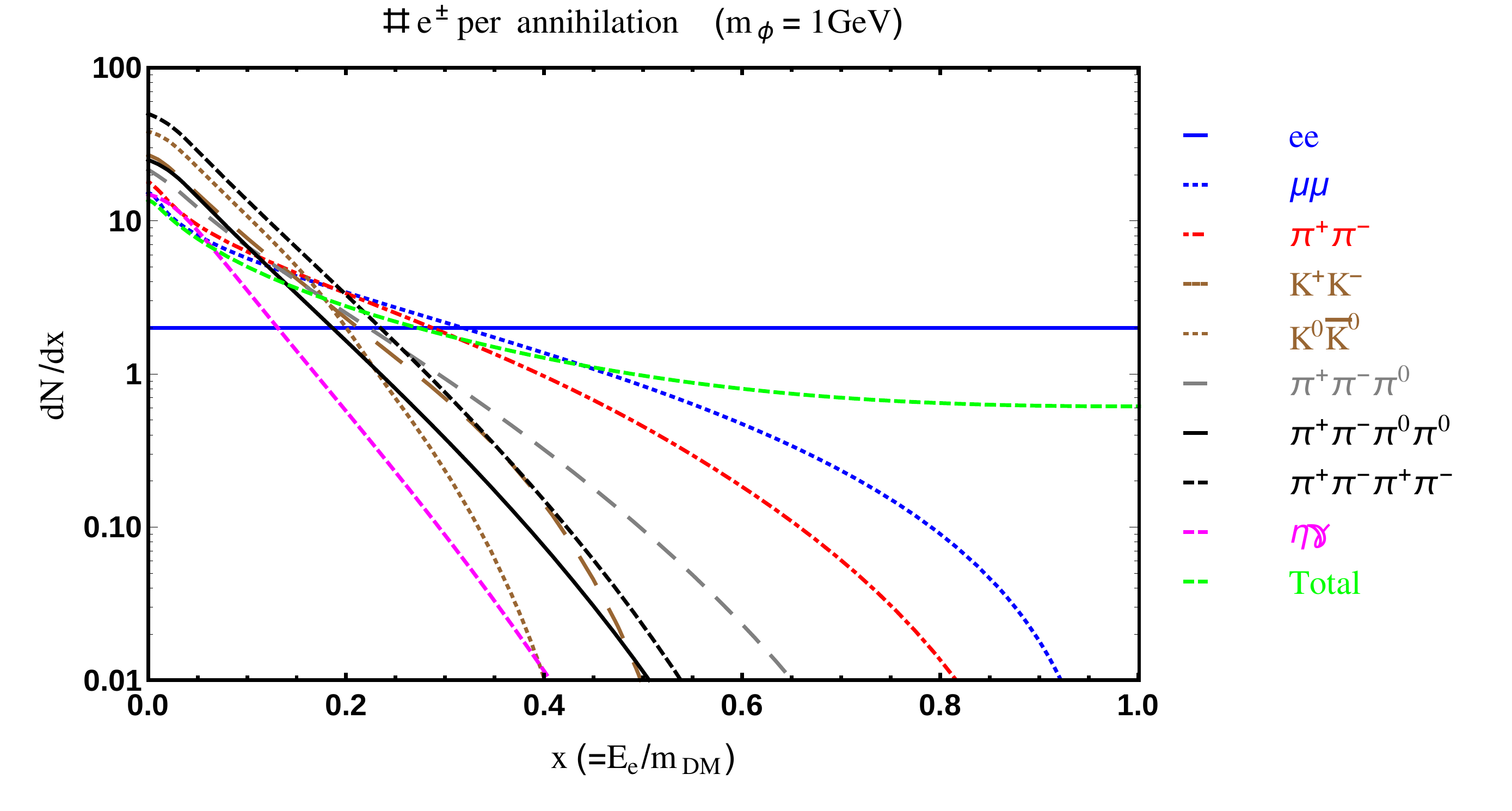} %
\includegraphics[width=0.45\textwidth]{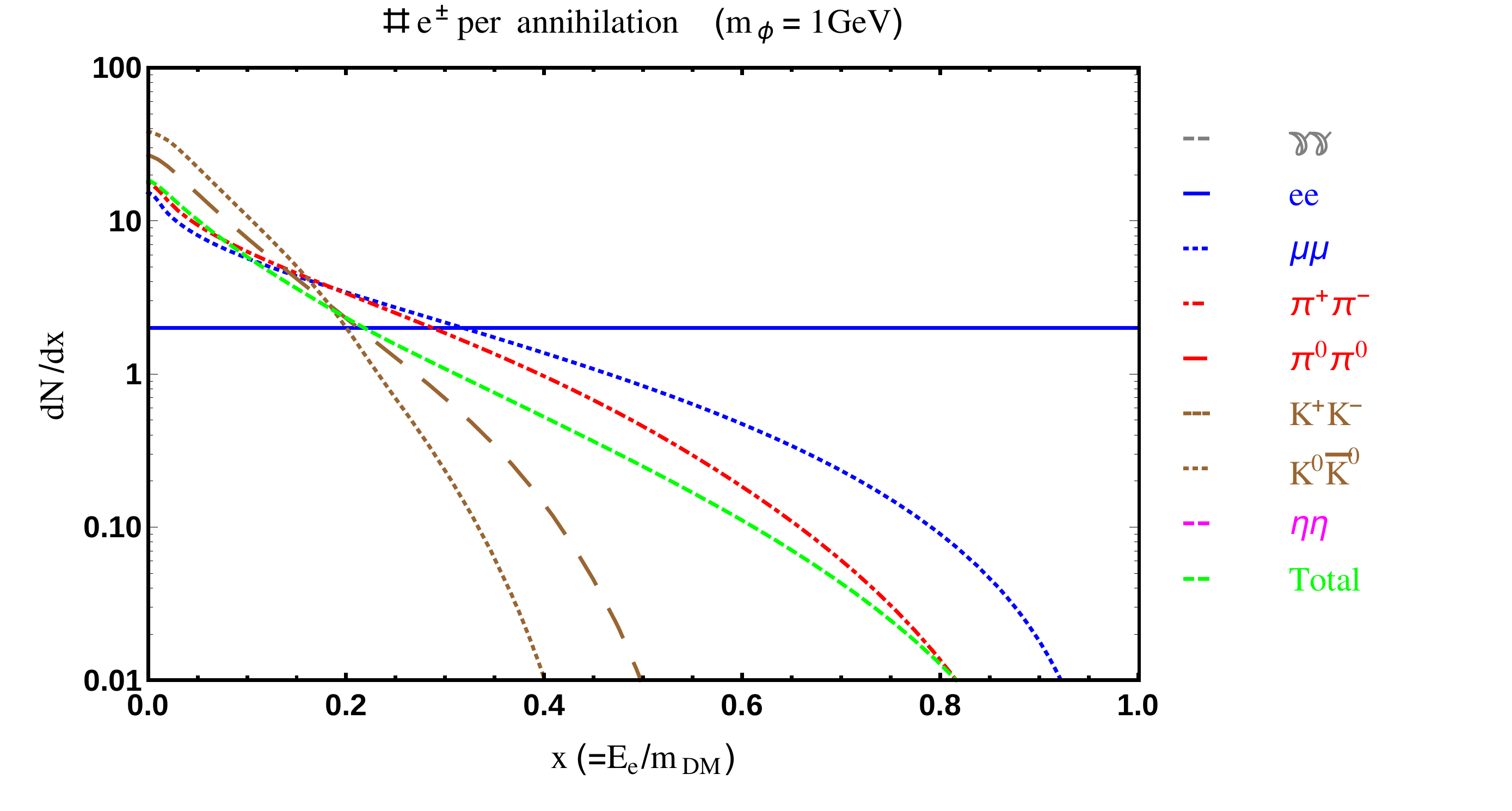} %
\\
\includegraphics[width=0.45\textwidth]{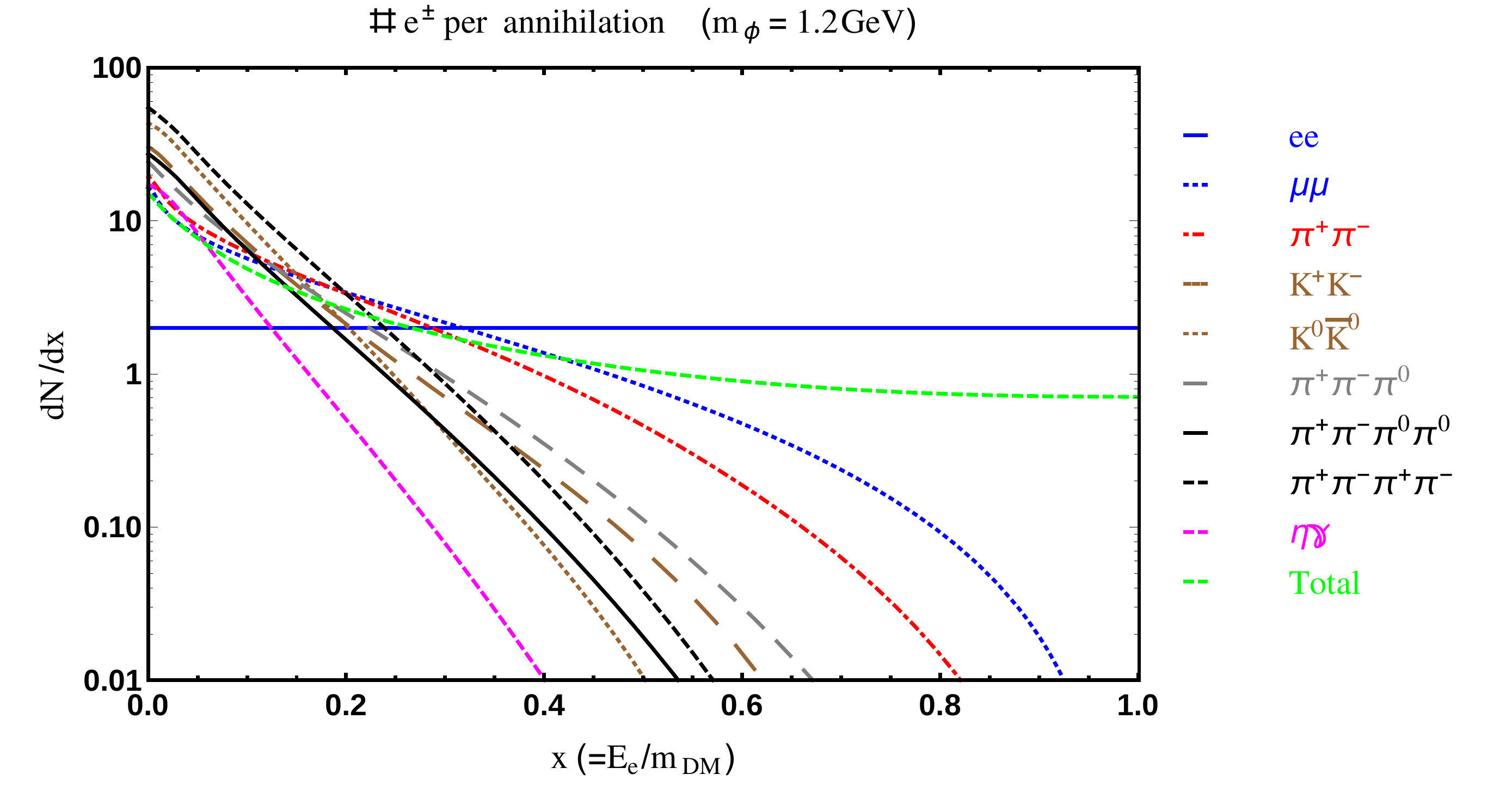} %
\includegraphics[width=0.45\textwidth]{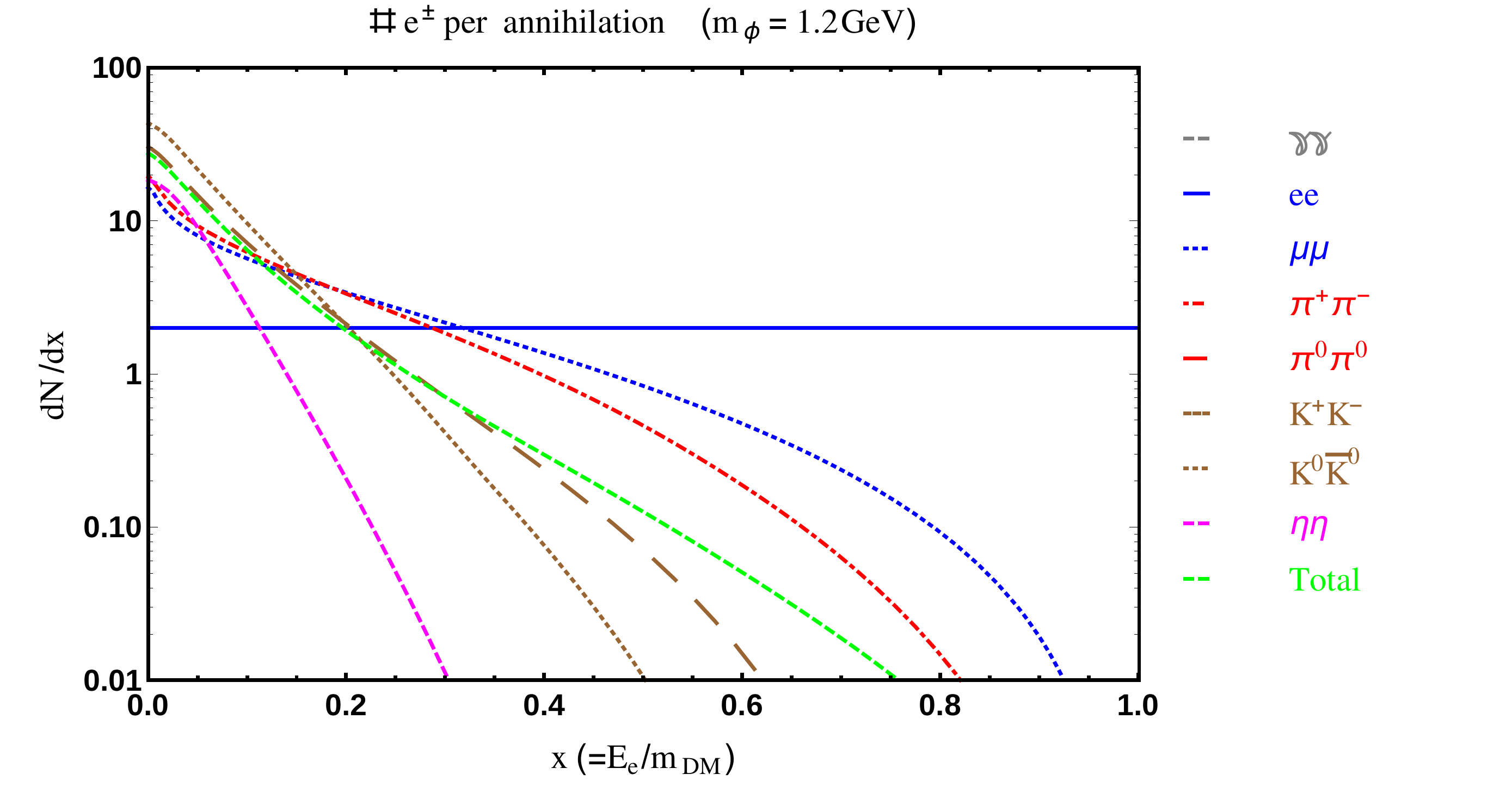} %
\caption{The electron distribution $dN/dx$ for dark photon (\textit{left panel}) and dark scalar (\textit{right panel}) in the lab frame. \label{fig:especdNdx}}
\end{figure*}

Then we briefly introduce how we get the electron spectrum for other particles. For ${\pi ^ +  }$, the decay to $\mu ^ +   + \nu _\mu$ is about $99.9877\%$, while the rest is to $e^ +   + \nu _e$. We boost the electron from muon and also add the electron from the direct decay into the electron spectrum. For ${\pi ^ 0  }$, the decay to $e^ +  e^ -  \gamma$ is quite small, about $1.17\%$. We neglect electron from ${\pi ^ 0  }$, because in most of the decay channels, ${\pi ^ 0  }$ are produced with ${\pi ^ \pm  }$ at similar rate or even smaller. For $K^{\pm}$, there are seven decay channels relevant for electron spectrum, with $\pi ^ \pm $, $\pi ^ 0 $, $\mu^{\pm}$ and $e^\pm$ in the final states. We properly boost all the electron from the daughter particles, except $\pi ^ 0 $ which is neglected in the calculation. For $K^0$ and $\eta$, the calculation is the same as $K^{\pm}$. For $3\pi$ and $4\pi$ final states, we use the natural phase space and only count the electrons from ${\pi ^ \pm  }$.

We plot the electron distribution $dN/dx$ for dark photon and dark scalar in the lab frame in Figure \ref{fig:especdNdx}. The Kaon channel has different electron spectrum for $1\gev$ and $1.2\gev$, due to dark mediator mass is close to two Kaon mass. The electron spectrum mainly comes from $e^+ e^-$ at high energy for dark photon, but not for dark scalar. The dark scalar has smaller electron spectrum than dark photon due to small $e^+ e^-$ BR.


\section{CMB Limits on thermal cross-section}
\label{sec:CMBthermal}

\begin{figure*}[t]
\includegraphics[width=0.45\textwidth]{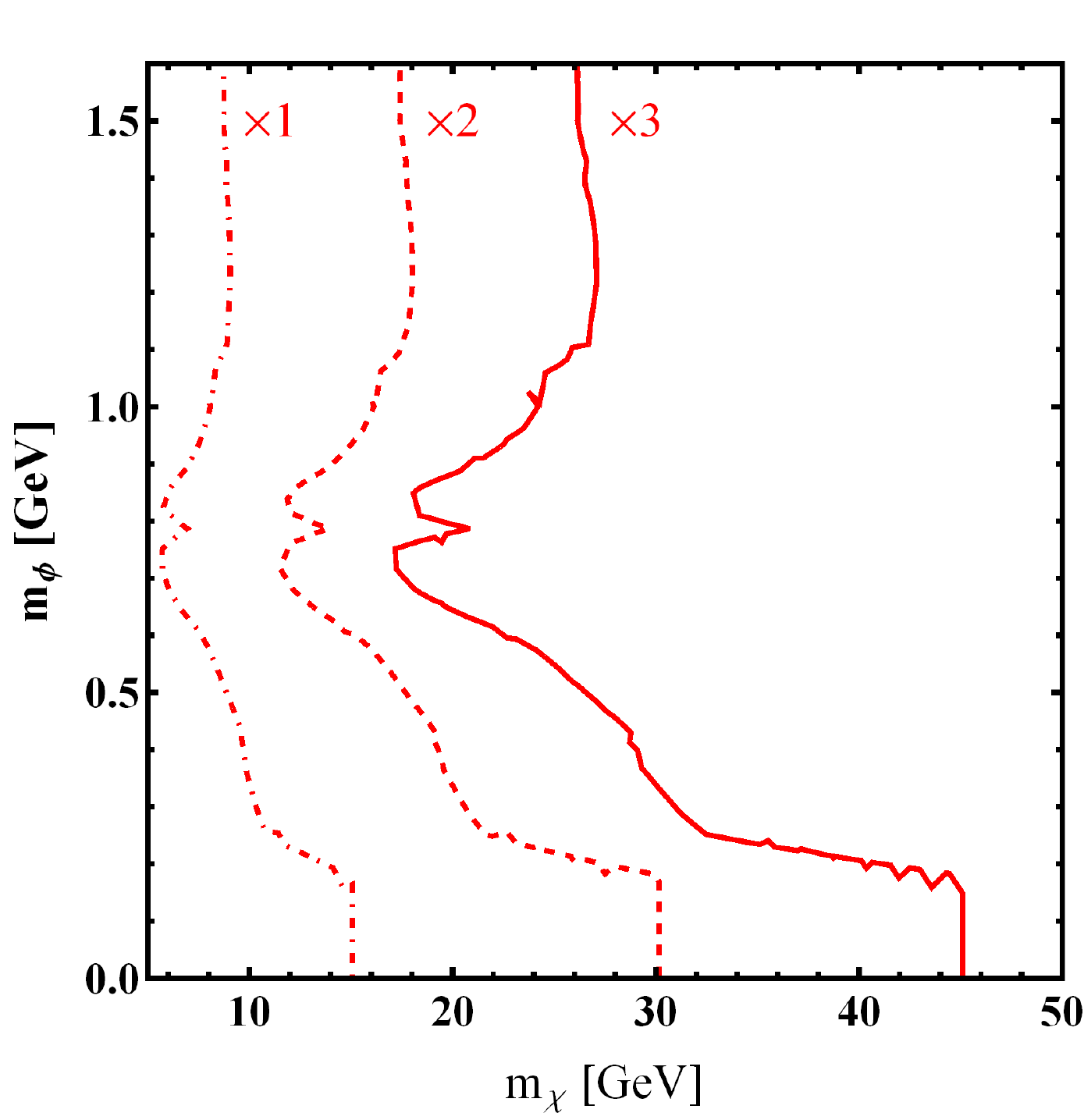} %
\includegraphics[width=0.45\textwidth]{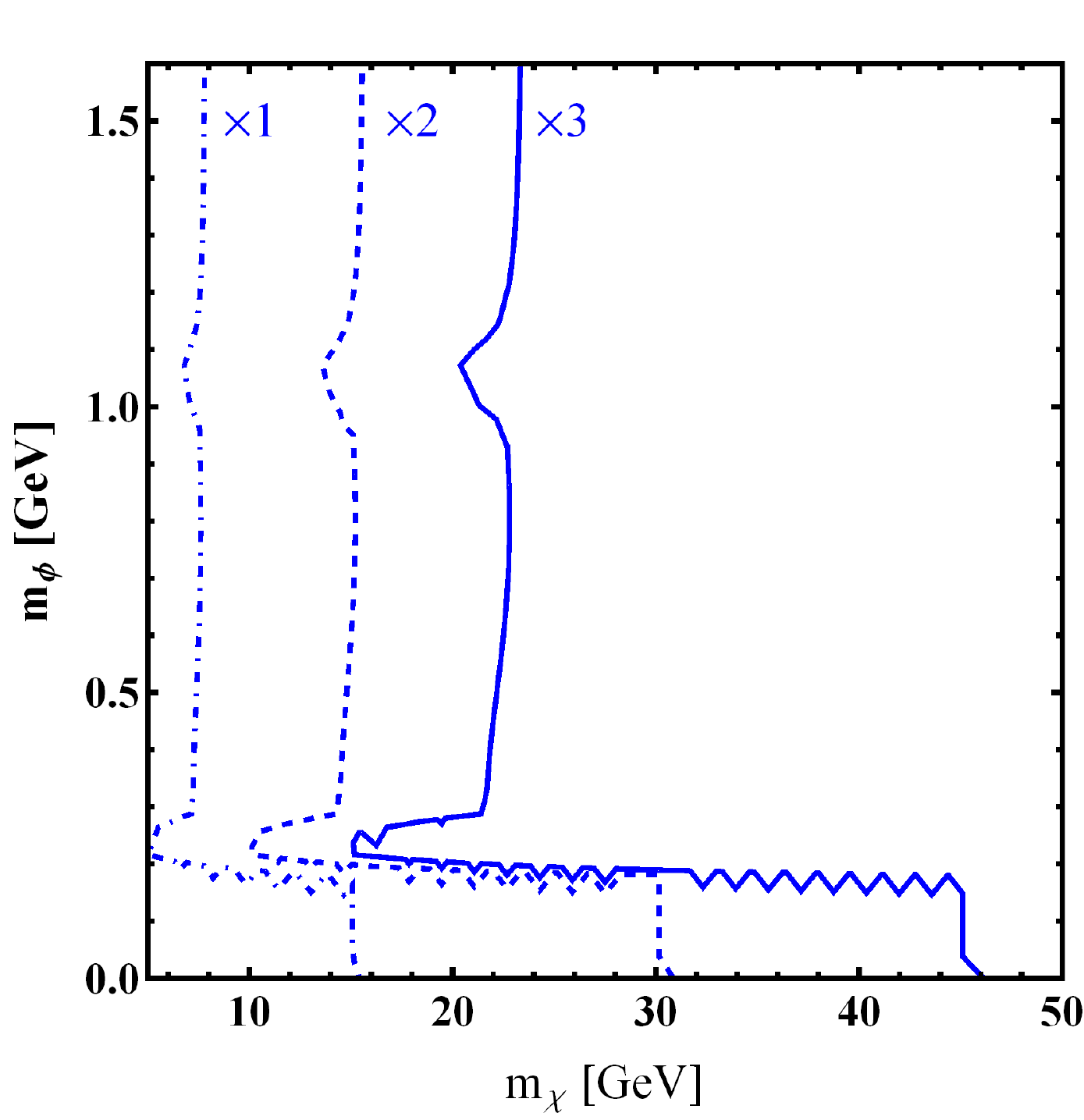} %
\caption{The contours of excluded annihilation cross-section at freeze-out from Plank as a function of $m_\chi$ and $ m_\phi$. The left side of the contour is excluded. $\times 1,2,3$ denotes annihilation cross-section in units of $10^{-26} cm^3/s$.
 \label{fig:CMBthermal1}}
\end{figure*}

We plot the contours of excluded annihilation cross-section at freeze-out from Plank as a function of $m_\chi$ and $ m_\phi$ in Fig.~\ref{fig:CMBthermal1}. The contours are calculated following the formula,
\begin{align}
\left\langle {\sigma v} \right\rangle f_{eff} \left| {_{Planck} (m_\chi  )} \right./f_{eff}^\phi  (m_\phi  ) = 3 \times 10^{ - 26} cm^3 /s \, ,
\end{align}
where $\left\langle {\sigma v} \right\rangle f_{eff} \left| {_{Planck} (m_\chi  )} \right.$ is the Planck excluded $\left\langle {\sigma v} \right\rangle f_{eff} $ and $f_{eff}^\phi $ is the efficiency factor for dark mediator model. It shows DM with thermal cross-section $3 \times 10^{-26} cm^3/s$ in the dark mediator models should be larger than $\sim 20$~GeV.

